\documentclass[a4paper,english,titlepage,10.5pt,colorlinks]{scrbook}
\usepackage{amsmath, amssymb, latexsym}
\usepackage[bookmarks=false]{hyperref}
\usepackage[T1]{fontenc}
\usepackage[latin1]{inputenc}
\usepackage{lmodern} 
\usepackage{grffile}
\usepackage{pdfpages}
\usepackage{lscape}
\usepackage{multicol}
\usepackage{ae}
\usepackage{color}
\usepackage{aecompl}
\usepackage[T1]{fontenc}
\usepackage[latin1]{inputenc}
\usepackage{amsmath}
\usepackage{graphicx}
\usepackage{subfig}
\usepackage{xcolor}
\usepackage{setspace}
\usepackage{appendix}
\usepackage{amssymb}
\usepackage{amsthm}
\usepackage{wrapfig}
\usepackage{bm}
{\theoremstyle{plain}
\newtheorem*{Pro}{Proposition}
\newtheorem*{Po}{Postulation}}
\usepackage{fancybox} 
\usepackage{ifthen}

\begin{document}

\begin{titlepage}
%\vspace*{-0.8in}
\setlength{\oddsidemargin}{1cm}
\vspace*{-1in}
\enlargethispage{\baselineskip}

\title{Strong Conservation Form \\ and Grid Generation \\ in 
Nonsteady Curvilinear Coordinates \\ 
\vskip 1.0cm
\large{Numerical Methods \\ 
for Implicit Radiation Hydrodynamics in 2D and 3D}\\ }

\hspace*{6cm}
\begin{figure}[ht]
 \centering
\subfloat{\includegraphics[height = 1.2cm]{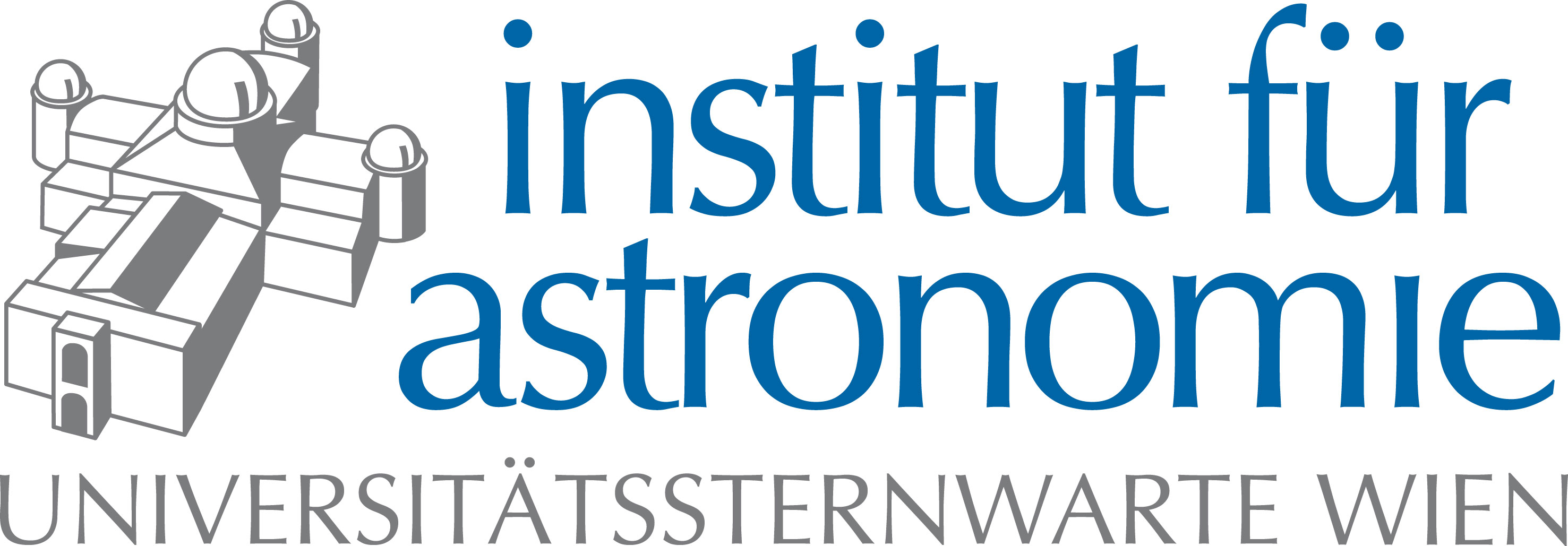}} \hspace{6cm}
\subfloat{\includegraphics[height = 1.2cm]{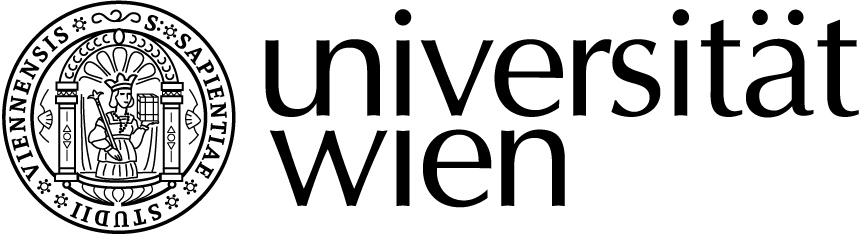}}
\end{figure}
  \vspace*{1.2cm}
 \begin{center} \sf
\Large 
\textbf{MAGISTERARBEIT}
 \\
 \vspace*{2cm}
 \normalsize Titel der Magisterarbeit \\
 \vspace*{0.5cm}
 \LARGE \bfseries  
 Strong Conservation Form \\ and Grid Generation \\ in 
Nonsteady Curvilinear Coordinates  
\vskip 0.5cm
\large{Numerical Methods \\ 
for Implicit Radiation Hydrodynamics in 2D and 3D}
 \vspace*{2.5cm}
\end{center}
\begin{center} \sf
% \small
\normalsize
 angestrebter akademischer Grad: \\
\large
 Magister der Naturwissenschaften (Mag.rer.nat.)\\
\end{center}
 \vspace*{2.5cm}
%\begin{flushleft}
 \begin{tabbing} \sf
 Verfasser: \qquad \qquad \= \sf Harald H\"oller, Bakk.rer.nat.\\
 \sf Matrikelnummer: \> \sf 0104177 \\
 \sf Studienrichtung: \> \sf Astronomie A 066 861 \\
 \sf Betreut von: \> \sf Ao.Univ.-Prof.Dr. Ernst Dorfi\\
 \end{tabbing}
%\end{flushleft}

 \hspace*{1cm} 
\begin{flushleft}
 \sf Wien, 25. M\"arz 2010
 \end{flushleft}
 \normalsize
 \vfill
\end{titlepage}

\thispagestyle{empty}
\cleardoublepage
\thispagestyle{empty}

\newpage 
\thispagestyle{empty}
\section*{Danksagung}
\begin{center}
\parbox{13cm}{
Allen Menschen, Tieren und Beh\"orden, die mich im Laufe des Studiums 
unterst\"utzt haben, m\"ochte ich herzlich danken und vielen davon 
ein Bussi geben. \\

\noindent 
Ich verneige mich vor meinem Betreuer Prof. Ernst Dorfi f\"ur die 
geduldigen und 
wegweisenden Diskussionen, seine fordernde Hartn\"ackigkeit und 
die vielen motivierenden Anekdoten aus der Welt der Wissenschaft. \\

Diese Arbeit ist \emph{Kollegen Nonius Skala} gewidmet. 

}
\end{center}

\newpage 
\thispagestyle{empty}
\section*{Erkl\"arung}

Ich erkl\"are ehrenw\"ortlich, dass ich die vorliegende Arbeit selbst\"andig 
und ohne fremde Hilfe verfasst, andere als die angegebenen Quellen nicht 
benutzt und die den benutzten Quellen w\"ortlich oder inhaltlich 
entnommenen Stellen als solche kenntlich gemacht habe. 
\vspace{1.5cm}
\begin{flushright}
  Wien, am 25. M\"arz 2010
\end{flushright}

\newpage 

\tableofcontents
\newpage

% ---------------------------------------------------------------------------- %
% CHAPTER 1 ------------------------------------------------------------------ %
% ---------------------------------------------------------------------------- %

\thispagestyle{plain}
\chapter*{Prologue}

\section*{Motivation}

In astrophysics a multitude of systems and configurations are described 
with concepts from hydrodynamics  
coupled with gravitation, radiation and magnetism. Mathematically radiation 
hydrodynamics (RHD) and magnetohydrodynamics (MHD) are systems of 
coupled nonlinear partial differential equations. 
Multiple fields in astrophysics have been adopting and developing sophisticated 
numerical methods for solving these PDEs in various applications. 

Explicit numerical schemes in computational fluid dynamics 
have been popular for many years and 
received substantial boosts due to advances in technology and 
parallelizing techniques. Riemann solvers and related methods are extensively 
adapted in 2D and 3D computations however they have one main disadvantage. 
An inherent limitation for time steps in explicit schemes
impedes applications where various time scales are of interest. 
Hence a majority of these calculations emphasize on small temporal and 
spatial scales but fail to treat phenomenons properly on very diverse scales. 

Techniques for implicit numerics are computationally more expensive 
since they are hardly 
parallelizable but do not limit the increments of the scheme intrinsically. 
Implicit RHD in 1D has proven favorable for describing astrophysical 
processes on large spatial and temporal scales like nonlinear 
pulsation or long term development of supernovae. 
The idea of this thesis was to study suitable generalizations of 
these concepts to 2D and 3D. Additional degrees of freedom in 
multi-dimensional implicit RHD 
would disclose the treatment of large scale convection and how it 
interacts with rotation, mixing, the coupling of 
rotation and pulsation, transport of angular momentum and mass loss, 
interactions of discs and stars during star formation processes etc. 
The target of this thesis 
 was to find the strong conservation form of the equations 
of radiation hydrodynamics in non-steady curvilinear coordinates and 
to study multi-dimensional adaptive grid generation. 

\newpage
\thispagestyle{plain}
\section*{Synopsis}

The Euler equations of hydrodynamics, the Maxwell equations as well as 
radiative transport equations are hyperbolic PDEs that connect certain
densities and fluxes via conservation laws. Generally they emerge 
from natural symmetries that constitute major principles in mathematical 
physics. The numerical implementation of these equations essentially needs 
to comprise these qualities. Applied mathematics have developed an 
articulate framework of numerical methods for conservation laws that 
ensure the conservation of mass, momentum, energy etc. if applied 
properly. The main challenge is to compute the fluxes correctly 
which indeed will be the issue of the first three chapters 
%\ref{Fund_Math}, \ref{Fund_Phys}  and \ref{Conservative Numerics} 
in this paper. 
Self gravitation is described via the Poisson equation that poses an extra 
challenge due to its elliptic nature. While 1D computations avoid solving this 
pure boundary value problem by considering an integrated form we will have to 
design a different approach to self gravitation in 2D and 3D.  

\thispagestyle{plain}
A generalization of implicit conservative numerics to multiple dimensions
requires advanced concepts of tensor analysis and differential geometry and 
hence a more thorough dedication to mathematical fundamentals than 
maybe expected at first glance. Hence we begin to discuss fundamental 
mathematics and physics of RHD with special focus on differential 
geometric consistency and study numerical methods for nonlinear 
conservation laws to gain a solid definition of the term \emph{conservative}. 
The efforts in tensor analysis will be needed when applying 
\emph{Vinokurs theorem} to gain the \emph{strong conservation form} 
for conservation laws in general curvilinear coordinates. 
Moreover, it will be required to slightly reformulate the 
\emph{artificial viscosity} for such nonlinear coordinates.  

Astronomical objects are characterized by fast flows and high propagation 
speeds on the one hand but {\sl astronomical} length and time scales 
on the other hand. Implicit numerical schemes are not affected by 
the \emph{Courant Friedrichs Levy condition} which limits explicit schemes 
to rather impracticably small time steps. Implicit methods however 
produce algebraic problems that require matrix inversion which is 
computationally expensive. In order to achieve viable resolution, 
\emph{adaptive grid techniques} have been developed. 
It is desired to treat processes on small 
length scales like shocks and ionization fronts as well as physics 
at the extent of the objects dimension itself like large scale convection 
flows and pulsations. 
The combination of implicit schemes and 
adaptive grids allows to resolve astrophysics 
appropriately at various scales.  

In the last chapter of this paper we study problem oriented 
\emph{adaptive grid generation in 2D and 3D}. 
We establish three main postulations 
for an ideal grid and analyze several feasible approaches.

% ---------------------------------------------------------------------------- %
% CHAPTER Fundamental Mathematics of RHD ------------------------------------- %
% ---------------------------------------------------------------------------- %

\chapter{Fundamental Mathematics of RHD} \label{Fund_Math}

The equations of radiation hydrodynamics (RHD) are a system of nonlinear 
second order partial differential equations with hyperbolic and
elliptic parts. Generally they are posed as an initial value problem with 
boundary conditions. The simultaneous solution of hyperbolic terms, as coming 
from the dynamics and the strictly elliptical Poisson equation 
constitute one major problem for finding suitable solution methods. 
In the following chapter we will retrieve the fundamental mathematics 
of hyperbolic conservation laws as they appear in hydrodynamics and radiation
transfer as well as some basic concepts for elliptic partial differential 
equations like the Poisson equation in order to motivate certain numerical 
concepts that will arise in chapter \ref{Conservative Numerics}. 
As we will see, accurate analysis provides expedient approaches to our 
numerical problem.

% SECTION 1 ------------------------------------------------------------------ %

\section{The Hyperbolic Part} \label{Fund_Math_Hyp}

Dynamical physical processes are mathematically 
described via hyperbolic partial differential 
equations\footnote{The Schr\"odinger equation of quantum mechanics 
is excepted.}. The wave equation, the Burgers' equation, the 
equations of hydrodynamics and the Dirac equation are prominent examples of 
hyperbolic evolution equations. They always characterize 
systems with finite propagation speeds, a quality often referred to 
as {\sl causality}. Information from a point in space and time is causally 
associated solely with a finite region, a domain of dependence.  In 
chapter \ref{Conservative Numerics} we will resort to this important 
issue when we discuss stability and convergence of numerical methods. 

In the following section we want to define some fundamental mathematical 
terms and present important concepts that will accompany the entire paper. 

\subsection{Hyperbolic Conservation Laws}

The Euler equations of hydrodynamics as well as the angular moment equations 
of radiative transfer can be derived from conservation laws 
originating from one of the most fundamental theorems of mathematical 
physics, the Noether Theorem (see e.g. Bhutani (1981) 
\cite{1981JMPS...15..225B}, Kambe (2007) \cite{Kambe200798}). 
In hydrodynamics, the conservation of mass, momentum 
and energy leads to the continuity equation, the equation of motion and 
the energy equation. Analog correspondences are derived for radiation. 
Mathematically the equations of radiation hydrodynamics
form a system of \emph{hyperbolic conservation laws} that
describe the interaction of a \emph{density function} $\mathbf{d}$ 
\begin{equation} \nonumber \mathbf{d}: \mathbb{R}^n \times \left[ 0,\infty 
\right) \rightarrow \mathbb{R}^m \end{equation} 
and its \emph{flux} $\mathbf{f}$ 
\begin{equation} \nonumber \mathbf{f}: U 
\rightarrow \mathbb{R}^m, \quad U \subseteq \mathbb{R}^m, \quad U \, 
\text{open.} \end{equation}

\noindent The temporal and spatial change of the integrated density in a 
connected space $\Omega \in \mathbb{R}^n$ then equals the flux over the 
boundary $\partial \Omega$, where $\mathbf{S}$ is the outward oriented 
surface element. 
\begin{equation} \label{erh} \partial_t \int\limits_{\Omega} 
\mathbf{d} \, dV + 
\int\limits_{\partial \Omega} \mathbf{f} \cdot d\mathbf{S} = \mathbf{0} 
\qquad \forall \, t > 0 
\end{equation}
\emph{Hyperbolicity} is granted 
if the Jacobian matrix associated with the fluxes 
$\nabla_{\mathbf{d}}\mathbf{f}$ has real eigenvalues and there exists a 
complete set of eigenvectors. Provided it is physically justified, 
demanding this quality on a set of equations even can help to formulate 
constraints to the solutions (see subsection \ref{radiation_closure_relation}). 

For the in section \ref{physics_of_hd} further motivated 
physical problem of hydrodynamics this system gains the following form as 
an example\footnote{Notation: Summations respectively projections of 
tensorial quantities are always denoted by $\cdot$ or $:$ 
whereas tensor products $\otimes$ are omitted, i.e. $\mathbf{u} \mathbf{u} = 
\mathbf{u} \otimes \mathbf{u}$.}. 
\begin{equation} \label{hd_cons}
\mathbf{d}(\mathbf{x},t) = \left( \begin{array}{cc} \rho(\mathbf{x},t) \\ 
\big(\rho\mathbf{u}\big)(\mathbf{x},t) \\
\big(\rho \epsilon \big) (\mathbf{x},t) \end{array} \right), 
\quad \mathbf{f}(\mathbf{d}) = 
\left( \begin{array}{cc} \rho\mathbf{u} \\ \mathbf{u} \rho \mathbf{u} + 
\mathbf{P} 
\\ \rho \epsilon \mathbf{u} + \mathbf{P} 
\cdot \mathbf{u}
\end{array} \right)
\end{equation}

\begin{eqnarray} \label{hd_equations_cons}
\nonumber \partial_t \int\limits_{\Omega} \rho \, dV +
\int\limits_{\partial \Omega} \rho \mathbf{u} \cdot d\mathbf{S} & = & 0 \\
\nonumber \partial_t \int\limits_{\Omega} \mathbf{u} 
\rho \, dV +\int\limits_{\partial \Omega} 
\big( \mathbf{u} \rho \mathbf{u} + \mathbf{P} \big) \cdot d\mathbf{S} 
& = & \mathbf{0} \\
\partial_t \int\limits_{\Omega} \rho \epsilon \, dV + 
\int\limits_{\partial \Omega} \big( \rho \epsilon \mathbf{u} + \mathbf{P} 
\cdot \mathbf{u} \big) \cdot d\mathbf{S} & = & 0 
\end{eqnarray}

Let $\mathbf{f}$ be a continuously differentiable 
function, then we can 
rewrite equation (\ref{erh}) via the divergence theorem as follows. 
\begin{equation} \label{cauchy_problem_integral}
\int\limits_t\int\limits_{\Omega} \big( \partial_t \mathbf{d} + 
\mathrm{div} \, \mathbf{f}(\mathbf{d})\big) \, dV dt = \mathbf{0} \qquad 
\forall \, t > 0, \Omega \subset \mathbb{R}^n
\end{equation}
Locally the system of partial differential equations 
\begin{equation} \label{cauchy_problem}
\partial_t \mathbf{d} + 
\mathrm{div} \, \mathbf{f}(\mathbf{d}) = \mathbf{0} 
\end{equation}
gives pointwise solutions for our vector field $\mathbf{d}$ that describes 
the conserved densities or state variables 
of mass, momentum and energy. With initial condition 
(or the initial state of the density function) $\mathbf{d}_0$ 
\begin{equation} 
\label{initial_condition} \mathbf{d}(\mathbf{x},0)=\mathbf{d}_0(\mathbf{x}) 
\quad \forall \, \mathbf{x} \in \mathbb{R}^n 
\end{equation}
Equation (\ref{cauchy_problem}) with initial condition 
(\ref{initial_condition}) is called \emph{Cauchy problem}. 
The main theorem 
for this class of partial differential equations associated with analytic
 initial conditions 
is the theorem of Cauchy-Kovalevskaja  
which states local existence and 
uniqueness also for nonlinear Cauchy problems. However, with analyticity 
demanded this theorem has hardly any practical relevance. 
Questions of local and global well-posedness for general nonlinear 
Cauchy problems require sophisticated concepts of functional analysis 
(see e.g. Adams \cite{Adams}, Evans \cite{Evans}). 
We want to mention some basic concepts of these
variational formulations 
of partial differential equations in an extent that they arise in the 
context of radiation hydrodynamics and accordingly numerics.

\subsection{Weak Solutions}

Since even the simplest examples of one-dimensional scalar conservation laws 
like the Burgers' equation have classical solutions only in some special 
cases, one has to broaden the considered function space of 
possible solutions (the adequate function spaces are 
Sobolev spaces, see Evans \cite{Evans}). 
For so called \emph{weak solutions}, we appeal to generalized functions 
where also discontinuities can be dealt with. The generalized concept of 
differentiation of distributions shifts operations to \emph{test 
functions} $\gamma (\mathbf{x},t)$. These test functions 
$\gamma: G \in \mathbb{R}^n \times \mathbb{R}^{+} 
\rightarrow \mathbb{R}$ have compact support, 
meaning that there exists a compact subset (here i.e. closed 
and bounded) $K$ such that $\gamma (\mathbf{x},t)=0$ 
$\forall x \, \in G \setminus K(\gamma)$. 
\begin{equation}
\int\limits_{t \geq 0} \int\limits_{\mathbb{R}^n} \big( 
\partial_t \mathbf{d} + \mathrm{div} \, \mathbf{f}(\mathbf{u}) \big) \gamma \, 
dV dt = 0 \qquad \forall \, \gamma \in G 
\end{equation}
To simplify matters, we let these 
test functions be continuously differentiable. Then with partial integration 
and our postulations for $\gamma$ 
\begin{eqnarray}
\nonumber \int\limits_{t \geq 0}  \partial_t \mathbf{d} \, \gamma \, dt & = & 
\gamma \, \mathbf{d} \Big|_0^{\infty}
- \int\limits_0^{\infty} \mathbf{d} \, \partial_t \gamma \, dt \\
\nonumber & = & \gamma(\mathbf{x},0) \, \mathbf{d}_0(\mathbf{x})
- \int\limits_0^{\infty} \mathbf{d} \, \partial_t \gamma \, dt \\
\int\limits_{\mathbb{R}^n} \mathrm{div} \, \mathbf{f}(\mathbf{d}) \, \gamma \, 
dV & = & \underbrace{\gamma \, \mathbf{f} \Big|_{\dots} 
\Big|_{-\infty}^{\infty} }_{=0} - 
\int\limits_{\mathbb{R}^n} \mathbf{f}\,  \mathrm{grad} \, \gamma \, dV
\end{eqnarray}
we derive the weak formulation of our conservation law 
(\ref{cauchy_problem_integral}). 
\begin{equation} \label{cauchy_weak_formulation}
\int\limits_{t \geq 0} \int\limits_{\mathbb{R}^n} \big( 
\mathbf{d} \, \partial_t \gamma + \mathbf{f}(\mathbf{d}) \, 
\mathrm{grad} \, \gamma  
\big) \, dV dt = - \int\limits_{\mathbb{R}^n} 
\gamma(\mathbf{x},0) \, \mathbf{d}_0(\mathbf{x}) \, dV 
\end{equation}
The function $\mathbf{d} \in \mathcal{L}^{\infty}$ is called \emph{weak
 solution of the Cauchy problem}
(\ref{cauchy_problem_integral}), if it satisfies (\ref{cauchy_weak_formulation}) 
and $\mathbf{d} \in U$ with $\mathbf{d}_0 \in \mathcal{L}^{\infty}$. 
This weak solution is not necessarily unique and normally further constraints 
have to be imposed.

\subsection{Rankine-Hugoniot and Entropy Conditions} \label{Rankine_and_Entropy}

For the most physical problems it will be sufficient to look for 
weak solutions in the function space of piecewise continuously 
differentiable functions. The physical variables $\mathbf{d}$ are weak 
solutions, if they are a classical solutions
wherever they are $\mathcal{C}^1$ and 
at discontinuities (shocks) they need to satisfy additional conditions. 
As we will see in chapter \ref{Fund_Phys}, these conditions posses 
direct physical relevance when applied to a physical system like 
radiation hydrodynamics. 
The full derivation of the one-dimensional \emph{Riemann-problem} for 
(\ref{cauchy_problem_integral}) considered can be found in the 
Appendix \ref{app_rankine_hugoniot}. 

We conclude,  
\begin{eqnarray}
\nonumber \int\limits_0^{\infty}\int\limits_{-\infty}^{\infty} 
\left( d \, \partial_t \gamma + f(d) \, \partial_x
\gamma \right) \, dx dt  =  -\int\limits_{-\infty}^{\infty} 
\gamma(x,0)\, d_0(x) \, dx +\\
\label{rankine_1} + \int\limits_0^{\infty} \gamma(x,x/u_s) 
\, \left( \frac{f(d_l)-f(d_r)}{u_s}-\Big( d_l-d_r \Big) \right) \, dx 
\end{eqnarray}
 is weak solution of the Riemann-problem, if 
\begin{equation} \frac{f(d_l)-f(d_r)}{u_s}-\Big(d_l-d_r \Big)=0 \end{equation}
respectively 
\begin{equation} \label{ranking_hugoniot_conditions} 
u_s = \frac{f(d_l)-f(d_r)}{d_l-d_r} .
\end{equation}
We call $u_s$ \emph{shock velocity} and (\ref{ranking_hugoniot_conditions})
\emph{Rankine-Hugoniot condition}.   

As already mentioned, weak solutions of the Cauchy problem are not necessarily 
unique and thus may be physically pointless. However, there is a way to pick 
 physically valuable solutions out of multiple. One method is to 
implement an artificial viscosity term 
\begin{equation}
\partial_t \mathbf{d}^* + 
\mathrm{div} \, \mathbf{f}(\mathbf{d}^*) = \varepsilon \nu \Delta \mathbf{d}^* , 
\qquad \varepsilon > 0
\end{equation}
and evaluate the equation for $\varepsilon \rightarrow 0$ as limit case of 
the original one. The idea is motivated from physical diffusion, which 
broadens sincere discontinuities to differentiable steep gradients. The 
physical solution of the weakly formulated problem has to be the 
zero diffusion limit of the diffusive problem. However, in analytical practice 
this limit is rather bulky 
to calculate, hence simpler conditions have to be found. 
The common technique to do this is motivated from continuum physics 
as well, where 
an additional conservation law holds for the entropy of the fluid flow as 
long as the solutions remain smooth. Moreover, it is known, that along 
admissible shocks 
this {\sl physical} variable never decreases, so the conservation law for the 
entropy can be reformulated as inequality. 

We regard a scalar entropy function $\sigma(\mathbf{d})$ and an 
entropy flux $\phi(\mathbf{d})$ which satisfy 
\begin{equation}  \label{cons_entro}
\partial_t \sigma(\mathbf{d}) + \mathrm{div} \, \phi(\mathbf{d}) = 0.
\end{equation} 
Assuming differentiable functions, we rewrite that conservation law 
via the chain rule and compare it to (\ref{cauchy_problem}) multiplied with 
 $ \nabla_{\mathbf{d}} \sigma$ to obtain 
\begin{eqnarray}
\nonumber \nabla_{\mathbf{d}} \sigma\, \partial_t \mathbf{d} + 
\nabla_{\mathbf{d}}\phi \, \nabla_{\mathrm{x}} \mathrm{d} & = & 0 \\
\nonumber 
\nabla_{\mathbf{d}} \sigma \, \partial_t \mathbf{d} + \nabla_{\mathbf{d}} 
\sigma \, \nabla_{\mathbf{d}} \mathbf{f} 
\, \nabla_{\mathrm{x}} \mathrm{d} & = & 0 \\
\nabla_{\mathbf{d}} \sigma \, \mathrm{div} \, \mathbf{f}  & = & 
\mathrm{div} \, \phi  
\end{eqnarray} 
where in the more dimensional case the appearing matrices of gradients 
have to fulfill some further 
constraints (see e.g. Godlewski and Raviart (1992) 
\cite{GodRav1992}). For scalar equations, it is 
always possible to find an entropy function of that kind. 
Furthermore we postulate convexity for the entropy 
function.\footnote{In physical entropy one would require an eventual
inequality with $\geq$ in place of $\leq$ but mathematical literature  
commonly chooses $\sigma''>0$.} 
\begin{equation} 
 \nabla_{\mathbf{d}}^2 \sigma > 0, \qquad \forall \, \mathbf{d}
\end{equation}  
To get our actual entropy condition, we rewrite (\ref{cons_entro}) in the 
viscous form 
\begin{equation}
\partial_t \sigma(\mathbf{d}^*) + \mathrm{div} \, \phi(\mathbf{d}^*) = 
\varepsilon \nu \nabla_{\mathbf{d}} \sigma(\mathbf{d}^*) 
\Delta \mathbf{d}^* 
\end{equation}
and integrate over an arbitrary time interval $\left[ t_0, t_1 \right]$. 
\begin{eqnarray}
\nonumber && \int\limits_{t_0}^{t_1} \int\limits_{\Omega} 
\Big( \partial_t \sigma(\mathbf{d}^*) + \mathrm{div} \, \phi(\mathbf{d}^*)
 \Big) \, dV dt =  \varepsilon \nu \int\limits_{t_0}^{t_1}  
\int\limits_{\Omega} \nabla_{\mathbf{d}} \sigma(\mathbf{d}^*) 
\Delta \mathbf{d}^* \, dV dt = \\
\nonumber & & = \varepsilon \nu \int\limits_{t_0}^{t_1}  
\int\limits_{\partial \Omega} \Big( \nabla_{\mathbf{x}} \mathbf{d}^* \, 
\nabla_{\mathbf{d}} \sigma(\mathbf{d}^*)\Big) \cdot \, d\mathbf{S} dt - 
 \varepsilon \nu \int\limits_{t_0}^{t_1} \int\limits_{\Omega} 
\nabla_{\mathbf{x},i} {\mathbf{d}^*} \, 
\underbrace{\nabla^2 \sigma(\mathbf{d}^*)}_{>0} \, 
\nabla_{\mathbf{x},i} \mathbf{d}^* \, dV dt \\
\end{eqnarray} 
When we now consider our non diffusive limit for $\varepsilon \to 0$, the 
first term on the right hand side vanishes without further restriction 
whereas the second term has to remain non positive. With partial 
integration and divergence theorem we get our entropy condition 
(\ref{entropy_condition}). 
\begin{eqnarray} \label{entropy_condition}
\nonumber \int\limits_{t_0}^{t_1} \int\limits_{\Omega} 
\Big( \partial_t \sigma(\mathbf{d}) + \mathrm{div} \, \phi(\mathbf{d})
 \Big) \, dV dt & \leq & 0 \\ 
\nonumber  \int\limits_{\Omega} \sigma(\mathbf{d}(\mathbf{x},t_1)) \, dV 
& \leq & \int\limits_{\Omega} \sigma(\mathbf{d}(\mathbf{x},t_0)) \, dV - 
\int\limits_{t_0}^{t_1} \int\limits_{\partial \Omega} 
\phi(\mathbf{d}) \, d\mathbf{S} dt \\
\label{entropy_condition}
\end{eqnarray}
For bounded, continuous pointwise solutions 
$\mathbf{d}^*$ with $\mathbf{d}^* \to \mathbf{d}$ for $\varepsilon \to 0$, the 
\emph{vanishing viscosity solution} $\mathbf{d}$ is weak solution of the initial 
value problem (\ref{cauchy_problem_integral}) and fulfills entropy condition 
(\ref{entropy_condition}). Generally spoken, applying the entropy condition 
systems with shock solutions unveils those propagation velocities that ensure 
that no characteristics rise from discontinuities which would be 
non-physical. For details and proofs we
refer to LeVeque \cite{LeVeque1990}. 

In chapter \ref{Conservative Numerics} we will discuss these analytic 
concepts in numerical context and work out its concrete applications 
for (radiation) hydrodynamics.

% SECTION 2 ------------------------------------------------------------------ %

\section{The Elliptic Part} \label{Fund_Math_Ell}

As already elaborated radiation hydrodynamics are partial differential 
equations of the hyperbolic type. However, the dynamics are only one part of 
the challenge emerging in astrophysical applications. Self gravitation in 
classical astronomical objects is described via the Poisson equation, an 
elliptical partial differential equation which emerges as pure
boundary value problem for the gravitational potential. 
Unlike hyperbolic PDEs elliptical problems mostly describe equilibrium 
physics and their solutions are typically quite smooth as we will see.  

\subsection{The Poisson Equation} 

The \emph{Poisson equation} 
describes the spatial distribution of a scalar field 
$\Phi$
\begin{equation}
\Phi : \bar{U} \subseteq \mathbb{R}^n \rightarrow \mathbb{R}
\end{equation}
induced by a source $\rho$ 
\begin{equation}
\rho : U \subseteq  \mathbb{R}^{n}
\rightarrow \mathbb{R}
\end{equation} 
and has the following form; the choice of the minus sign originates 
from the customary definition for ellipticity of differential operators.
\begin{equation} \label{poisson_equation}
- \Delta \Phi = \rho . 
\end{equation} 
When it comes to our physical application of self-gravitation,
we will pull the sign into 
the definition of the gravitational potential and 
consider $\Delta \Phi_{\text{grav}}
 = 4 \pi G \rho_{\text{m}}$. 

The Poisson equation is posed with Dirichlet or Neumann 
boundary conditions which means either boundary values or their derivatives 
are given. Often it is useful to impose mixed boundary conditions, where 
the boundary of the region $\partial \Omega$ is separated in a part 
with Dirichlet conditions $\partial \Omega^{\mathrm{Dir}}$ and a part 
with Neumann boundary conditions $\partial \Omega^{\mathrm{Neu}}$ and 
$\partial \Omega = \partial \Omega^{\mathrm{Dir}} \cup
\partial \Omega^{\mathrm{Neu}}$. 
\begin{eqnarray}
\nonumber \Phi  & = & \psi^{\mathrm{Dir}} \quad \mathrm{in} 
\, \, \partial \Omega^{\mathrm{Dir}} \\
\mathbf{n} \cdot \nabla \Phi & = & \psi^{\mathrm{Neu}} \quad \mathrm{in} 
\, \,  \partial \Omega^{\mathrm{Neu}}
\end{eqnarray}
The Poisson equation can be understood as the inhomogeneous generalization 
of the \emph{Laplace equation}
 $\Delta \tilde{\Phi} = 0$. Physically, latter equation 
describes a solenoidal, irrotational vector field 
$\mathbf{u}=\nabla \tilde{\Phi}$ with
$\mathrm{div}\,\mathbf{u} = \mathrm{rot}\, \mathbf{u} = 0$. 
This circumstance implies that $\tilde{\Phi}$ must be invariant 
under rotation and 
can be used to derive the fundamental solution. With Green's function and 
convolution of the inhomogeneous term $\rho$ we derive the analytical 
solution of the Laplace equation\footnote{
$\tilde\Phi(\mathbf{x}) = \tilde\Phi(r)$, 
$\partial_i r = \frac{x^i}{r}$ $\dots $ $\rightarrow$ 
Green's functions of the fundamental solution of the Laplace 
equation in 3D is $\Delta^{(3)} 
\left(-\frac{1}{4 \pi r}\right) = \delta^{(3)}(\mathbf{x})$ and 
$\Phi = \int G(\mathbf{x},\mathbf{x}') \rho(\mathbf{x}') \, dV$ is solution 
of $\Delta \Phi = \rho$. }. In the light of these presumptions, we can 
investigate fundamental attributes of solutions of the Laplace equation in 
order to learn something about the solutions of the Poisson equation.

\subsection{Harmonic Functions and the Maximum Principle}

One important aspect of the Laplace operator as it emerges in Poisson 
and Laplace equation is the fact that it acts 
regularizing; the according theorem follows. 
Hence it will be adequate to deal with differentiable 
functions when we look for solutions of (\ref{poisson_equation}). 
Any function satisfying $\Delta \tilde{\Phi} = 0$ in $U$ 
and $\mathcal{C}^2$ 
is called \emph{harmonic function}. 

Given $\tilde{\Phi}$ is harmonic, we can derive a mean-value formula which 
states that on a spherical subset\footnote{
Closed $n$-dimensional ball $B(\mathbf{x},r)$ with center 
$\mathbf{x}$, radius $r>0$.} $B(\mathbf{x},r) \subset U$, a solution 
$\tilde{\Phi}(\mathbf{x})$ equals both 
the average of $\tilde{\Phi}$ over the sphere $\partial B$ and the 
average of $\tilde{\Phi}$ on the entire ball $B$. 
\begin{equation}
\tilde{\Phi}(\mathbf{x}) = \frac{\Gamma(\frac{n}{2}+1)}{n\pi^{n/2}r^{n-1}}
\int\limits_{\partial B(\mathbf{x},r)} 
\tilde{\Phi} \, dS =  \frac{\Gamma(\frac{n}{2}+1)}{\pi^{n/2} r^n}
\int\limits_{B(\mathbf{x},r)} \tilde{\Phi} 
\, dV
\end{equation}
In 3 dimensions this simply yields
$\tilde{\Phi}(\mathbf{x}) = \frac{1}{4 \pi r^2}
\int \tilde{\Phi} \, dS =  \frac{3}{4\pi r^3}
\int \tilde{\Phi} 
\, dV$ with corresponding integration ranges. 
In the light of these conclusions,
we deduce some interesting properties of harmonic functions. 

\noindent The \emph{(strong) maximum principle} states for a $\tilde{\Phi} 
\in \mathcal{C}^2(U) \cap \mathcal{C}(\bar{U})$ harmonic within $U$, 
\begin{enumerate}
\item $\max_{\bar{U}} \tilde{\Phi} = \max_{\partial U} \tilde{\Phi}$. 
\item Furthermore, if $U$ is connected and $\exists \, x_0 \in U$ such that 
$u(x_0)=\max_{U} \tilde{\Phi}$, it follows that $\tilde{\Phi}$ is constant 
within all $U$. 
\end{enumerate}
Assertion \emph{\small{i}.)} implies 
that the maximum will be found at the boundary of the 
domain and \emph{\small{ii}.)} states that if a maximum is on 
the inside of the boundary the function $\tilde{\Phi}$ consequently is
constant within $U$. Another important conclusion of this theorem concerns the
Poisson equation. It can be shown that for the boundary value problem 
\begin{eqnarray}
\nonumber
- \Delta \Phi & = & \rho \quad \text{in} \, U \\ 
\Phi & =  & g \quad \text{in} \, \partial U
\end{eqnarray}
there exists 
at most one solution $\Phi \in \mathcal{C}^2(U)  \cap \mathcal{C}(\bar{U})$. 
With this \emph{uniqueness theorem} we gain at least some comfort for
the gravitational part of our numerical problem. As a further side note 
we mention the \emph{regularity theorem}. It reveals that if 
$\tilde{\Phi} \in \mathcal{C}^2$ is harmonic then it automatically is 
infinitely differentiable, i.e. $\tilde{\Phi} \in \mathcal{C}^{\infty}$. 

Without dwelling on the Laplace equation any further, we recapitulate 
that mathematical analysis of the elliptical part of our system of 
equations brings forward less sophisticated problems than the 
hyperbolic part. Obviously, uniqueness and regularity are soothing qualities 
when it comes to solving partial differential equations. All theorems and 
proofs of this section can be found in Lawrence C. Evans (1998), p.17 ff.
\cite{Evans}.

% ---------------------------------------------------------------------------- %
% CHAPTER Fundamental Physics of RHD ----------------------------------------- %
% ---------------------------------------------------------------------------- %

\chapter{Fundamental Physics of RHD}  \label{Fund_Phys}

Astronomical objects are basically self-gravitating aggregations of (gaseous) 
material with magnetic fields and radiation. Radiation hydrodynamics 
as we will discuss them, factor out magnetic effects and 
solely consider radiative contributions through photons. 

The fundamental equations of hydrodynamics describe the motion of fluids in 
in terms of continuous quantities. The dynamics and kinetics of 
fluid particles are described by macroscopic functions like the 
mass density $\rho$, its velocity $\mathbf{u}$, internal energy and gaseous 
pressure as primary variables mostly.  
In this model, infinitesimal volume elements of the fluid are still big enough 
that intermolecular forces can be neglected, meaning that they contain 
a sufficient number of molecules. 
The thermodynamic variables intrinsically are assumed to be well defined 
(equilibrium thermodynamics) and characterized by an equation of state. 

The radiative transfer equation describes energy transfer in form of 
electromagnetic radiation, affected by absorption, emission and scattering. 
The fundamental quantity to describe the radiative field is the spectral 
intensity $I_{\gamma}$, which can be understood as macroscopic quantity to 
describe the photonic energy flow. The mass density $\rho$ is also source for 
the gravitational potential $\Phi$. 
In the following sections we will physically motivate the set of 
equations of radiation hydrodynamics with gravitation bit by bit. 

One major intention in the upcoming paragraphs 
will be rigorous and consistent (in terms 
of differential geometry) definitions of the physical variables 
which will be crucial for the numerical 
methods in general cuvilinear coordinates investigated 
in section \ref{ConsGeneral}.

% SECTION HYDRODYNAMICS------------------------------------------------------- %

\section{Hydrodynamics} \label{physics_of_hd}

As anticipated, the equations of hydrodynamics can be 
derived from conservation laws of mass, momentum and energy. Since 
the fundamentals of hydrodynamics are more than well documented 
(e.g. in Landau, Lifschitz \cite{LandauLif}) we confine 
our portrayal to a brief synopsis and concentrate on 
a few interesting details. 

\subsection{Continuity Equation} 

The conservation of mass is physically justified as long as no mass is generated 
or annihilated
\begin{equation} 
\frac{d m}{dt} = \frac{d}{dt}\int\limits_{V(t)} \rho \, dV = 0 
\end{equation}
and leads to the continuity equation. 
\begin{eqnarray}
\nonumber 
\int\limits_{V(t)} \Big( \partial_t \rho + 
\mathrm{div} \,  \big( \rho \mathbf{u} \big) \Big) \, dV & = & 0 \\
\label{cont_eq_def}
\int\limits_{V(t)} \partial_t \rho + \int\limits_{\partial V(t)} 
 \big( \rho \mathbf{u} \big) \cdot d\mathbf{S} & = & 0
\end{eqnarray}
Clearly the domain $\Omega$ of the conservation law, introduced in 
section \ref{Fund_Math_Hyp} 
is an $n$-dimensional Volume that is a bounded subset of the 
$\mathbb{R}^n, \, n=1,2,3$. The mass density $\rho$ is a non negative 
scalar field $\rho(\mathbf{x},t): \mathbb{R}^{n+1} \to \mathbb{R}^{+}$ and 
the velocity $\mathbf{u}$ a vector field $\mathbf{u}(\mathbf{x},t): 
\mathbb{R}^{n+1} \to \mathbb{R}^{n+1}$. 
Solutions of these primary variables $\rho, \mathbf{u}$, that is
densities of the conservation laws, will emerge as 
piecewise continuous functions. 

As a side note the flow is called 
stationary if $\partial_t \rho = 0$ and incompressible if 
$\mathrm{div} \mathbf{u}=0$. 
However, in astrophysical applications these special 
cases will fail to describe the physics appropriately.

\subsection{Equation of Motion}

In order to yield the equation of motion from the conservation of momentum 
it is required to specify which forces act upon the fluid. Now we are concerned 
with a non-scalar conservation law and as we will see in chapter 
\ref{Conservative Numerics}, in this case we will have to avail ourselves of 
general tensorial notation. Thus, at this point we introduce one main tensorial 
quantity (and secondary physical variable), the non-relativistic 
viscous stress tensor. 
\begin{eqnarray}
\nonumber \mathbf{T} & = & \rho \mathbf{u} \mathbf{u} + \mathbf{P} - 
\mathbf{\sigma} \\
\label{visc_stress_def}
T^{i j} & = & \rho u^i u^j + P^{ij} - \sigma^{ij}
\end{eqnarray}
$\mathbf{T}(\mathbf{x},t): \mathbb{R}^{n+1} \times \mathbb{R}^{n+1} \to 
\mathbb{R}^{n+1} \times \mathbb{R}^{n+1}$. 
This symmetric\footnote{$\mathbf{T}$ is symmetric, if conservation of angular 
momentum $ \rho \mathbf{x}\times\mathbf{u}$ is invoked, see derivation 
in the Appendix \ref{cons_ang_mom}. } 
tensor ($T^{ij} = T^{ji}$) describes contributions from momentum 
$\rho u^i$ in $j$-direction, generally anisotropic pressure $P^{ij}$ and 
friction $\sigma^{ij}$
forces. Diagonal elements express pressure and off-diagonal elements shear 
stresses. The equation of motion in then yields\footnote{In 
general relativity the conservation of momentum
can be found in its most 
elegant form $\nabla \cdot \mathbf{T} = 0$. In this case 
all variables are defined in a four-dimensional spacetime, 
where the 0-component of the 
covariant nabla operator contains the temporal derivative.}
\begin{eqnarray}
\nonumber \int\limits_{V(t)} \Big( \partial_t \big( \rho  \mathbf{u} \big) +
\mathrm{div} \,  \mathbf{T} \Big) \, dV & = & 0 \\
%\nonumber 
%\int\limits_{V(t)} \Big( \partial_t \big( \rho  \mathbf{u} \big) +
%\mathrm{div} \,  \big( \rho \mathbf{u} \mathbf{u}^T + \mathbf{P} - 
%\mathbf{\sigma} \big) \Big) \, dV & = & 0 \\
\int\limits_{V(t)} \partial_t \big( \rho  \mathbf{u} \big) \, dV + 
\int\limits_{\partial V(t)} \big( \rho \mathbf{u} \mathbf{u} + \mathbf{P} - 
\mathbf{\sigma} \big) \cdot d\mathbf{S} & = & 0 . \label{cons_EOM}
\end{eqnarray}
Special cases for the energy-stress tensor in fluid dynamics are e.g. the
dust model where no particle interaction is assumed $\mathbf{P}=\sigma=0$. 
Anyway, in numerous physical applications the pressure forces can be 
expected to be isotropic. If so, the pressure tensor simplifies 
to a diagonal tensor 
which consists of the scalar gaseous pressure $p$ and a {\sl measure 
of uniformity}, the the metric tensor $\mathbf{g}$, which yields 
$\mathbf{P}=p\mathbf{g}$. The divergence of that tensor can be further 
simplified to the well known gradient of the gaseous pressure. 
We will extensively occupy ourselves with the viscous 
pressure tensor $\sigma$ in succession, 
especially in section \ref{Artificial_Viscosity}. 
At this point it shall only be mentioned that we neglect physical viscosity 
since the gas of a star is easily approximated as ideal fluid. As 
indicated in section \ref{Rankine_and_Entropy}, 
we will impose artificial dynamic viscosity when we concern ourselves  
with mathematical methods for solving nonlinear conservation laws. 

Implicitly we already have defined a thermodynamic equation of state 
which describes the state of the fluid in terms of pressure $p$, 
volume $V$ and temperature $T$ by imposing an ideal fluid. 
Since stars are largely composed of hot, low-pressure hydrogen gas, it is 
quite reasonable to apply this model.

\subsection{Energy Equation}

The energy balance of a fluid includes kinetic and pressure parts as well 
as inner energy. Latter is once more a thermodynamic quantity which is 
associated with the equation of state. 
We define a new non-negative scalar field, the specific inner energy 
$\epsilon(\mathbf{x},t):\mathbb{R}^{n+1} \to \mathbb{R}^+$ 
which in case of an ideal fluid equals its thermic energy. The conservation 
of the total energy of a bounded fluid element in this strictly 
hydrodynamic case yields\footnote{ Using $\nabla(\frac{1}{2}\mathbf{u}^2) = 
\mathbf{u} \nabla \cdot \mathbf{u} $.}
\begin{eqnarray}
\nonumber 
\int\limits_{V(t)} \Big( \partial_t \big( \frac{1}{2}\rho \mathbf{u}^2 + 
\rho \epsilon \big) + \mathrm{div} \,  \big( \mathbf{T} \cdot \mathbf{u} + 
\rho \epsilon\mathbf{u} \big) \Big) \, dV 
& = & 0 \\
\nonumber 
\int\limits_{V(t)} \partial_t \big( \frac{1}{2}\rho \mathbf{u}^2 + 
\rho \epsilon \big) dV + \int\limits_{\partial V(t)} \Big( \big(  
\frac{1}{2}\rho \mathbf{u}^2 + \rho \epsilon \big) \mathbf{u} + 
\big( \mathbf{P} - \sigma \big) \cdot \mathbf{u} \Big) \cdot d\mathbf{S}
& = & 0 . \\
\label{energy_eq_def}
\end{eqnarray}
This \emph{total energy equation} can be dissected into a mechanical and an 
{\sl internal} 
part, where former describes the change of kinetic energy of a fluid as 
the rate of work done on an element by pressure and stress forces. 
The gas energy part can be formulated as entropy conservation 
($\partial_t (\rho s) + \mathrm{div} \,  (\rho s \mathbf{u}) = 0$) and 
hence implies, that the flow is adiabatic.

% SECTION RADIATION ---------------------------------------------------------- %

\section{Radiation}

The fundamentals of radiative transfer and its moments can be found in 
several appropriate text books (see e.g. Mihalas and Mihalas 
\cite{Mihalas}), therefore we will concentrate 
on some noteworthy highlights in this section as well. In the previous 
paragraph we treated the physics of fluids which we now want to extend to 
radiating fluids {\sl in a nutshell}. 

Radiation contributes to the total energy, 
momentum, stress and pressure in the fluid. 
In succession we will work out means 
of coupling among these physical quantities in the material. 
When we concerned ourselves with the fundamental mathematics of conservation 
laws, we denoted that also the fundamental dynamics of radiation will 
appear as hyperbolic conservation laws. In fact, the crucial equation 
to describe radiation is the radiation transfer equation 
(RTE), which basically expresses conservation of energy. However,
 we shall begin 
to define the fundamental dynamical properties of the radiation field.

\subsection{The Radiation Field and its Momenta}

The macroscopic quantities of the radiation field are based on statistical 
physics of a microscopic description of ultra-relativistic massless particles, 
i.e. photons. To establish the nexus
to the continuum view which we will pursuit, we define the scalar 
non-negative photon distribution function $f_{\gamma}(\mathbf{x},t;
\mathbf{n},\nu) : \mathbb{R}^{3+1} \times \mathbb{R}^3 \times \mathbb{R} 
\to \mathbb{R}^+$
which can be interpreted as photon density in the corresponding phase 
space\footnote{ The momentum of a photon, traveling in direction 
$\mathbf{n}$ is given by $\mathbf{p}_{\gamma}=\frac{h \nu}{c} \mathbf{n}$. 
$\nu$ is the frequency, $c$ the speed of light and $h$ the Planck constant. 
$f_{\gamma}$ is invariant under Lorentz transformations. The subset $\gamma$ 
is chosen to discern mere radiation (here photonic) variables. }. 
The 
central physical quantity for the upcoming description of the radiation field 
is the spectral intensity $I_{\gamma}(\mathbf{x},t;\mathbf{n},\nu)$ defined 
by 
\begin{equation} 
I_{\gamma} = \frac{h^4 \nu^3}{c^2} f_{\gamma} . 
\end{equation} 
That implicitly describes a radiative energy transport in the following sense 
\begin{equation}
d \mathcal{E}_{\gamma} = I_{\gamma} \, d\mathbf{S} 
\cdot \mathbf{n} \, d\omega d\nu dt . 
\end{equation}
The spectral intensity $I_{\gamma}$ at position $(\mathbf{x},t)$, 
traveling in direction $\mathbf{n}$ with frequency $\nu$ is defined to 
account for the energy $\mathcal{E}_{\gamma}$ transported by radiation of 
frequencies $d\nu$ across a surface element $d\mathbf{S}$ in time $dt$ 
into a solid angle $d\omega$ around $\mathbf{n}$. Further useful properties 
are the angular moments of the radiation field, that is the mean intensity 
$J_{\gamma}$ (zeroth moment), the radiation flux $\mathbf{F}_{\gamma}$ 
(first moment) and the evidently symmetric radiation pressure 
tensor $\mathbf{P}_{\gamma}$ 
(second moment). The velocity of our particles will simply be given by the 
speed of light and their direction of motion $\mathbf{u}= c \mathbf{n}$, which 
also illustrates the link to  
the procedure we carried out for deriving the equations of 
hydrodynamics.  
\begin{eqnarray}
\nonumber J_{\gamma}(\mathbf{x},t;\nu) & = & \frac{1}{4\pi} \int I_{\gamma}
 \, d\omega \\ 
\nonumber \mathbf{F}_{\gamma}(\mathbf{x},t;\nu) & = & \int I_{\gamma} 
\mathbf{n} \, d\omega \\ 
\mathbf{P}_{\gamma}(\mathbf{x},t;\nu) & = & \frac{1}{c} \int I_{\gamma} 
\mathbf{n}\mathbf{n} \, d\omega 
\end{eqnarray}
Analogously to our fundamental equations of hydrodynamics, we can derive 
dynamic equations for the radiation field by considering the conservation of 
{\sl density} $J_{\gamma}$ and {\sl momentum} $\mathbf{F}_{\gamma}$. 
\begin{eqnarray}
\nonumber \int\limits_{V(t)} \Big( \partial_t J_{\gamma} +
\mathrm{div} \,  \big(c \mathbf{F}_{\gamma} \big) \Big) \, dV & = & 0 \\
\label{cons_radiation}
\int\limits_{V(t)} \Big( \partial_t \mathbf{F}_{\gamma} +
\mathrm{div} \,  \big( c^2 \mathbf{P}_{\gamma} \big) \Big) \, dV & = & 
\mathbf{0} 
\end{eqnarray}
From the viewpoint we pursued in section \ref{Fund_Math_Hyp} discussing 
hyperbolic conservation laws we   
define a density function $\mathbf{d}(\mathbf{x},t;\nu) = 
(J_{\gamma},\mathbf{F}_{\gamma})$ and a 
flux $\mathbf{f}(\mathbf{d})= c(\mathbf{F}_{\gamma}, c \mathbf{P}_{\gamma})$. 
Equations (\ref{cons_radiation}) can hence be interpreted as 
homogeneous outlook to the 
moments of the radiation transfer 
equation with collision terms (\ref{RTE_lab}) 
in the forthcoming section. 

% SUBSECTION INTERACTION WITH MATTER ------------------------------------------%

\subsection{Interaction with Matter}

For the treatment of radiation hydrodynamics it will be necessary to describe 
the interaction of the radiation field with the fluid. Again the macroscopic 
approach chosen here 
to describe this interaction can be understood canonically with 
concepts from statistical physics and quantum mechanics. Moreover, 
it will be required  to process the variables of the radiation 
field into the fluid frame of the material correctly via Lorentz 
transformation and account for the frequency change with respect to the lab 
frame that comes with Doppler shift.   

When radiation passes through material, energy generally will be removed 
depended on a material property, the opacity or extinction coefficient 
$\chi_{\gamma}$. 
On the other hand, the material releases radiation in terms of an 
emission coefficient $\eta_{\gamma}$ which will be isotropic 
in its rest frame. Both properties are defined as non-negative 
scalar functions of position, time, radiation direction and frequency 
$(\chi_{\gamma},\eta_{\gamma})(\mathbf{x},t;\mathbf{n},\nu) : 
\mathbb{R}^{3+1} \times \mathbb{R}^3 \times \mathbb{R} 
\to \mathbb{R}^+$. Both processes, extinction and emission, can be further 
divided in thermal and scattering events. Thermal extinction and emission 
converts radiation energy to thermal and vice versa, whereas scattering 
in principle changes direction and frequency of traveling photons. The amount 
of energy removed along a material element of length $dl$ is 
\begin{equation}
d \mathcal{E}_{\gamma} = \big( \chi_{\gamma}^{\text{therm}} + 
\chi_{\gamma}^{\text{scat}} \big) I_{\gamma} \, dS dl d\omega d\nu dt  
\end{equation}
and the amount of radiation energy released by emission is 
\begin{equation}
d \mathcal{E}_{\gamma} = \big( \eta_{\gamma}^{\text{therm}} + 
\eta_{\gamma}^{\text{scat}} \big) \, dS dl d\omega d\nu dt . 
\end{equation}
The conservation of spectral intensity (\ref{cons_radiation}) considering 
emission and extinction yields 
\begin{equation} \label{RTE_lab}
\int\limits_{V(t)} \Big( \frac{1}{c} \partial_t I_{\gamma} +
\mathrm{div} \, \big( I_{\gamma} \mathbf{n} \big) - \eta_{\gamma} + 
\chi_{\gamma} I_{\gamma} \Big) \, dV = 0 
\end{equation}
and is called \emph{radiation transfer equation} in the laboratory frame. 
In practice of RHD, upper equation and its momenta have yet to be transformed 
into the fluid frame as mentioned and simplified in several ways. 

Firstly the monochromatic quantities $I_{\gamma}, \chi_{\gamma}, \dots$   
will solely be considered with their \emph{collective} effects by integrating 
them over all frequencies\footnote{ Many authors use a notation, where 
the frequency $\nu$ is written as subset, instead of in the list of arguments
$I_{\nu}(\mathbf{x},t;\mathbf{n})$ and the frequency integrated variables 
lack of that subset $I, \chi, \dots$.}. 
\begin{eqnarray}
\nonumber \bar{J}_{\gamma}(\mathbf{x},t) & = & \frac{c}{4\pi} \bar{E}_{\gamma} 
 =  \frac{1}{4\pi} \int I_{\gamma} \, 
d\omega d \nu \\ 
\nonumber \bar{\mathbf{F}}_{\gamma}(\mathbf{x},t) & = &
4 \pi \bar{\mathbf{H}}_{\gamma}(\mathbf{x},t)  =  \int I_{\gamma} 
\mathbf{n} \, d\omega d\nu  \\ 
\bar{\mathbf{P}}_{\gamma}(\mathbf{x},t) & = & \frac{4\pi}{c} 
\bar{\mathbf{K}}_{\gamma}(\mathbf{x},t)  =  \frac{1}{c} \int I_{\gamma} 
\mathbf{n}\mathbf{n}^T \, d\omega  d\nu 
\end{eqnarray}
Another commonly used simplification to determine the proportion of 
extinction and emissivity by a source function $\bar{S}_{\gamma}$ is to 
impose local thermal equilibrium (LTE) where the radiation field is 
characterized by one single variable, the temperature $T$ 
(black body radiation). In this case, the specific intensity depends only on 
the gas temperature and is given by Planck's law 
\begin{equation}
I_{\gamma, \text{LTE}}(\mathbf{x},t;\nu) = \dfrac{2 h \nu^3}{c^2}
\dfrac{1}{e^{\dfrac{h\nu}{kT}}-1}
\end{equation}
and the collective, frequency integrated form yields the source 
function $\bar{S}_{\gamma}(\mathbf{x},t) : \mathbb{R}^{3+1} \to
\mathbb{R}^+$
\begin{equation} 
\bar{S}_{\gamma} = \frac{4 \pi \bar{\eta}_{\gamma}}{\bar{\chi}_{\gamma,P}} = 
\int\limits_0^{\infty} I_{\gamma, \text{LTE}} \, d\nu 
= \frac{8 \pi^5 k^4}{15 c^3 h^3} T^4 . 
\end{equation}
With $\bar{\chi}_{\gamma,P}$ we introduce a Planck mean opacity 
$\bar{\chi}_{\gamma,P}$ respectively the Planck mean opacity per mass 
$\bar{\kappa}_{\gamma,P}$ 
\begin{equation}
\bar{\kappa}_{\gamma, P} \rho = \bar{\chi}_{\gamma,P} = 
\frac{\int\limits_0^{\infty} \chi_{\gamma} 
I_{\gamma, \text{LTE}} \, d\nu}{\int\limits_0^{\infty} 
I_{\gamma, \text{LTE}} \, d\nu}
\end{equation}
and with $\bar{\chi}_{\gamma,R}$ a Rosseland mean which will emerge in the 
radiation energy equation. 
\begin{equation}
\frac{1}{\bar{\kappa}_{\gamma, R} \rho} = 
\bar{\chi}_{\gamma,R} = \frac{\int\limits_0^{\infty} \frac{1}{\chi_{\gamma}}
\dfrac{ \partial I_{\gamma, \text{LTE}}}{\partial T} \, d\nu}{ 
\int\limits_0^{\infty} \dfrac{ \partial I_{\gamma, \text{LTE}}}{\partial T} 
\, d\nu}
\end{equation}
At this point we have defined all necessary variables and motivated a list 
of simplifications for our system of radiation hydrodynamics equations. 
However, two important steps are to be provided yet. First, one has to deal 
with the problem that in particular equations (\ref{cons_radiation}) 
but respectively all $n$ moment equations contain in each case $n+1$ momenta 
of the radiation field which is known as \emph{closure problem}. 
The last step to the full set of equations of RHD will be the Lorentz 
transformation into the fluid frame. 

% SUBSECTION RADIATION CLOSURE RELATION ---------------------------------------%

\subsection{Radiation Closure Relation} \label{radiation_closure_relation}

In section \ref{Fund_Math_Hyp} it was mentioned, that imposing 
hyperbolicity on the conservation law can help to narrow possible solutions 
in some way. A conservation law is called hyperbolic if 
the Jacobian matrix associated with the fluxes $\nabla_{\mathbf{d}} \mathbf{f}$ 
has real eigenvalues and there exists a complete set of corresponding 
eigenvectors. As we have seen, interaction with matter adds source terms to 
our radiation conservation laws, we abbreviate them from now on 
with $S^0_{\gamma}$ and $\mathbf{S}^1_{\gamma}$. 
\begin{eqnarray}
\nonumber \partial_t J_{\gamma} +
\mathrm{div} \,  \big(c \mathbf{F}_{\gamma} \big)  & = & 
\underbrace{\int \big( \eta_{\gamma} - 
\chi_{\gamma} I_{\gamma} \big) \, d\omega}_{S^0_{\gamma}} \\
\partial_t \mathbf{F}_{\gamma} +
\mathrm{div} \,  \big( c^2 \mathbf{P}_{\gamma} \big) & = & 
\underbrace{\int \big( \eta_{\gamma} - \chi_{\gamma}
 I_{\gamma} \big) \mathbf{n} \, d\omega}_{
\mathbf{S}^1_{\gamma}} \label{rad_coupling_terms}
\end{eqnarray}
However, these terms do not influence if the system is hyperbolic or not since 
they appear without derivatives. We define the two \emph{Eddington factors}
(actually they are non-negative tensors) 
\begin{equation} 
\mathcal{F} = \frac{\mathbf{F}_{\gamma}}{J_{\gamma}}, \qquad 
\mathcal{P} = \frac{\mathbf{P}_{\gamma}}{J_{\gamma}}
\end{equation}
which can be interpreted as normalized flux and pressure and therefore 
are a {\sl measure of the geometry} of the radiative field.  It is reasonable 
to assume that $\mathcal{P}$ is only a fuction of $\mathcal{F}$ and henceforth
we write $d\mathcal{P}/d\mathcal{F}=\mathcal{P}'$. 
Likewise we define the 
norms of the Eddington factors  
$f = \left|\mathcal{F} \right|$ and $p = \left| 
\mathcal{P} \cdot \mathcal{F}/f \right|$. Now we 
express the density function and the flux of the conservation law by 
means of the Eddington factors which represents our choice of 
dependend and independend variables for the Jacobian. 
\begin{equation} \label{rad_hyp_sys}
\mathbf{d}_{\gamma} = \big( J_{\gamma}, c \mathcal{F} J_{\gamma} \big) 
, \qquad 
\mathbf{f}_{\gamma} =  \big( c \mathcal{F} J_{\gamma}, c^2 \mathcal{P}
J_{\gamma} \big)
\end{equation}
The Jacobian of the radiative conservation law yields\footnote{
Using $\left . \frac{\partial c \mathcal{F} J_{\gamma}}
{\partial J_{\gamma}}\right|_{( \mathcal{F} J_{\gamma})}=0$  and $\partial \frac{1}{\mathcal{F}} = -\frac{1}{\mathcal{F}^2} 
\partial \mathcal{F}$. }
\begin{equation}
\nabla_{\mathbf{d}_{\gamma}}\mathbf{f}_{\gamma} = 
\frac{\partial f^i}{\partial d^j} = 
\left( \begin{array}{cc} 0 & 1 
\\ c^2\big( \mathcal{P} - \mathcal{F}
\mathcal{P}' \big) & 
c\mathcal{P}' \end{array} \right) .
\end{equation}
For upper Jacobian we obtain the following 
characteristic equation 
\[
\lambda = 
\frac{1}{2} \left(c p' \pm \sqrt{c^2 \left(4 p+p' 
\left(-4 f+p'\right)\right)}\right) 
\]
and  system (\ref{rad_hyp_sys}) is hyperbolic if the 
discriminant is not negative i.e. 
$\left( p' - 2f\right)^2 + 4\left( p - f^2 \right) \geq 0$
respectively as long as the constraint $f^2 \leq p $ holds. 
Moreover, $f$ and $p$ as normalized moments obey $(p,f) \leq 1$ 
and we derive the important conclusion  
\begin{equation}
\label{schwarz_inequality} f^2 \leq p \leq 1 .
\end{equation}
Equation (\ref{schwarz_inequality}) can be interpreted as flux limiter, 
which ensures causal values for the flux function, namely that it can not get 
greater than the energy density $J_{\gamma}$ times the speed of light $c$. 
Physically it is clear that $f$ and $p$ take the value $1$ for transparency 
which is often called the free-streaming limit. In this case, radiation 
propagates freely and the total amount of intensity is transported by the flux 
function. The characteristic speeds in this regime evidently yield 
$\lambda= \pm c$. 

On the other end of the opacity scale, the optically thick regime, photons 
have very small mean free paths $l_{\gamma}$ and underlie 
random-walk processes through the material. In this case, the radiation 
field is totally isotropic, the net flux vanishes $\mathbf{F}_{\gamma}=0$ 
and the radiation stress tensor 
$\mathbf{P}_{\gamma}=\frac{1}{3} \mathbf{1} J_{\gamma}$ 
gets diagonal (for arguments see Appendix \ref{diffusion_approx_appendix}).  
Consequently the eigenvalues of the Jacobian yield $\lambda= \pm c \sqrt{1/3}$ 
which are the characteristic propagation speeds in this case. 

When we will discuss numerical high resolution methods for such nonlinear 
conservation laws in section \ref{Artificial_Viscosity_and_Flux} 
we will encounter the concept of flux limiting again. 
  To learn more about 
possible closure relations associated with the hyperbolicity approach, see 
Pons, Ibanez,  Miralles (2000) \cite{HyperbolicCharacter}.

% SUBSECTION FLUID FRAME ------------------------------------------------------%

\subsection{Fluid Frame} \label{Fluid_Frame}

In order to account for the local radiation properties of a fluid, we need to
consider the interaction of matter and radiation 
in a common frame, ideally the comoving frame of the fluid flow. 
The transformation however not only has to describe advection effects 
(i.e. change in photon number density), but also 
impacts on momentum, energy and directions due to the Doppler shift 
of the photons. 
For the purpose of simplicity we will ignore pure scattering terms and 
assume local thermal equilibrium (LTE) again. 

An adequate procedure, especially with possibly relativistic fluid flows 
allowed is to apply Lorentz transformations. Introductions 
to Minkowski geometry and Lorentz transformation can be found in 
several appropriate text books like Ray d'Inverno \cite{RayD}. 
At this point we just note that the relation between 
a four-vector $\mathbf{x}=(ct,\mathbf{r})$
in the lab frame and the moving frame $\mathbf{x} 
\rightarrow \mathbf{x}_0$ is given via 
\begin{equation}
\label{matrixlor} 
\left( \begin{array}{cc} ct_0 \\ {\mathbf{r}}_0 \\ \end{array} \right) =
\underbrace{\left( \begin{array}{cc} \gamma & -\frac{\gamma}{c} \mathbf{u} \\ 
-\frac{\gamma}{c} \mathbf{u} & \mathbf{1}  + (\gamma-1) 
\frac{\mathbf{u}\mathbf{u}}{u^2}\\ \end{array} \right)}_{\mathbf{L}}
\left( \begin{array}{cc} ct \\ \mathbf{r} \\ \end{array} \right) \\
\end{equation}
respectively ${x^{\alpha}}_0  = {L^{\alpha}}_{\beta} x^{\beta}$ with
$(\alpha, \beta)=0,1,2,3$. 
The subset $0$ denotes variables in the 
comoving frame, the gamma factor is defined by $\gamma^{-1} =
\sqrt{1-|\mathbf{u}|^2/c^2}$ where $\mathbf{u}$ is the fluid 
relative velocity. When we apply these rules to the 
photon four-momentum $p^{\alpha}_{\gamma}
= \frac{h\nu}{c}(1,n^{\alpha})$, we  obtain the required
relations for the photon propagation in the fluid frame. 
\begin{eqnarray}
\nonumber
\nu_0 & = & \gamma \nu \left( 1 - \frac{\mathbf{n} \cdot 
\mathbf{u}}{c}\right) \\
\mathbf{n}_0 & = & \frac{\nu}{\nu_0} \Big( \mathbf{n} - \gamma 
\frac{\mathbf{u}}{c} \big( 1- \frac{ \gamma \frac{\mathbf{n} \cdot 
\mathbf{u}}{c}}{\gamma+1} \big)\Big) 
\end{eqnarray}
These formulae for redshift (respectively blueshift) and aberration often 
are yet simplified for practical astrophysical applications, 
arguing that terms $\mathcal{O}(\frac{u}{c})^2$ can be 
neglected since fluid velocities are at least one magnitude
smaller than the speed of light. This reduction also implies that the 
outstanding transformation is mere spatial as with 
$\gamma= \frac{\partial t}{\partial t_0} =1$ follows $t=t_0$. 

When we intend to transform the radiative quantities like the 
spectral intensity and its derived variables 
we need to occupy ourselves with relativistic 
kinetics respectively with a relativistic form of the Boltzmann transport 
equation. To wit, the spectral intensity
itself is not Lorentz invariant but the 
photon distribution function $f_{\gamma}$ defined earlier. 
These considerations date back to Thomas (1930) \cite{THOMAS01011930} 
respectively Lindquist (1966) \cite{Lindquist1966487}, who drafted
a general relativistic formulation. 

The connection between the spectral intensity, emissivity 
and absorption in the lab frame and in the comoving frame 
accordingly yield
\begin{equation} 
I_{\gamma} = \left(\frac{\nu}{\nu_0}\right)^3 {I_{\gamma}}_0, 
\quad \eta_{\gamma} = \left(\frac{\nu}{\nu_0}\right)^2 {\eta_{\gamma}}_0, 
\quad \chi_{\gamma} = \frac{\nu}{\nu_0} {\chi_{\gamma}}_0 .
\end{equation}
The frequency integrated moments of the radiation field occur as 
elements of the radiation energy-stress tensor which is defined by 
\begin{equation} T^{\alpha \beta}_{\gamma} = c^2 \int 
\frac{d\mathbf{p}}{\epsilon_{\gamma}}
p^{\alpha} p^{\beta} f_{\gamma} \end{equation}
and can easily be rewritten to a more familiar form\footnote{Using 
$p^{\alpha}_{\gamma}= \frac{h\nu}{c}(1,n^{\alpha})$ and 
$\epsilon_{\gamma}= h \nu$.}. 
\begin{equation} 
\label{TEFP} \mathbf{T}_{\gamma} =  \frac{1}{c} \int \int 
\left( \begin{array}{cc} 1 & \mathbf{n} \\ 
\mathbf{n} & \mathbf{n}\mathbf{n} \end{array} \right) 
I_{\gamma} \, d\nu d\Omega = 
\left( \begin{array}{cc} \bar{E}_{\gamma} & \bar{\mathbf{F}}_{\gamma}/c \\ 
\bar{\mathbf{F}}_{\gamma}/c & \bar{\mathbf{P}}_{\gamma} 
\end{array} \right) 
\end{equation}
The conservation law associated with radiation transfer 
gains its most elegant form in this notation, 
namely $\nabla \cdot \mathbf{T}_{\gamma} = \mathbf{S}_{\gamma}$ with source 
terms $\mathbf{S}_{\gamma}=(S^0_{\gamma},\mathbf{S}^1_{\gamma})$ 
as defined before. 
The Lorentz transformation of $\mathbf{T}_{\gamma}$ yields 
${\mathbf{T}_{\gamma}}_0 = \mathbf{L}
\mathbf{T}_{\gamma} \mathbf{L}$ as the operator
is unitary and thus $\mathbf{L}=\mathbf{L}^T$.s When we confine ourselves to 
small velocities again and discard terms higher $\mathcal{O}(\frac{u}{c})^2$, 
we would receive the applicable approximation formulae.  

What is left to do, is to mark the moments of the transport 
equation in this classical limit. For forcible 
deduction we refer to 
Buchler (1979) \cite{Buchler1979293} and (1983) \cite{RobertBuchler1983395}
who elaborated a fluid frame description in arbitrary geometry inclusive 
of terms $\mathcal{O}(\frac{u}{c})$. The frequency integrated 
radiation energy equation and the radiation flux equation  
in the comoving frame accordingly yield 
\begin{eqnarray}
\nonumber
\partial_t \bar{J}_{\gamma 0} + \mathrm{div} \, \big(  \mathbf{u} 
\bar{J}_{\gamma 0}
 \big) + c \, \mathrm{div} \, \bar{\mathbf{H}}_{\gamma 0} + 
\bar{\mathbf{K}}_{\gamma 0} : \mathrm{grad} \, \mathbf{u} 
- c \bar{\chi}_{\gamma,P 0} \big( \bar{J}_{\gamma 0} 
- \bar{S}_{\gamma 0}\big) & = & 0 
\\ \nonumber 
\partial_t \bar{\mathbf{H}}_{\gamma 0} + \mathrm{div} \, \big( \mathbf{u} 
\bar{\mathbf{H}}^T_{\gamma 0} \big) +
c \, \mathrm{div} \, \bar{\mathbf{K}}_{\gamma 0} + \bar{\mathbf{H}}_{\gamma 0}
 \cdot \mathrm{grad} \, \mathbf{u} + c \bar{\chi}_{\gamma, R 0}
\bar{\mathbf{H}}_{\gamma 0} & = &  0 . \\
\label{rad_en_flux_def}
\end{eqnarray}
Notation: from now on, the neat radiation variables 
$(J,\mathbf{H}, \dots)$ shall be read 
as photonic, frequency integrated functions in the comoving frame 
$(\bar{J}_{\gamma 0},\bar{\mathbf{H}}_{\gamma 0},\dots)$.

% SECTION GRAVITATION ---------------------------------------------------------%

\section{Gravitation}

In section \ref{Fund_Math_Ell} we highlighted some mathematical basics for 
the Laplace and the Poisson equation in context of elliptical 
partial differential equations. The Poisson equation for the gravitational potential, a scalar 
function $\Phi_{\text{grav}} (\mathbf{x},t): 
\mathbb{R}^{n+1} \to \mathbb{R}$ yields 
\begin{equation} \label{Poisson_grav}
\Delta \Phi_{\text{grav}} = 4 \pi G \rho_{\text{m} }
\end{equation}
where $\rho_{\text{m} }(\mathbf{x},t) : \mathbb{R}^{n+1} \to \mathbb{R}^+$ is the 
time dependend mass density again in $n=1,2,3$ spatial dimensions. 

However, most applications of 
radiation hydrodynamics with gravitation avoid solving the Poisson equation 
for the gravitational potential per se, especially in spherical symmetric 
applications\footnote{With radial symmetry, the gravitational acceleration 
$\mathbf{G}(r)$ is simply determined by the integrated mass 
$m(r)=\int_0^r 4\pi \rho_{\text{m} }(r') dr'$ 
via the well known relation $\mathbf{G}(r) = - \partial_r \Phi_{\text{grav}} = 
- \frac{G m(r)}{r^2}$, 
an integrated form of the Poisson equation.}.
The main challenge associated with self gravitation in 2D and 3D is 
that it requires the numerical solution of the Poisson equation in each 
time step which is computationally expensive. 
Moreover, it is not trivial to find appropriate boundary 
conditions for this elliptical problem in astrophysical context since 
most objects are not bounded naturally but artificially respectively 
numerically.

\subsection{Self-Gravitation in 2D and 3D} 

There have been efforts in finding and applying 
solution methods for self-gravitating 
flows in multiple dimensions since decades. Differential algorithms 
solve the Poisson equation in the form (\ref{Poisson_grav}) whereas 
integral methods rather consider the theory of Greens 
functions as briefly mentioned in section \ref{Fund_Math_Ell} where 
the formal solution in 3D yields 
\[
\Phi_{\text{grav}} = -
G \int \dfrac{\rho(\mathbf{x})}{\mathbf{x}-\mathbf{x}'} \, d\mathbf {x} .
\] 

With latter family of algorithms the grids boundary causes no problems 
and the integration could principally be conducted with standard 
methods. However, the number of grid cells to be summed ($\sim N^2$) and 
therewith the computational cost get huge in multiple dimensions. 
Several authors thus suggest to expand the integrand e.g. in 
spherical harmonics in order to reduce operations to an order $\sim N$.
M\"uller and Steinmetz (1995) \cite{Mueller199545} developed an 
efficient Poisson solver based on such a separation ansatz. 
Cosmological hydrodynamic simulations have affiliated these ideas and 
implemented them in a number of cosmological codes. 

Spingel (2009) \cite{Springel2009} suggests an interesting method to 
realize self gravitation in a conservative way by 
reformulating the total energy equation. He 
adds a {\sl gravitational source term} 
on the right hand side of equation (\ref{energy_eq_def})  
whereby it gains the form of a
conservation law that ensures the total energy to remain constant. 
The total energy of the system consequently yields the sum over all 
integrated cell energies. In stars however this modification is rather 
inapplicable since the contribution by potential energy exceeds 
all other energy terms by far and this term would numerically dominate 
the energy equation which is unfit in practice. 
Hence we stick with the formulation where 
changes in potential energy directly contribute to the momentum 
equation (see the full set of equations \ref{full_set}). 
Alternatively we suggest a {\sl non-static} equation for the gravitational 
force in multiple dimensions. 

\subsection{Non-Static Gravitation Equation} \label{grav_idea}

The gravitational potential $\Phi_{\text{grav}}$ is not needed explicitly 
at any point of the computation since coupling is accomplished 
via the gravitational force respectively the gradient of the 
potential i.e. $\mathrm{grad} \, \Phi_{\text{grav}}$. 

We define the 
gravitational force vector $\mathbf{G}(\mathbf{x},t) : 
\mathbb{R}^{n+1} \to \mathbb{R}^{n+1}$ in the familiar way  
\begin{equation}
\mathbf{G} = - \mathrm{grad} \, \Phi_{\text{grav}}
\end{equation}
hence the Poisson equation can be written 
$\mathrm{div} \, \mathbf{G} = - 4 \pi G \rho$.
We differentiate with respect to $t$ and use the continuity equation 
(\ref{cont_eq_def}) to obtain
\begin{eqnarray}
\nonumber
\partial_t \, \mathrm{div} \, \mathbf{G} & = & - 4 \pi G \, \partial_t \rho \\
\nonumber 
& = &  4 \pi G \, \mathrm{div} \, \big( \rho \mathbf{u} \big) . 
\end{eqnarray}  
Since $\Phi_{\text{grav}}$ is $\mathcal{C}^{\infty}$ 
it is reasonable to assume that the derivatives with respect to 
$t$ exists and is continuous as well, the operations can be 
interchanged. 
We consider the conservative, integral form of 
these equations again and use the divergence theorem. 
\begin{eqnarray}
\nonumber 
\int\limits_{V(t)} \mathrm{div} \, \partial_t \, \mathbf{G} \, dV & = & 
\int\limits_{V(t)} 4 \pi G \, 
\mathrm{div} \, \big( \rho \mathbf{u} \big) \, dV \\
\label{int_dyn_grav}
\int\limits_{\partial V(t)} \partial_t \, \mathbf{G} \cdot d\mathbf{S} & = & 
4 \pi G \int\limits_{\partial V(t)}\! \big( \rho \mathbf{u} \big)
\cdot d\mathbf{S} 
\end{eqnarray}
We obtain an equation for the gravitational force in form of a 
pure initial value problem. 
\begin{equation} \label{dyn_grav}
\partial_t \, \mathbf{G} = 4 \pi G\, \rho \mathbf{u} 
\end{equation}
This way we avoid solving the Poisson equation in multiple dimensions by 
solely considering its gradient, the gravitational force. 
Equation (\ref{dyn_grav}) describes gravitation
non-statically by understanding temporal changes of gravitational 
forces as mass transport. If, respectively in what extent, this formulation is 
numerically preferable will have to be studied in future computations. 

Self-gravitation expressed via (\ref{dyn_grav}) would not have to be 
solved at any time of the evolution but only initially contrary to 
Poisson equation implementations. 
Hence it is supposedly especially suited for explicit and
parallelized methods.

\newpage 
% SECTION Full Set of Equations -----------------------------------------------%

\section{Full Set of RHD Equations with Self-Gravitation } \label{full_set}

The system of equations of radiation hydrodynamics with gravitation is 
now presented in its for our purposes somewhat simplified formulation. 
The following system 
has been basis for a number of implicit RHD computations like 
Dorfi (1999) \cite{Dorfi1999153} and St\"okl (2006) \cite{Stoekl:Diss}. 
In latter PhD thesis physical assumptions and simplifications are 
illustrated and their consequences executed more detailed than in this paper. 

\begin{description}

\item[Continuity Equation] 
\begin{equation}
\partial_t \rho + \mathrm{div} \, (\mathbf{u} \rho) =
0 
\end{equation}

\item[Equation of motion]
\begin{equation}
\partial_t \big( \rho  \mathbf{u} \big) + 
\mathrm{div} \, \big( \rho \mathbf{u} \mathbf{u} + \mathbf{P} - 
\mathbf{\sigma} \big)  - \rho \mathbf{G} - \frac{4\pi}{c} 
\chi_R \mathbf{H}  = 0  
\end{equation}

\item[Equation of Internal Energy] 
\begin{equation}
\partial_t \big( \rho \epsilon \big) + \mathrm{div} \, \big(
 \rho \epsilon \mathbf{u} + 
\big( \mathbf{P} - \sigma \big) \cdot \mathbf{u} \big)
 - 4 \pi \chi_P (J-S) = 0 
\end{equation}

\item[Equation of Radiation Energy] 
\begin{equation}
\partial_t {J} + \mathrm{div} \, \big( \mathbf{u} 
{J} \big) + c \, \mathrm{div} \, {\mathbf{H}} + 
{\mathbf{K}} : \mathrm{grad} \, \mathbf{u} 
- c {\chi}_{P} \big( {J} - {S}\big)  =  0 
\end{equation}

\item[Radiation Flux Equation] 
\begin{equation}
\partial_t {\mathbf{H}} + \mathrm{div} \, \big( \mathbf{u} 
{\mathbf{H}} \big) +
c \, \mathrm{div} \, {\mathbf{K}} + {\mathbf{H}} \cdot 
\mathrm{grad} \, \mathbf{u} + c {\chi}_{R}
{\mathbf{H}} =  0
\end{equation} 

\item[Poisson Equation]
\begin{equation}
\Delta \phi = 4 \pi G \rho 
\end{equation}
respectively 

\item[Non-Static Gravitation Equation]
\begin{equation} 
\partial_t \, \mathbf{G} = 4 \pi G\, \rho \mathbf{u} 
\end{equation}
\end{description}
Basically we collected equations (\ref{cont_eq_def}), (\ref{cons_EOM}), 
(\ref{energy_eq_def}) plus (\ref{rad_en_flux_def}) and added
gravitation. 
The viscosity $\sigma$ will 
be replaced by an artificial viscous pressure $-\mathbf{Q}$ which we establish 
in section \ref{Artificial_Viscosity_and_Flux}.  
In section \ref{Covariant_RHD_Spher} we present this system of equations 
in strong conservation form for non-steady curvilinear 
coordinates.

% ---------------------------------------------------------------------------- %
% CHAPTER Conservative Numerics ---------------------------------------------- %
% ---------------------------------------------------------------------------- %

\chapter{Conservative Numerics} \label{Conservative Numerics}

%\begin{center}
%\parbox{13cm}{\sl(\ref{cons_entro})
In this chapter we want to develop some numerical methods for conservation 
laws after having motivated and anticipated relevant mathematical concepts in 
the previous sections. Concepts for error functions, convergence and 
stability for linear and nonlinear equations will be 
studied in form of an overview. 
A particular focus will lie on conservative methods 
for nonlinear problems in general curvilinear coordinates which will 
require some tensor analysis and differential geometry.  

In sections \ref{ConsGeneral} 
and  \ref{Covariant_RHD_Spher} we will present the mathematical framework 
for strong conservative numerics in non-steady curvilinear 
coordinates and exemplarily 
sketch the system of RHD in spherical and polar coordinates. 
An articulate introduction 
and motivation to computational methods in astrophysical context can 
be found in LeVeque, Mihalas, Dorfi, M\"uller \cite{DorfiSaas-Fee}. 
%}
%\end{center}

% SECTION Numerical Methods for Conservation Laws -----------------------------%

\section{Numerical Methods for Conservation Laws}

Standard numerical methods for partial differential equations are 
established under the assumption of (classical) differentiability. 
Routine finite difference schemes of first order usually smear or 
{\sl smoothen} the solution 
in the vicinity of discontinuities, since they come with intrinsic numerical 
viscosity. Standard second order methods show something to the effect of 
the Gibbs phenomenon, where we see oscillations around shocks. 
In the past decades so called {\sl high resolution methods}
have been developed 
in order to achieve proper accuracy and resolution for nonlinear, discontinuous 
problems as they appear also in radiation hydrodynamics. 
In the following sections we will introduce some basic technical terms 
in discrete numerical methods for conservation laws and present some 
important concepts for nonlinear problems.

\subsection{Introduction to Discretization Methods}

The basic idea of discretization methods for PDEs is to represent 
functions by a finite amount of data and to approximate 
differential operators by difference operators. To simplify matters, we
confine ourselves to $1+1$ dimensional problems and constant increments 
in the upcoming introductory sections. Our toy model for analysis shall 
be the time-dependent Cauchy problem (\ref{cauchy_problem}), namely
$\partial_t \mathbf{d} + 
\mathrm{div} \, \mathbf{f}(\mathbf{d}) = \mathbf{0} $
but in linear form and one space dimension\footnote{
We maintain the {\sl tensor notation} 
in boldface $\mathbf{d}$ for densities and fluxes for distinction purposes
als in 1+1 dimensions.} 
$\partial_t \mathbf{d}+ 
A \, \partial_x \mathbf{d} = 0$ 
with constant $A$. Concerning structure and nomenclature we basically
follow LeVeque \cite{LeVeque1990} in the following sections. 

For the spatial part we introduce a mesh width $h=\Delta x$ and a 
time step $k=\Delta t$ for the temporal discretization. 
Any point in this computational domain can be attained 
by multiple steps in $t$ and $x$, that is $(x_j,t_n)=(jh,nk)$, where 
$j \in \mathbb{Z}$ and $n \in \mathbb{N}$ with 
starting point or initial condition $\mathbf{d}(x_j,t_0)$. 
What we want to obtain is 
the approximate solution of the exact
$\mathbf{d}(x,t)$ at discrete grid points $\mathbf{d}(x_j,t_n)$
and we characterize these approximation as $\mathbf{D}^n_j$. 

The difference scheme is somewhat arbitrary and its design 
depends on the 
concrete problem to be approximated. The essential requirement is that in
the limit of $(j,n) \to 0$ the difference operator reproduces the
differential operator. The method is called explicit when the function 
at the new time step is calculated directly via function values at prior 
time steps. We introduce a special notation for such operators with 
$\mathcal{H}(\mathbf{D} ^n;j)=\mathbf{D} _j^{n+1}$ which 
reproduces the vector of approximations 
at the new time step $\mathbf{D} ^{n+1}_j$. 
Some well known explicit schemes for our toy model are 
for example the so called 
\emph{Upwind}, \emph{Lax-Friedrichs}, \emph{Leapfrog} and \emph{Lax-Wendroff}
 method. 
\begin{eqnarray}
\nonumber \mathcal{H}(\mathbf{D} ^n;j) & = & 
{\mathbf{D} }_j^n-\frac{k}{h}{A}({\mathbf{D} }_j^n-{\mathbf{D} }_{j-1}^n) \\
\nonumber & = & \frac{1}{2}({\mathbf{D} }_{j-1}^n+{\mathbf{D} }_{j+1}^n)-
\frac{k}{2h}{A}({\mathbf{D}}_{j+1}^n-{\mathbf{D} }_{j-1}^n )\\
\nonumber & = & {\mathbf{D} }^{n-1}_j-
\frac{k}{h}{A}({\mathbf{D} }_{j+1}^n-{\mathbf{D} }_{j-1}^n) \\
\nonumber 
& = & {\mathbf{D} }_j^n-\frac{k}{2h}{A}({\mathbf{D} }_{j+1}^n-
{\mathbf{D} }_{j-1}^n)+
\frac{k^2}{2h^2}{A}^2({\mathbf{D} }_{j+1}^n-2 
{\mathbf{D} }_j^n+{\mathbf{D} }_{j-1}^n) \\
\label{expl_schemes} 
\end{eqnarray}

\subsection{Error and Stability in the Linear Case}

There is a number of criteria that reveal the quality of a numerical 
method, however the most obvious is to define an error function which 
states the difference between the exact and the numerical 
approximate solution. 
In this sense the pointwise error is defined by
\begin{equation}
\mathbf{E}_j^n = \mathbf{D}_j^n - \mathbf{d}_j^n .
\end{equation}
In conjunction with conservation laws the numerical solution is often 
seen as an approximation to a cell average of $\mathbf{d}(x_j,t_n)$, 
defined by 
\begin{equation} \label{cell_average}
{\bar{\mathbf{d}}}_j^{n} = \frac{2}{x_{j+1} - x_{j-1}} 
\int\limits_{x_{j-1/2}}^{x_{j+1/2}} \mathbf{d}(x',t_n) \, dx' \mathrm{,} 
\end{equation}
where the half step is simply $x_{j+1/2}=x_j+h/2$
and the pointwise error in terms of cell averages yields
\begin{equation}
\bar{\mathbf{E}}_j^n = \bar{\mathbf{D}}_j^n - \bar{\mathbf{d}}_j^n .
\end{equation}
Let be $\mathbf{D} _k(x,t)=\mathbf{D} _j^n$ for $(x,t) \in [x_{j-1/2},x_{j+1/2} )
\times [t_n,t_{n+1})$, then the \emph{error function} in terms of 
piecewise constant functions is defined by
\begin{equation} \mathbf{E}_k(x,t)= \mathbf{U}_k(x,t)-\mathbf{u}(x,t). 
\end{equation}
We have already mentioned that the difference operator has to approximate 
the differential operator properly. Regarding the temporal iteration we 
want that smaller time steps cause smaller errors. 
We call the numerical method 
\emph{convergent} (in a norm\footnote{
For conservation laws the a practicable norm is the 1-norm 
$\| \dots \| = \int_{-\infty}^{\infty} | \dots | \, dx$, since 
more rigorous demands come into conflict with approximation 
of discontinuities; $\| \mathbf{D}_k(.,t_n) \|_1 = 
h\sum_j | \mathbf{D}_j^n | $.}), 
if for arbitrary initial condition 
$\mathbf{d}_0$,  $\forall \, t >0$ 
\begin{equation} 
\lim_{k\rightarrow0} \| \mathbf{E}_k( \cdot ,t) \|=0 .
\end{equation}
Assuming smooth solution we always can expand the exact solution into a 
Taylor series and compare the leading terms to the numerical scheme. 
By that means we get a measure of how well the difference equation
reproduces the difference equation locally. The \emph{local 
truncation error} is therefore defined by 
\begin{equation}
\mathbf{L}_k(x,t) = \frac{1}{k} \Big( \mathbf{d}(x,t+k) -
\mathcal{H}\big( \mathbf{d}(., t); x\big)  \Big) 
\end{equation}
and we call the numerical method \emph{consistent}, if 
\begin{equation} 
\lim_{k \rightarrow 0} \| \mathbf{L}_k(.,t)\| = 0 .
\end{equation}
The method is of \emph{order} $p$, if for all sufficiently smooth initial 
data with compact support, there is some constant $C_L$, such that
\begin{equation} 
\label{order} \| \mathbf{L}_k(.,t)\| \leq C_{L} k^p, \qquad
\end{equation}
for all $k < k_0$ and $t < T$. The restriction to a local $T$ falls 
for stable methods, where the order remains the same for any $t$ or $k$. 

There is a profound stability theory for linear methods; 
its details 
can be found in several relevant text books (e.g. Richtmeyer, Morton 
\cite{Richtmeyer1994}). Nevertheless, at this point we shall
mention the \emph{Lax-Richtmeyer stability} which states that a method is 
stable if for each time $T$ there is a constant $C_S$ and a value 
$k_0 > 0$ such that 
\begin{equation}
\left\| \mathcal{H}^n_k \right\| \leq C_S 
\end{equation}
for all $nk \leq T$, where $k< k_0$. 
The important \emph{Lax equivalence theorem} claims 
that for consistent linear 
methods stability is necessary and sufficient for convergence.

In this context we also should 
mention the \emph{Courant, Friedrichs, Lewy (CFL) condition}, which 
seizes the fact that hyperbolic equations stand out due to a bounded
\emph{domain of dependence}. The solution at a point $\mathbf{d}(x,t)$ is 
determined by initial data within some finite distance to this point. 
This range of influence is related to the fact, that hyperbolic 
equations always have finite propagation speed, as briefly mentioned in 
the introduction to section \ref{Fund_Math_Hyp}. When Courant, Friedrichs 
and Lewy studied convergence of numerical schemes for hyperbolic 
conservation laws, they observed that one necessary condition for convergence 
is that the domain of dependence of the finite difference scheme includes 
the domain of dependence of the PDE. In other words, data from one time point 
must not propagate faster than the characteristic speed 
of the original equation. Mathematically worded, the domain of 
dependence $\mathcal{D}(\overline{x},\overline{t})$ is a set of points $(x,t)$ 
with the property that a disturbance of $\mathbf{d}(x,t)$ 
influences the solution at $(\overline{x},\overline{t})$. 
The numerical domain of influence $\mathcal{D}_k(\overline{x},\overline{t})$ is
analogously defined. For the satbility of an (explicit) method it is 
necessary that for $h,k \rightarrow 0$ the numerical domain of dependence 
$\Omega_k(\overline{x},\overline{t})$ contains asymptotically the 
domain of dependence the corresponding initial value problem. Practically, 
an inevitable (but not sufficient) stability condition for such methods 
is represented by a limited ratio of time step to mesh size. For a scalar 
advection equation with propagation speed $a$, the CFL condition yields 
$0 \geq \frac{ak}{h} \geq 1$ as an example.

\subsection{Conservative Methods and Lax Wendroff Theorem}

In the linear case, the Lax equivalence theorem states that a consistent 
and stable discretization scheme automatically converges to the 
weak solution of the conservation law. For nonlinear equations, 
consistent methods do not necessarily converge to the weak solution of the 
conservation law\footnote{E.g. the CIR method, a nonlinear 
generalization of the Upwind 
scheme, is not applicable for discontinuous solutions, since it is not 
conservative. }. We have already mentioned that nonlinear conservation 
laws can have several weak solutions but we are only interested in those, 
that also satisfy the entropy condition. So it is not surprising that 
it is required to make sure, the numerical scheme satisfies some 
discrete analogon of the entropy condition (\ref{cons_entro}). 

The proper postulation to the numerical scheme that prevents convergence 
to non-solutions is that the difference method has \emph{conservation form}, 
i.e. 
\begin{equation} \label{conservation_form}
\mathbf{D}_j^{n+1} = \mathbf{D}_j^n - \frac{k}{h} \Big( 
\mathbf{F} \big( \mathbf{D}_{j-p}^n, \dots, \mathbf{D}_{j+q}^n \big) - 
 \mathbf{F} \big( \mathbf{D}_{j-p-1}^n, \dots, \mathbf{D}_{j+q-1}^n \big) \Big) 
\end{equation}
for a \emph{numerical flux function} $\mathbf{F}$ of $p+q+1$ arguments. 
This numerical flux can be interpreted as averaged flux 
through $x_{j+1/2}$ and $[t_n, t_{n+1} ]$ in the sense that $\mathbf{F}
\cong \frac{1}{k} \int_{t_n}^{t_{n+1}} \mathbf{f} 
\big( \mathbf{d} (x_{j+1/2},t) \big) \, dt$. In section \ref{ConsGeneral} 
we will turn our intensive 
attention to this numerical quantity when we concern ourselves with 
conservative methods in general coordinates.  

We also reformulate the concept of \emph{consistency for conservative schemes}. 
The conservative method is consistent with the original conservation law, if 
the numerical flux function $\mathbf{F}$ reproduces the true flux $\mathbf{f}$ 
in the case of constant flow $\bar{\mathbf{d}}$ so that 
\begin{equation}
\mathbf{F}(\bar{\mathbf{d}}, \dots, \bar{\mathbf{d}}) = f(\bar{\mathbf{d}}) .
\end{equation}
Moreover, we require as the arguments of $\mathbf{F}$ approach a common value
$\bar{\mathbf{d}}$, that the value of $\mathbf{F}$ meets 
$\mathbf{f}(\bar{\mathbf{d}})$ smoothly\footnote{
The requirement is Lipschitz continuousity which is exigently given 
for any smooth and consequently differentiable flux function $\mathbf{f}$.}. 
So for each $\bar{\mathbf{d}}$ 
there may exist	 some constant $K > 0$ so that 
\begin{equation} 
\left\| \mathbf{F} \big( \mathbf{D}_{j-p}, \dots, \mathbf{D}_{j+q} \big)  -
 \mathbf{f}(\bar{\mathbf{d}}) \right\| \leq \, K \! \max_{-p \leq l \leq q} 
\left\| \mathbf{D}_{j+l} - \bar{\mathbf{d}} \right\| .
\end{equation}

Before we can establish a nonlinear convergence theorem we first have to 
define a \emph{total variation} of an arbitrary $\mathcal{L}^{\infty}$ 
function $\mathbf{v}$. 
\begin{equation}
\mathrm{TV}(\mathbf{v}) = \lim\sup_{\varepsilon \to 0} \frac{1}{\varepsilon}
\int\limits_{- \infty}^{\infty}
\left\| \mathbf{v}(x) - \mathbf{v}(x - \varepsilon) \right\|
\, dx
\end{equation}
which reduces to a more familiar look for differentiable $\mathbf{v}$
respectively appropriate distribution derivatives, namely 
$\mathrm{TV}(\mathbf{v})=
\int_{- \infty}^{\infty} \left\| \mathbf{v}' \right\|\, 
dx$. We call a sequence of solutions $\mathbf{D}_i$ \emph{convergent} 
to a solution $\mathbf{d}$, if for every bounded set $[a, b] \times [0,T]$ 
\begin{equation}
\int\limits_0^T \int\limits_a^b \left\| \mathbf{D}_i (x,t) - 
\mathbf{d}(x,t)\right\|\, dx dt \, \to 0 \quad \text{for} \quad i \to \infty
\end{equation}
and there exists an upper bound $R$ for the total variation such that 
\begin{equation}
\mathrm{TV}\big(\mathbf{D}_i(\cdot, t) \big) < R 
\end{equation}
for $0 \leq t \leq T$ and $i \in \mathbb{N}$. 

This leads us to the essential 
\emph{Lax-Wendroff theorem} which averts the following. Let $\left\{ 
\mathbf{D}_i(x,t) \right\}$ be a sequence of solutions of a 
conservative and consistent difference method 
for the one dimensional conservation law, 
where the mesh parameters $(k,h) \to 0$ for $i \to \infty$. Suppose the
numerical approximation at the $i$-th grid 
$\left\{ \mathbf{D}_i(x,t) \right\}$ 
converges to a function $\mathbf{d}(x,t)$ in the sense made 
precise before; then $\mathbf{d}(x,t)$ is a weak solution of the 
conservation law. 

\noindent Still, the consistent conservative scheme can converge to a 
weak solution that does not satisfy the entropy condition 
(\ref{cons_entro}). We add a comparable 
condition for the numerical method in form of the 
\emph{Lax-Wendroff corollary}. Let $\sigma$ entropy function and $\phi$ 
entropy flux and let be $\Phi$ the consistent (in the above defined sense)
corresponding numerical flux function. Given the Lax-Wendroff theorem 
is satisfied and 
\begin{equation}
\sigma\big(\mathbf{D}_j^{n+1}\big) \leq \sigma\big(\mathbf{D}_j^n\big) 
- \frac{k}{h} \Big( \Phi \big( \mathbf{D}^n,j\big) - \Phi \big( 
\mathbf{D}^n,j-1 \big)\Big)
\end{equation}
holds, then $\mathbf{d}(x,t)$ satisfies entropy condition 
(\ref{entropy_condition}).

As in the linear case, the picture is not drawn completely until we 
clarify stability of the numerical method. All we know now is, that 
if a sequence of approximation converges, the limit is a weak solution. 
However, we require intrinsic 
stability of the method in order to ensure convergence of the 
difference scheme. Without going into details,  
\emph{total variation stability} is the suitable concept for nonlinear 
methods, for which the Lax equivalence theorem is not applicable. 
It can be shown that TV-stability, consistency and conservativeness 
guarantee nonoscillatory convergence of the advection scheme. 

As a further remark to the topic of methods for nonlinear conservation 
laws, we want to present an important class of methods 
that ensure TV-stability. 
The idea goes back to Harten (1983) \cite{Harten1997260}, who was 
studying high resolution methods for hyperbolic conservation laws and 
proposed so called \emph{total variation diminishing (TVD)} 
(originally \emph{nonincreasing}) methods. We call a numerical scheme 
$ \mathcal{H}(\mathbf{D} ^n;j)$ total variation diminishing if 
\begin{equation}
\mathrm{TV}( \mathbf{D}^{n+1} ) \leq \mathrm{TV} (\mathbf{D}^n)
\end{equation}
for all grid functions $\mathbf{D}^n$ at all time steps $n$. 
Of course, the true solution of the conservation law must be compatible with 
this postulation and indeed it can be shown, that any weak solution 
$\mathbf{d}(x,t)$ satisfies $\mathrm{TV}( \mathbf{d}(\cdot,t_{n+1}) ) 
\leq \mathrm{TV} (\mathbf{d}(\cdot, t_n))$ for any $n>0$. 

In the upcoming two subsections we want to present some 
sophisticated high resolution numerical methods incorporating these 
techniques for nonlinear conservation laws. 
The emphasis will lie on concepts and techniques relevant for our 
RHD problem as they have been implemented in applications like 
Dorfi (1999) \cite{Dorfi1999153} 
respectively Dorfi et al. (2004) \cite{Dorfi2006771}

\subsection{Artificial Viscosity} 
\label{Artificial_Viscosity_and_Flux}

The TVD approach is not the only one to find entropy-satisfying weak solutions 
with numerical schemes. One of the simplest techniques are monotone 
methods\footnote{The more general concept is called $l_1$-contracting 
methods, see e.g. Harten et al. (1976) \cite{Harten1976}.}, 
however the have one main disadvantage. Monotone methods are at most 
first order accurate. In the past decades major efforts were made in 
developing numerical methods for nonlinear hyperbolic conservation laws 
that are at least second order. 

One patent attempt to finding such a high resolution method is to adapt 
a well known high order method like Lax-Wendroff for nonlinear problems. 
As we have contemplated in section \ref{Rankine_and_Entropy} we can add an 
\emph{artificial viscosity} term to the conservation law in a way that 
the entropy condition is satisfied. Numerically we are keen to design this 
viscosity in a manner that it affects discontinuities but vanishes 
sufficiently elsewhere so that the order of accuracy can be maintained those 
regimes where the solution is smooth. Interestingly this idea 
was inspired by physical dissipation 
mechanisms and dates back 
more than half a century to VonNeumann and Richtmeyer 
(1950) \cite{vonneumann:232} who investigated new methods to treat shocks 
in hydrodynamic simulations. Bearing in mind that we have to 
provide conservation form, 
the numerical flux function gets modified by a an artificial viscosity 
$\mathbf{q}(\mathbf{D};j)$ for instance in the following way.
\begin{equation}
\mathbf{F}_{\text{visc}} ( \mathbf{D};j ) = \mathbf{F} 
( \mathbf{D};j ) - h \mathbf{q} \big( \mathbf{D};j ) ( 
\mathbf{D}_{j+1} - \mathbf{D}_j \big)
\end{equation}
Since the original design of this additional viscous 
pressure in the form $q = c_2 \rho (\Delta \mathbf{u})^2$, $c_2 \in
\mathbb{R}$, for one dimensional advection $\partial_t \mathbf{d} + a 
\partial_x \mathbf{d} = q \partial_{xx} \mathbf{d}$ it underwent a number 
of modifications and generalizations. It turned out to be numerically 
prerferable to add a linear term, see Landshoff (1955) 
\cite{Landshoff} in order to control 
oscillations, generalizations to multi-dimensional flows mostly 
retained the original analogy to physical dissipation 
and reformulated the velocity term accordingly, 
e.g. Wilkins (1980) \cite{Wilkins1980281}. 

In any case artificial viscosity 
broadens shocks to steep gradients at some characteristic length scale 
but on the other hand shall not cause unintentional smearing. 
The concrete composition and implementation of this artificial viscosity 
coefficient $\mathbf{q}$ depends on the application. 
Tscharnuter and Winkler (1979) \cite{Tscharnuter1979171} pointed out 
that the viscous pressure in 3D radiation hydrodynamics 
has to unravel normal stress, quantified by the 
divergence of the velocity field  and shear stress, which 
is expressed by the symmetrized gradient of 
the velocity field according to the general theory of viscosity. 
It is designed to {\sl switch on} only in case of compression ($\mathrm{div}
\, \mathbf{u} < 0$) yielding  
\begin{equation} \label{Q_Tscharnuter}
\mathbf{Q} = 
- q_2^2 l_{\text{visc}}^2
\rho \max(-\mathrm{div} \, \mathbf{u}, 0)  \left( \Big[
\nabla \mathbf{u} \Big]_s  - \frac{1}{3}  \mathbf{e} \, 
\mathrm{div} \, \mathbf{u}
\right) , 
\end{equation} 
and is set zero otherwise. In section \ref{Artificial_Viscosity} 
we will elaborate, why we have 
to modify this definition for general geometries by replacing the unit 
tensor $\mathbf{e}$ by a metric tensor $\mathbf{g}$. In section 
\ref{Covariant_RHD_Spher} we will 
specify co- and contravariant components of the 
artificial viscosity tensor in spherical coordinates.

\subsection{Flux Limiters} 

When we discussed the radiation closure relation in section 
\ref{radiation_closure_relation}, the approach was analytical 
respectively physical. Hyperbolicity of a conservation law implies 
real eigenvalues of the corresponding Jacobian 
$\nabla_{\mathbf{d}} \mathbf{f}$ which leads to characteristic 
propagation speeds which in case of radiation are connected 
with the speed of light. Pons, Ibanez,  Miralles (2000) 
\cite{HyperbolicCharacter} studied flux limiting in context of Gudonov-methods
based on these constraints on propagation speeds, 
Levermore (1984) \cite{Levermore1984149} studied flux limiters and Eddington
factors in anisotropic cases.  

\begin{figure}
\includegraphics[width=7.5cm]{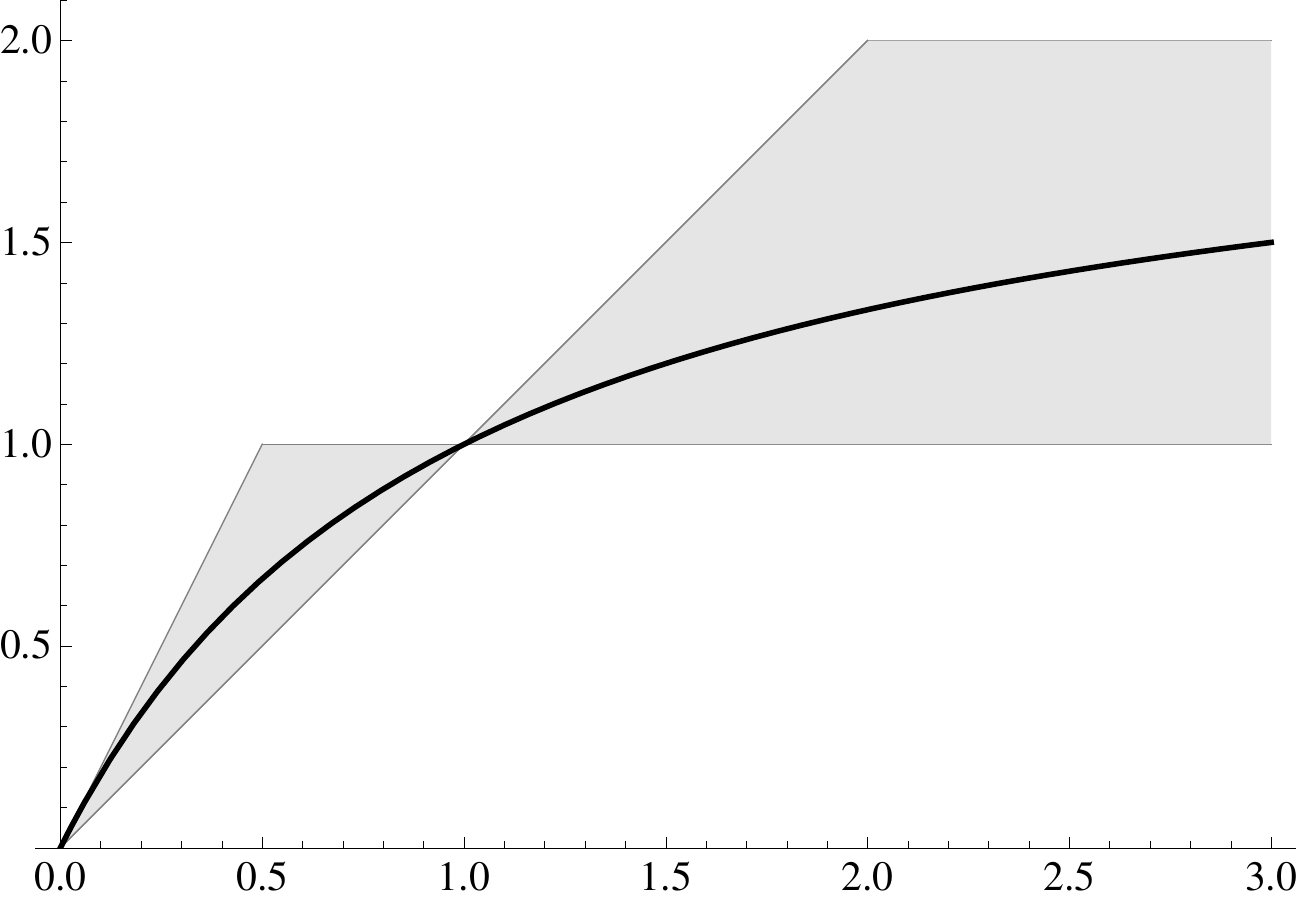} \centering
\caption{Van Leer Limiter} \label{van_leer}
\end{figure}

Flux limiters (or slope limiters) 
basically restrain solution gradients near shocks and thereby 
avoid oscillation in high resolutions methods (such as second order TVD). 
With flux limiting, the advection scheme gets split up in 
high order fluxes in smooth regions and lower order schemes 
(like Upwind) in the vicinity of discontinuities. With a limiter 
function $\Phi(\mathbf{D};j)$ the flux yields 
$\mathbf{F} = \mathbf{F}_{\text{low}} + \Phi\big( \mathbf{F}_{\text{high}}
-  \mathbf{F}_{\text{low}} \big)$ respectively in a more pracitical 
formulation 
\begin{equation}
\mathbf{F}(\mathbf{D};j) = \mathbf{F}_{\text{high}}(\mathbf{D};j) - 
(1- \Phi(\mathbf{D};j)) \big(
\mathbf{F}_{\text{high}}(\mathbf{D};j) - \mathbf{F}_{\text{low}}(\mathbf{D};j)
 \big) .
\end{equation}
There is a number of different ways in implementing such a flux 
limiter albeit they are typically functions of the steepness of the gradient. 
One ambiguity lies in this {\sl measure of smoothness} in discrete 
terms, in this context commonly denominated $\theta_j$. 
One passable way is the ratio of consecutive gradients 
\begin{equation}
\theta_j = \frac{\mathbf{D}_j - \mathbf{D}_{j-1}}{\mathbf{D}_{j+1} 
- \mathbf{D}_j}
\end{equation}
and is obviously near $1$ for smooth data. Several flux limiter methods, 
i.e. different functions $\Phi(\theta_j)$ have been suggested and 
tested extensively (e.g. Van Leer (1979) \cite{vanLeer1979101}). 
In implicit RHD computations like Dorfi et al. (2004) \cite{Dorfi2006771} 
and previous papers implement the Van Leer limiter which is 
constructed in the following way. 
\begin{equation}
\Phi(\theta_j) = \left\{ \begin{array}{ll}
\dfrac{2 \theta}{1+\theta} & \text{if } \theta \geq 0 \\
0 & \text{otherwise } 
\end{array} \right. 
\end{equation}
The Van Leer limiter describes a smooth function, a somewhat 
distinct characteristic in comparison to other feasible methods. 
In figure \ref{van_leer} the gray region denotes TVD; 
for details we refer to LeVeque \cite{LeVeque1990} once again. 

% SECTION Conservative Methods in General Coordinates  ------------------------%

\section{Conservative Methods in General Curvilinear 
Coordinates} \label{ConsGeneral}

In the previous section we presented necessary premises for conservative 
difference schemes, particularly the conservation form for a numerical 
method (\ref{conservation_form}). The numerical flux function 
$\mathbf{F}$ we defined can be interpreted as flux averaged over a cell 
(\ref{cell_average}) and a time interval $\mathbf{F}
\cong \frac{1}{k} \int_{t_n}^{t_{n+1}} \mathbf{f} 
\big( \mathbf{d} (x_{j+1/2},t) \big) \, dt$. 
Descriptively written, the conservation form ensures that whatever 
fraction of the density gets transported from one to another cell,  
arrives there. In other words, if supposedly all density of one cell gets 
advected to an other, we need to find the same {\sl physical} amount in there
afterwards. However, if the cell volumes differ and the flux is a cell 
integrated quantity it is required to assess the cell volume itself
properly. 
Hence in order to 
conserve a certain density function $\mathbf{d}(\mathbf{x},t)$ 
in general coordinates where non non-steady, nonlinear volume 
elements occur, we need 
to investigate the numerical flux more thoroughly in terms of 
differential and integral calculus. 

Mathematically the fluxes that occur in hyperbolic conservation laws are 
generated via spatial differential operators which are obtained by applying 
the divergence theorem to a differential volume, 
bounded by a surface (\ref{cauchy_problem_integral}). In the upcoming 
subsections we want to elaborate how differential geometric 
considerations lead to a reformulation of these differential operators, 
Gradient and Divergence in a {\sl geometrically conservative} form.

\subsection{Tensor Calculus and Elementary Differential Geometry} 
\label{Tensor_Calculus}

To begin with, we want to recapitulate some relations from 
tensor analysis and elementary differential geometry  
and define linear and nonlinear coordinate systems, coordinate transformations, 
base vectors and co- and contravariant vector components.  
In the following section we will use Einstein summation convention, 
indices in braces do not designate components but coordinate sytems.  
Consequent argumentations and proofs can be found in 
Lichnerowicz \cite{Lichnero}, 
Schutz \cite{Schutz} and Liseikin \cite{LiseikinAComp}. 

Let $\mathcal{M}$ be an affine space with corresponding vector space 
$\mathcal{V}$ and $\Sigma_{(\alpha)}$ a \emph{linear coordinate system}. 
The coordinate system maps points $\mathbf{x} 
\in \mathcal{M}$ (linear, bijective) to a $d$-tuple of numbers $x^{i}$. 
$\Sigma_{(\alpha)} : 
\mathcal{M} \rightarrow \mathbb{R}^d, \quad \mathbf{x} \mapsto
({x_{(\alpha)}}^{i} )$.
Coordinate lines of a linear coordinate system are straight 
lines on the affine space. 
Two linear coordinate systems $\Sigma_{(\alpha)}$ and $\Sigma_{(\beta)}$ 
are connected via a \emph{coordinate transformation}
\begin{equation}
{x_{(\alpha)}}^{i} = {\Lambda_{(\alpha \beta)}}^{i}_{j} \, {x_{(\beta)}}^{j} +
{\lambda_{(\alpha \beta)}}^{i} . 
\end{equation}
Such affine transformations 
imply translation (${\lambda_{(\alpha \beta)}}^{i}$)
as well as rotation and shearing (${\Lambda_{(\alpha \beta)}}^{i}_{j}$) 
of coordinate lines. 
The natural \emph{base vectors}
of any coordinate system are 
tangential vectors to the coordinate lines and thus 
constant in the linear case. 
\begin{equation} \label{def_basis}
\mathbf{e}_{(\alpha) i} = \frac{\partial 
\mathbf{x}}{\partial {x_{(\alpha)}}^{i}}
\end{equation}
The index $i$ in this case denotes the $i$-th base vector. 
The dual base is implicitly defined via the inner product 
${{\mathbf{e}_{(\alpha)}}}^{i} \cdot \mathbf{e}_{(\alpha) j} =
{\delta^{i}}_{j}$; the dual base ${\mathbf{e}_{(\alpha)}}^{i} \in 
\mathcal{V}^*$ is a linear map $\mathcal{V}^* \to \mathbb{R}$, which 
maps the base vectors  ${\mathbf{e}}_{(\alpha) j} \in \mathcal{V}$ 
to zero or one. 
The natural and the dual base vectors transform unequally to a new 
coordinate system $\Sigma_{(\beta)}$. 
\begin{eqnarray}
\nonumber && \mathbf{e}_{(\beta) i}  = 
{\Lambda_{(\alpha \beta)}}^{j}_{i} \,
\mathbf{e}_{(\alpha)
j} = \frac{\partial {x_{(\alpha)}}^{j}}{\partial {x_{(\beta)}}^{i}} \,
\mathbf{e}_{(\alpha) j} \\ 
 && {\mathbf{e}_{(\beta)}}^{ i} =  {\Lambda_{(\alpha \beta)}}^{i}_{j} \,
{\mathbf{e}_{(\alpha)}}^{j} = \frac{\partial {x_{(\alpha)}}^{i}}{\partial
{x_{(\beta)}}^{j}}
\, {\mathbf{e}_{(\alpha)}}^{j} 
\end{eqnarray}
The matrix $\Lambda_{(\alpha \beta)}$ 
is referred to as \emph{Jacobian} of the 
coordinate transformation.  

A transformation to a \emph{curvilinear coordinate system}
is no longer necessarily linear nor bijective. 
With nonlinear coordinate systems, the base vectors are not
constant but functions of position.  
\begin{equation} \label{nonlinear_base_def}
\mathbf{e}_{(\alpha) i}(\mathbf{x}) = \frac{\partial \mathbf{x}}{\partial
{x_{(\alpha)}}^{i}}
\end{equation} 
Of course, the entries of the 
Jacobian $\Lambda_{(\alpha \beta)}$ are now functions too 
\begin{equation}
{\Lambda_{(\alpha \beta)}}^{i}_{j}(\mathbf{x}) = \frac{\partial
{x_{(\alpha)}}^{i}}{\partial {x_{(\beta)}}^{j}}
\end{equation}
and $
{\mathbf{{e}}_{(\alpha)}}^{i}(\mathbf{x}) \cdot \mathbf{{e}}_{(\alpha)
j}(\mathbf{x}) =
{\delta^{i}}_{j}$ is valid pointwise. 
All vectorial and tensorial quantities $T_{\dots }^{ \dots }$ in this 
domain are defined via their components with respect to the local base in 
the following sense. 
\begin{equation} 
{T_{(\alpha)}}_{\rho \dots \sigma}^{i \dots j}(\mathbf{x}) =
T({\mathbf{e}_{(\alpha)}}^{i}(\mathbf{x}), \dots,
{\mathbf{e}_{(\alpha)}}^{j}(\mathbf{x}),
\mathbf{e}_{(\alpha) \rho}(\mathbf{x}), \dots 
\mathbf{e}_{(\alpha)
\rho}(\mathbf{x}) )
\end{equation}
We define the \emph{covariant components of the 
metric tensor} $\mathbf{g}$ as local measure of 
distances in the coordinate system 
\begin{equation} \label{metric_def}
{\mathbf{{e}}}_{i}(\mathbf{x}) \cdot \mathbf{{e}}_{j}(\mathbf{x}) =
g_{i j}(\mathbf{x}) .
\end{equation}
The metric tensor is obviously  
symmetrical due to commutativity of the scalar product and implicates 
a symmetrical bilinear form $\mathbf{x} \cdot \mathbf{y} = 
g_{ij}x^i y^j$ on the Euclidean vector space $\mathcal{V}$. 
From now on lower indices designate \emph{covariant} and upper indices 
\emph{contravariant} components. Together with the contravariant metric tensor 
$\mathbf{e}^{i}(\mathbf{x}) \cdot \mathbf{e}^{j}(\mathbf{x}) =
g^{i j}(\mathbf{x})$ we gain the following transformation rules 
between co- and contravariant components of a tensor. 
\begin{equation} \label{raislower}
v_{i} = g_{i j}v^{j}, \quad \quad v^{i} = g^{i
j}v_{j}
\end{equation}
Raising and lowering indices in linear coordinates just means to 
transpose respectively to interchange rows and columns of higher rank 
tensors. With curvilinear coordinate systems the dual space is also 
{\sl metrically} different. 

Next we want to liaise the recently defined geometrical items with 
differential calculus. Vectorial and tensorial quantities 
feature a \emph{native} construction in differential 
geometric terms. 
Vector components that are generated via a Gradient of a scalar are 
covariant by definition, e.g. $\partial_i \phi = p_i$ whereas 
coordinates $x^i$ are contravariant. Higher rank tensors 
are constructed by differentiation or tensorial products, e.g. 
$\mathbf{t} = \mathbf{p} \mathbf{q}$\footnote{
We generally omit explicit 
tensor products $\mathbf{p} \mathbf{q} = \mathbf{p} \otimes \mathbf{q}$ 
respectively $\mathbf{e}^i \mathbf{e}^j = 
\mathbf{e}^i \otimes \mathbf{e}^j$.}, and its native 
depiction in components is ${t_i}^j = p_i q^j$. The explicit form of such a 
tensor $\mathbf{t}$ in coordinates $x^l$ is 
\begin{equation}
\mathbf{t}(\mathbf{x}) = {t_i}^j (\mathbf{x}) 
\mathbf{e}^{i} (\mathbf{x}) \mathbf{e}_{j} 
(\mathbf{x}) . 
\end{equation}
For this reason, the tensorial physical quantities in the  
equations of radiation hydrodynamics will have to be 
considered firstly in their native form as well. 
When performing operations at and between them it will be necessary to 
consider a consistent denotation (e.g. contravariant components) and 
raising and lowering indices according to (\ref{raislower}) will 
be required. 

Evidently when differentiating a tensor 
we have to consider not only its 
components as functions but also the generally non constant base vectors 
in the coordinate system $\Sigma_{(\alpha)}$. 
Without forceful motivation we 
define a \emph{covariant derivative} of a vector field
$\nabla_{(\alpha)} \mathbf{v}$ 
and the \emph{Christoffel symbols} $\Gamma_{(\alpha)}$ in the following way. 
Let $\nabla$ be a map that transforms a 
tensor of rank $(m,n)$ to a tensor of rank $(m,n+1)$, satisfying $\nabla 
\phi =  \partial \phi$ for any scalar $\phi$. Let it furthermore  
satisfy all congenital demands on a derivative like the 
Leibnitz rule, additivity, commutativity and torsion-freeness 
and it shall be associated with the partial derivative via 
\begin{equation}
\nabla_{i} v^j = \partial_{i} v^j + 
{\Gamma^j}_{ik} v^k
\end{equation}
where the Christoffel symbols compensate for the locality of the 
base vectors.  
\begin{equation}
{\Gamma^j}_{ik}(\mathbf{x}) {\mathbf{e}}_{j}(\mathbf{x})
 = \partial_i \mathbf{e}_{k}(\mathbf{x})
\end{equation}
We postulate that this covariant derivative provides the proper 
{\sl affine connection} am implicitly defines the correct
transformations between general coordinate systems and can be 
expanded to tensors of higher rank and arbitrary composition 
of co- and contravariant components. 
\begin{eqnarray}
\nonumber \nabla_i v^i & = & \partial_i v^i + {\Gamma^i}_{ik} v^k  =
\mathrm{div} \, \mathbf{v} \\
\nonumber \nabla_i v_j & = &  \partial_i v_j - {\Gamma^k}_{ij} v_k \\
\nonumber 
\nabla_i {t_j}^k & = & \partial_i {t_j}^k + {\Gamma^k}_{l i} {t_j}^l - 
{\Gamma^l}_{j i} {t_l}^k \\
\label{covar_derivative}  \dots & = & \dots 
\end{eqnarray}
For linear coordinate systems the Christoffel symbols vanish trivially and 
the covariant derivative is identical to the partial derivative. 
Moreover, the covariant derivative can be called {\sl metrical} in terms 
of compatibility with the metric tensor $\mathbf{g}$. The 
\emph{Ricci Lemma} for the fundamental tensor states that the 
covariant derivative of the covariant components of $\mathbf{g}$ vanish, which 
can be checked quickly.  
\begin{equation}  \label{ricci_lemma}
\nabla_i g_{jk} = \partial_i g_{jk} - {\Gamma^l}_{ij} g_{lk} - 
{\Gamma^l}_{ik} g_{jl} = 0 
\end{equation}
When we contract upper equation with $g^{jk}$ we obtain 
$g^{jk} \partial_i g_{jk} - {\Gamma^j}_{ij} - {\Gamma^k}_{ik} = 0$
and gain an interesting formula
\begin{equation}
{\Gamma^j}_{jk} = \frac{1}{2}g^{jk} \partial_i g_{jk}
\end{equation}
which reminds of the derivative of the determinant of the metric tensor 
$\mathrm{det} \, \mathbf{g} = g$ which is 
\begin{equation}
\partial_i g = g g^{jk} \partial_i g_{jk}. 
\end{equation}
This yields to the following important relation between the 
Christoffel symbols and the metric. 
\begin{equation}
{\Gamma^k}_{ki} = \frac{1}{2} \frac{\partial_i g}{g} = 
\frac{\partial_i \sqrt{|\mathbf{g}|}}{\sqrt{|\mathbf{g}|}}
\end{equation}
We recognize the left hand side as a geometrical term in the divergence of a 
vector field (\ref{covar_derivative}) which leads to a meaningful 
reformulation of this derivative. 
\begin{equation} \label{cons_div}
\mathrm{div} \, \mathbf{v} = \frac{1}{\sqrt{|\mathbf{g}|}} 
\partial_i \left[ \sqrt{|\mathbf{g}|} v^i \right]
\end{equation}
Immediately we see the connection to the divergence theorem 
from integral calculus when we remember 
that the Jacobian determinant defines the volume element 
$dV_{(\alpha)} = | \Lambda_{(\alpha)} | dx^i$ and we know that 
$| \Lambda_{(\alpha)} | = \sqrt{|\mathbf{g}_{(\alpha)}|}$. 
\begin{eqnarray}
\nonumber
\int \mathrm{div} \, \mathbf{v} \, dV & = & \int \frac{1}{\sqrt{|\mathbf{g}|}} 
\partial_i \left[ \sqrt{|\mathbf{g}|} v^i \right] \sqrt{|\mathbf{g}|} 
\, dx^1 \dots dx^n \\
\label{cons_div_mot}
& = & \int \partial_i \left[ \sqrt{|\mathbf{g}|} v^i \right] 
\, dx^1 \dots dx^n 
\end{eqnarray}
Latter expression equals the flux of the vector field $\mathbf{v}$ over 
the boarders of the volume, a factor which shall be 
crucial for our geometrically 
conservative formulation of the equations of RHD.
Unfortunately the clue (\ref{cons_div_mot}) is confined to scalar 
equations which is expressed in \emph{Vinokur's theorem} published by 
Vinokur (1974) \cite{Vinokur1974105} and proven elegantly by 
Bridges (2008) \cite{Bridges2008882}. It states, how \emph{strong 
conservation form} (i.e. geometrically conservative in our diction) 
can be maintained regardless of what coordinate 
system used also for vectorial conservation laws.

\subsection{Strong Conservation Form}  \label{Strong_Cons_Form}

The numerical flux function we introduced in (\ref{conservation_form}) 
as {\sl integral} quantity
has to be understood not solely as a function at nodes but as a 
function of cells and their geometries, that is to say 
\begin{equation} \label{flux_geo}
\mathbf{F} = \mathbf{F}(\mathbf{D},\sqrt{|\mathbf{g}|}, \mathbf{e}^i) .
\end{equation}

Before we postulate how we obtain geometrically conservative fluxes 
or in other word, how we need to rephrase the covariant derivative 
(\ref{covar_derivative}) generally, 
we gain insight to another eclectic concept in 
differential geometry. 
The theory of \emph{differential forms}
 is an approach to differential calculus in 
more dimensions that is independent of coordinates. Again we have 
to refer to relevant literature like Bishop and Goldberg  
\cite{goldberg:bishop:1980} 
for stringent motivation and definitions. 
For us, differential forms shall just be {\sl objects that 
emerge under integrals}, like $w dx, \mathbf{f}
 d\mathbf{S}, \phi dV, \dots$
%\footnote{
%We generally omit wedge products products and boldface notations in this 
%very brief paragraph which means $dx^i dx^j = d\mathbf{x}^i d\mathbf{x}^j = 
%d\mathbf{x}^i \wedge d\mathbf{x}^j$.} 
and in that manner 
we have seen them consistently. We are going to like them because 
our (numerical) fluxes are objects that emerge under integrals as well. 
Moreover, we accept that 
there is a peculiar generalized concept of differentiating such 
differential forms which basically simplifies to the covariant derivatives when 
we settle for particular coordinates. 

We introduce a new notation, 
the \emph{wedge product} of two vectors $\mathbf{p}$
and $\mathbf{q}$ as the following antisymmetric tensor\footnote{
Of course we imply tensor products $\mathbf{p} \mathbf{q} 
= \mathbf{p} \otimes \mathbf{q}$ again. }  
\begin{equation}
\mathbf{p} \wedge \mathbf{q} = \mathbf{p} \mathbf{q} - \mathbf{q} \mathbf{p} .
\end{equation}
Tensor analytically this wedge product is motivated from the observation 
that any antisymmetric tensor $\mathbf{t}$ can be decomposed to 
linear combinations $\mathbf{t} = t^{ij}(\mathbf{e}_i \mathbf{e}_j - 
\mathbf{e}_j \mathbf{e}_i)$, $(i < j)$ respectively in terms 
of the wedge product $\mathbf{t} = t^{ij} \mathbf{e}_i \wedge \mathbf{e}_j$.

We define a \emph{$p$-form} (i.e. a differential form of degree $p$) by 
\begin{equation}
\omega = {w_{i_1 \dots i_p}} \, dx^{i_1} \wedge dx^{i_2} \wedge 
\ldots \wedge dx^{i_p}
\end{equation} 
which is obviously totally antisymmetric in its indices. 
The \emph{exterior 
derivative} $d \omega$ of a $p$-form $\omega = w_{i_1 \dots i_p} 
 \, dx^{i_1} \wedge \ldots \wedge dx^{i_p}$ is 
\begin{equation}
d\omega = \partial_j w_{i_1 \dots i_p} \, dx^j \wedge dx^{i_1} 
\wedge \ldots \wedge dx^{i_p}
\end{equation}
which yields
 a $(p+1)$-form. As an example, the exterior derivative of a $1$-form 
$\omega = w_i \, dx^i$ yields 
$d \omega = \partial_i w_j \, dx^i dx^j$. 
The general \emph{Stokes theorem} shows the connection of the exterior 
derivative on differential forms to integration theory, namely 
\begin{equation}
\int\limits_{\Omega} d \omega = \int\limits_{\partial \Omega} \omega  .
\end{equation}
The generalized concept of the \emph{covariant derivative} maps 
a $p$-form to a $(p-1)$-form and is defined by 
\begin{equation}
\delta \omega = (-1)^{p+1} \ast d \ast \omega 
\end{equation} 
where $\ast$ is the \emph{Hodge star operator} which could be 
understood as generalization 
of the Levi-Civita tensor $\varepsilon$ with metrical weight.
\begin{equation}
\ast \big( dx^{i_1} \wedge \ldots \wedge dx^{i_p} \big) = 
dx^{i_{p+1}} \wedge dx^{i_{p+2}} \wedge \ldots \wedge dx^{i_n}
\end{equation} 
The Hodge operator is the source for the volume form $\omega_{\mathrm{Vol}}
= \ast 1 = \sqrt{|\mathbf{g}|} \, 
dx^1 \wedge \ldots \wedge dx^n$ which we already encountered in the divergence 
theorem (\ref{cons_div_mot}). 

The benefit of dealing with differential forms shows when we 
associate them with well known concepts of tensor calculus. 
The exterior derivative of an $1$-form $\omega$ in $\mathbb{R}^3$ 
yields the Curl, to wit $d(w \, d\mathbf{x}) = 
\mathrm{curl}\, w \, d\mathbf{S}$ which in component notation 
is often expressed by 
$d(w \, d\mathbf{x}) = \partial_i w^j \varepsilon_{ijk} dS^k$. 
Moreover, as pointer to the nativeness of dealing with 
differential forms we mention that we mathematically defined the 
Curl of a vector field via a closed line integral (in a 
plane orthogonal to a normal vector $\mathbf{n}$) in the limit of 
vanishing path, i.e. $ (\mathrm{curl} \, \omega) \cdot \mathbf{n} = 
\lim_{A \to 0} \frac{1}{A} \oint \omega \cdot d\mathbf{r}$\footnote{
Let be $\omega = a \, dx + b \, dy + c \, dz$, then $d\omega =
(\partial_y a \, dy + \partial_z a \, dz) \wedge dx + \dots = 
(\partial_y c - \partial_z b) \, dy \wedge dz + 
(\partial_z a - \partial_x c) \, dz \wedge dx 
+ (\partial_x b - \partial_y a) \, dx \wedge dy = \mathrm{rot} \omega$ where 
$dy \wedge dz = {\mathbf{e}}^1$, $dz \wedge dx = {\mathbf{e}}^2$ etc.}. 
The main difference to tensorial concepts 
is, that with operations on differential forms we map {\sl something 
that emerges under an integral} to {\sl something that emerges under 
an integral}. Without further excursions we claim that we now have the 
munition to design geometrically conservative differential operators 
for general coordinates. 

The first elaborate insight to geometrically 
conservative differential operators can be 
found in Thompson, Warsi, Mastin \cite{ThompsonWarsi}, although 
they completely exclude differential forms and play extensively 
with tensor analysis what makes the book a little delicate to comprehend. 
However, differential form theory perfectly endorses their conclusions and also 
Bridges' proof (2008) \cite{Bridges2008882} of Vinokur's theorem 
resorts to differential forms as virtually 
inherent description for conservation laws. 
We conclude the connection between covariant derivatives in terms
of (\ref{covar_derivative}) and differential forms. 
The covariant derivative on a $1$-form $\omega= w_i \, dx^i$ yields explicitly 
\begin{eqnarray}
\nonumber 
\delta \omega & = & \ast d \ast \omega  \\
\nonumber & = & \ast d \left( \frac{1}{(3-1)!} 
\omega_i g^{ij} \varepsilon_{jkl} 
\sqrt{|\mathbf{g}|} \, dx^k dx^l \right)  \\
\nonumber & = & \ast \left( \frac{1}{2} \varepsilon_{jkl} \, \partial_m 
\left[ w_i g^{ij} \sqrt{|\mathbf{g}|} \right] \, dx^m dx^k dx^l 
\right) \\
\nonumber & = & \frac{1}{2} \varepsilon_{jkl} \, \partial_m 
\left[ w_i g^{ij} \sqrt{|\mathbf{g}|} \right] \frac{1}{0!} \, 
\underbrace{g^{mr} g^{ks} g^{lt} \varepsilon_{rst}}_{|\mathbf{g}|^{-1} 
\varepsilon^{mkl}} \sqrt{|\mathbf{g}|} \\
% \nonumber & = & \frac{1}{\sqrt{|\mathbf{g}|}} \partial_m \left[ 
% w_i g^{im} \sqrt{|\mathbf{g}|} \right] \\
& = &  \frac{1}{\sqrt{|\mathbf{g}|}} \partial_m \left[ 
 \sqrt{|\mathbf{g}|} w^m \right] 
\end{eqnarray}
which we recognize as result (\ref{cons_div}), namely $\mathrm{div} \, w$. 
For the \emph{strong conservation form} of the equations of radiation 
 hydrodynamics we will basically need the following three operations. 
\begin{description}
\item[Gradient of a scalar $\phi$]
\begin{equation} \label{grad_strong}
\mathrm{grad} \, \phi = \nabla_i \phi \, {\mathbf{e}}^i = 
\frac{1}{\sqrt{|\mathbf{g}|}}
\partial_i \left[ \sqrt{|\mathbf{g}|} {\mathbf{e}}^i \phi \right] 
\end{equation}
\item[Divergence of a vector field $\mathbf{v}$] 
\begin{equation} \label{div_strong}
\mathrm{div} \, \mathbf{v} =  \nabla_i v^i = 
\frac{1}{\sqrt{|\mathbf{g}|}}
\partial_i \left[ \sqrt{|\mathbf{g}|} {\mathbf{e}}^i \cdot 
\mathbf{v} \right] 
\end{equation}
\item[Divergence of a second rank tensor $\mathbf{t}$] 
\begin{equation} \label{div_t_strong} \mathrm{div} \, \mathbf{t} = 
\nabla_i t^{ij}  {\mathbf{e}}_j = 
\frac{1}{\sqrt{|\mathbf{g}|}}
\partial_i \left[ \sqrt{|\mathbf{g}|} {\mathbf{e}}^i \cdot 
\mathbf{t} \right] 
\end{equation}
\end{description}
Of course these differential operators can be translated to 
the well known covariant derivatives (\ref{covar_derivative}) 
with tensor analysis as well. 
\begin{eqnarray}
\nonumber \frac{1}{\sqrt{|\mathbf{g}|}} \partial_{\mu}
\left[ \sqrt{|\mathbf{g}|} \phi \, \mathbf{e}^{\mu} \right] & = &
{\Gamma^{\nu}}_{\nu \mu}\mathbf{e}^{\mu} \phi + \partial_{\mu}
\mathbf{e}^{\mu}
\phi + \mathbf{e}^{\mu} \partial_{\mu} \phi \\
\nonumber & = & {\Gamma^{\nu}}_{\nu \mu}\mathbf{e}^{\mu}
\phi + \partial_{\mu}(g^{\mu \rho}\mathbf{e}_{\rho})  \phi  +
\mathbf{e}^{\mu} \partial_{\mu} \phi \\
%\nonumber & = &  {\Gamma^{\nu}}_{\nu \mu}\mathbf{e}^{\mu}
%\phi + ( - {\Gamma^{\mu}}_{\mu \sigma}g^{\sigma \rho}  \mathbf{e}_{\rho} -
%{\Gamma^{\rho}}_{\mu \sigma} g^{\mu \sigma} \mathbf{e}_{\rho} + g^{\mu 
%\rho} {\Gamma^{l}}_{\rho \mu} \mathbf{e}_{l}) \phi +
%\mathbf{e}^{\mu} \partial_{\mu} \phi \\
\nonumber & = & {\Gamma^{\nu}}_{\nu \mu}\mathbf{e}^{\mu}
\phi - {\Gamma^{\mu}}_{\mu \sigma} \mathbf{e}^{\sigma} \phi +
\mathbf{e}^{\mu} \partial_{\mu} \phi \\
& = & \mathbf{e}^{\mu} \partial_{\mu} \phi 
\end{eqnarray}

The canonic shape of tensorial conservation laws 
$\int \big( \partial_t \mathbf{d} + 
\mathrm{div} \, \mathbf{f}(\mathbf{d})\big) \, dV dt = \mathbf{0}$ in 
strong conservation form for non-steady coordinates\footnote{In 
the following section we will 
argue, why the volume element $\sqrt{|\mathbf{g}|}$ also appears 
in the time derivative.} yields  
\begin{equation}
 \partial_t \Big[ \sqrt{|\mathbf{g}|} \, \mathbf{d} \Big] + 
\partial_i \Big[ \sqrt{|\mathbf{g}|} \, \mathbf{f}(\mathbf{d}) 
\cdot \mathbf{{e}}^i \Big] = 0 
\end{equation}
where of course $\mathbf{d}$ and $\mathbf{f}$ still have to be decomposed 
according to their native tensorial components. 

We substantiate upper proposition by an example of a vector conservation law, 
the equation of motion (\ref{cons_EOM}) in $n$ dimensions. 
The differential expression 
\begin{equation} 
\label{strong_cons_EOM}
 \partial_t \Big[ \sqrt{|\mathbf{g}|} \rho \mathbf{u} \Big] + \partial_i \Big[
\sqrt{|\mathbf{g}|} \left( \rho \mathbf{u} \mathbf{u} + \mathbf{P} + 
\sigma \right) \cdot \mathbf{{e}}^i \Big] = 0, \quad i=1, \dots, n
\end{equation}
implicates the integral form of conservation of momentum which would be 
considered numerically for non-steady coordinates. 
The important difference to a component-wise  
structure is that in this case  
undifferentiated terms would arise in $n$ equations (see 
Vinokur (1974) \cite{Vinokur1974105}). 

\noindent Neither the 
covariant derivative form for the divergence 
$\nabla (.) = \partial (.) + \Gamma (.)$, 
nor the $k$-th component of the expression $\partial_i 
[\sqrt{|\mathbf{g}|} \, f^{ik}]$ 
which would even be analytically true only for antisymmetric tensors 
$\mathbf{f}$ anyway,
satisfies the conservation law adequately. 
Since the base vectors $\mathbf{{e}}^i$ are functions of 
position for general coordinates, a separation as implied would 
introduce undifferentiated terms. However, the $k$ Cartesian components 
of equation (\ref{strong_cons_EOM}) are geometrically conservative 
which means they satisfy the strong conservation form that we 
demand\footnote{Here we simply used $t^{jl}\mathbf{e}_j \mathbf{e}_l 
\cdot \mathbf{e}^i = t^{jl} \mathbf{e}_j {\delta_l}^i = t^{ji} \mathbf{e}_j$.}. 
\begin{equation}
\nonumber \left\{ \partial_t \Big[ \sqrt{|\mathbf{g}|} 
\rho u^l \mathbf{{e}}_l \Big]
+ \partial_i \Big[
\sqrt{|\mathbf{g}|} \left( \rho u^iu^l + p^{il} + \sigma^{il} \right)
\mathbf{{{e}}}_l \Big] \right\}^k  = 0
\end{equation}
By now we have established a mathematical framework to deal with 
nonlinear conservation laws in curvilinear coordinates.  
For practical astrophysical applications though we will have to dissect 
this framework a little further 
to time dependent coordinate systems in order to work with 
adaptive grids as we motivated them for implicit numerical methods.

\subsection{Adaptive Grids} \label{Adaptive Grids}	

In fluid dynamics we basically distinguish two main reference 
systems that suit unequally for various applications. The 
\emph{Eulerian frame} is the fixed reference system of an external 
observer in which the fluid moves with velocity $\mathbf{u}$
whereas the \emph{Lagrangian specification} describes the 
physics in the rest frame of the fluid. 
The transformation of 
an advection term for a density $\mathbf{d}$ between these 
two systems that obviously move with relative velocity $\mathbf{u}$ 
is given via the \emph{material derivative} 
$D_t = 
\partial_t \mathbf{d} + \mathbf{u}\cdot\nabla \mathbf{d}$.
Hence, when we work with comoving frames, the coordinate system 
respectively the computational grid is time dependent. 
There is a number of purposes where strict Eulerian or Lagrangian 
grids are suboptimal and we need to generalize the concept 
of the comoving frame to ubiquitous grid movements. We 
will come back to this topic in chapter \ref{Grid_Generation} when we
deal with generation of moving grids in multiple dimensions. 

For now we want to evolve the strong conservation form for time 
dependent general coordinate systems. 
The time derivative of a density $\mathbf{d}$ in the coordinate 
system
$\Sigma_{(\beta)}$ relative to a (e.g. static) coordinate system
$\Sigma_{(\alpha)}$ is given by ${\partial_t \mathbf{d}}_{(\beta)} = 
{\partial _t\mathbf{d}}_{(\alpha)} + 
\mathbf{\nabla}_{(\alpha)}\mathbf{d} \, {\partial_t \mathbf{x}}_{(\beta)}$
and from the view point of system $\Sigma_{(\alpha)}$, the time derivative 
yields   
\begin{equation}
{\partial_t \mathbf{d}}_{(\alpha)} = {\partial _t\mathbf{d}}_{(\beta)} - 
\dot{\mathbf{x}} \mathbf{\nabla}_{(\alpha)}\mathbf{d} .
\end{equation}
With $\dot{\mathbf{x}}$ we introduced the \emph{grid velocity} and the 
second term on the right side we call \emph{grid advection}, see e.g.
Warsi (1981) \cite{Warsi1981} and 
Thompson, Warsi, Mastin \cite{ThompsonWarsi}. 
Now let us consider an inhomogeneous advective term of a conservation law 
$\mathbf{K} = \mathbf{d}_t + \mathrm{div} \, (\mathbf{u} \mathbf{d})$
referring to a fixed coordinate system. 
\begin{equation}
\mathbf{K} = \mathbf{d}_t - \dot{\mathbf{x}} \cdot 
\mathbf{\nabla}\mathbf{d} + \mathrm{div} \,
(\mathbf{u}\mathbf{d}) 
\end{equation}
When we apply transformation rules (\ref{grad_strong}), 
(\ref{div_strong}), (\ref{div_t_strong}) to the spatial derivatives, we 
gain the following form. 
\begin{equation} \label{conv_cons}
\mathbf{K} = \mathbf{d}_t - \dot{\mathbf{x}} \cdot \frac{1}
{\sqrt{|\mathbf{g}|}}\partial_{i}
\Big[
\sqrt{|\mathbf{g}|} \mathbf{{e}}^{i} \mathbf{d} \Big] + 
\frac{1}{\sqrt{|\mathbf{g}|}}\partial_{i} \Big[
\sqrt{|\mathbf{g}|} \mathbf{u} \cdot \mathbf{{e}}^{i} \mathbf{d} \Big] .
\end{equation}
Admittedly this is not yet geometrically conservative, since it is not of 
an {\sl integral} structure. Following the idea of the 
Reynolds transport theorem 
we consider the temporal derivative of the volume respectively 
the determinant of the time dependent metric tensor 
$\sqrt{|\mathbf{g(\mathbf{x},t)}|}$ in order to study the conservation of a density 
function in variable volumes and obtain  
the \emph{strong conservation form for time dependent coordinate 
systems}\footnote{A derivation of the strong conservative form for adaptive 
grids can be found in Appendix \ref{app_adaprive}}. 
\begin{equation} \label{flux_cons} 
\sqrt{|\mathbf{g}|} \mathbf{K}  = \partial_t
\Big[\sqrt{|\mathbf{g}|} \mathbf{d} \Big] + \partial_{i}
\Big[\sqrt{|\mathbf{g}|} \mathbf{{{e}}}^{i} \cdot (\mathbf{u} -
\dot{\mathbf{x}}) \mathbf{d} \Big]
\end{equation}
If we define the contravariant velocity components relative to the moving 
grid by $U^{i} = \mathbf{{{e}}}^{i} \cdot (\mathbf{u} - \dot{\mathbf{x}})
$ the upper equations yields 
\begin{equation}
\sqrt{|\mathbf{g}|} \mathbf{K} = \partial_t
\Big[ \sqrt{|\mathbf{g}|} \mathbf{d} \Big] + \partial_{i}
\Big[ \sqrt{|\mathbf{g}|} U^{i} \mathbf{d} \Big] .
\end{equation}

\subsection{Artificial Viscosity} \label{Artificial_Viscosity}

We introduced artificial viscosity in form of a three dimensional 
viscous pressure tensor in 
section \ref{Artificial_Viscosity_and_Flux} according to 
Tscharnuter and Winkler (1979) \cite{Tscharnuter1979171} by equation
(\ref{Q_Tscharnuter}) but also mentioned that we will have to 
adapt it slightly for curvilinear coordinates. 

This has to do with one previous comment to the {\sl native} form of 
tensorial quantities as they are generated in differential 
geometry. The crucial term in 
(\ref{Q_Tscharnuter}) is the symmetrized velocity gradient that accounts 
for shear stresses. The symmetrization rule is defined for lower indices 
and yields in component formulation ${[\nabla \mathbf{u}]_s}_{ij} = 
\frac{1}{2}(\nabla_i u_j + \nabla_j u_i)$ which comes into conflict 
with the demand of vanishing trace $\mathrm{Tr} \mathbf{Q} = 0$ when 
the divergence term contains the unit tensor $\mathbf{e}$. 

\begin{Pro} The correct form of the viscous pressure tensor 
in general coordinates is 
\begin{equation} \label{art_visc_corr}
\mathbf{Q} = - 
q_2^2 l_{\text{visc}}^2 \rho \max(-\mathrm{div} \,  \mathbf{u}, 0)  
\left( \Big[ \nabla \mathbf{u} \Big]_s - \frac{1}{3} \mathbf{g} \, \mathrm{div}
\,  \mathbf{u} \right) .
\end{equation}
\end{Pro}
Since this is an important result, we want to deduce this explicitly step 
by step. The viscous pressure tensor must be symmetrical by definition, 
i.e. $Q_{ij}=Q_{ji}$ and as already stated $\mathrm{Tr} \mathbf{Q} = 
{Q^i}_i = 0$ so we first consider its {\sl native} covariant components 
\begin{equation} 
Q_{ij} =  - \mu_2 
\max(- \mathrm{div} \, \mathbf{u}, 0)
 \left( \frac{1}{2}
(\nabla_i u_j + \nabla_j u_i ) - \frac{1}{3} g_{ij} \, \mathrm{div}
\,  \mathbf{u} \right)
\end{equation}
where we renamed $q_2^2 l_{\text{visc}}^2 \rho = \mu_2$. In order to 
proof that the trace vanishes, we need to raise an index with the metric
and use the essential identity $g^{li}g_{lj}={g^i}_j ={\delta^i}_j$. 
\begin{equation} \nonumber 
{Q^i}_j = Q_{lj}g^{li} = - \mu_2 \max(- \mathrm{div}
\, \mathbf{u}, 0)
 \left( \frac{1}{2} 
g^{li}(\nabla_l u_j + \nabla_j u_l ) - \frac{1}{3} {\delta^i}_j \, 
\mathrm{div}
\,  \mathbf{u} \right)
\end{equation}
The Ricci Lemma (\ref{ricci_lemma}) 
for the fundamental tensor naturally also holds for the 
contravariant components and we can permute $g$ into the derivatives 
$\nabla_l g^{li} u_j$ and $\nabla_j g^{li} u_l$ which yields 
twice the Divergence $\nabla_i u^i= \mathrm{div} \,
\mathbf{u}$ when we conduct the contraction $j \to i$. 
In three dimensions the summation ${\delta^i}_i=3$ and we obtain 
our desired result 
\begin{equation}
{Q^i}_i = \dots 
 \left( \frac{1}{2} 
(2 \, \mathrm{div} \, \mathbf{u}) - \frac{1}{3}3 \, 
\mathrm{div} \, \mathbf{u} \right) = 0 .
\end{equation}
Tscharnuter and Winklers form of $\mathbf{Q}$ is not compatible 
with our differential geometric definitions, since the symmetrization is only 
defined for lower indices whereas the unit element $\mathbf{e}$ of a metric 
space is only defined for mixed indices, meaning there is no 
such thing as $\delta_{ij}$. However, they and other authors 
like Mihalas and Mihalas \cite{Mihalas} did not notice that little 
inconsistency since they considered mixed indices from the start
respectively 1D flows at last. Corrections have been 
suggested by Benson (1993) \cite{Benson1993} albeit they do not 
epxlicitly concern curvilinear coordinates and
were derived based on a TVD approach. 
The usefulness of our correction might not show until even more 
general coordinate systems respectively grids are considered. In chapter 
\ref{Grid_Generation} we will discuss nonorthogonal grids and 
concern ourselves with the question
at what step in the calculation geometric operations 
like raising or lowering indices should be made ideally. 

In the upcoming section we will 
recapitulate our general results for spherical coordinates 
without explicitly performing the discretization and 
marking down numerical fluxes according to (\ref{conservation_form}) 
respectively  Dorfi et al. (2004) \cite{Dorfi2006771}. 
It will become comprehensible why we suggest generating these 
discretized equations with {\sf Mathematica} as well.

% Section Covariant Formulation in Spherical Coordinates --------------------- %

\section{Strong Conservation Form of RHD in Spherical and Polar Coordinates}
\label{Covariant_RHD_Spher}

When we think of problem oriented coordinate systems
 in astrophysical applications spherical and polar coordinates  
are a first natural choice. In the following section we present the 
geometrically conservative form of the equations of radiation hydrodynamics 
in these cases exemplarily. We consider the transformation 
$(x,y,z) \to (r \in \mathbb{R}^+,\vartheta \in [0, \pi],\varphi \in 
[0,2\pi])$ with 
$x = r \sin \vartheta \cos \varphi$, $y = r \sin \vartheta \sin \varphi$ 
and $z = r \cos \vartheta$ as spherical coordinates. 
The map $(x,z) \to (r \in \mathbb{R}^+, \vartheta \in [0,2\pi])$ 
shall be our polar coordinates $x =r \sin \vartheta$ and $z = 
r \cos \vartheta$. This convention was chosen in favor of somewhat simplified 
generation of the equations components for which we used 
the computer algebra system {\sf Mathematica}. Some of the equation 
generating files as well as some computational tests on consistency like 
${Q^i}_i=0$ and $Q_{ij}=Q_{ji}$ 
can be found in the Appendix \ref{Mathematica_Files}. 

For the sake of completeness we mention some important 
geometrical attributes of our choice. 

\begin{description}

\item[Spherical Coordinates]
The covariant components of the metric tensor 
for these spherical coordinates are $g_{ij} = \mathrm{diag}
(1,r^2, r^2\sin \vartheta)$ and due to their orthogonality the 
contravariant components simply yield $g^{ij} = \mathrm{diag}
(1,1/r^2, 1/(r^2\sin \vartheta))$ and the volume element 
$\sqrt{|\mathbf{g}|} = r^2 \sin\vartheta$. 
For the contravariant base vectors we obtain 
\begin{eqnarray}
\nonumber 
% \mathbf{e}_r = (\cos\varphi \sin\vartheta,
% \sin\varphi \sin\vartheta,\cos\vartheta) && 
\mathbf{e}^r & = & 
(\cos\varphi \sin\vartheta, \sin\varphi \sin\vartheta ,\cos \vartheta) \\
\nonumber
% \mathbf{e}_{\vartheta} = (r \cos\varphi \cos\vartheta, r \cos\vartheta 
% \sin\varphi, -r \sin \vartheta) &  & 
\mathbf{e}^{\vartheta} & = & \left(
\frac{\cos \varphi \cos\vartheta}{r}, \frac{\cos\vartheta \sin\varphi}{r},
-\frac{\sin\vartheta}{r} \right) 
\\ 
% \mathbf{e}_{\varphi} = (-r \sin\varphi \sin\vartheta, r 
% \cos\varphi \sin\vartheta,0) && 
\mathbf{e}^{\varphi} & = & \left(-\frac{\csc\vartheta \sin\varphi}{r},
\frac{\cos\varphi \csc\vartheta}{r},0\right) 
\end{eqnarray}
and the non vanishing Christoffel symbols 
\begin{eqnarray}
\nonumber 
&& {\Gamma^r}_{\vartheta \vartheta} = -r, \quad {\Gamma^r}_{\varphi \varphi} = 
-r \sin^2 \vartheta\\
\nonumber 
&& {\Gamma^{\vartheta}}_{r \vartheta} = {\Gamma^{\vartheta}}_{\vartheta r} = 
\frac{1}{r}, \quad {\Gamma^{\vartheta}}_{\varphi \varphi} = -\cos \vartheta 
\sin \vartheta \\
\label{spher_christof}
&& {\Gamma^{\varphi}}_{r \varphi} 
= {\Gamma^{\varphi}}_{\varphi r} = \frac{1}{r}, 
\quad {\Gamma^{\varphi}}_{\varphi \vartheta} = \cot \vartheta .
\end{eqnarray}

\item[Polar Coordinates] Also polar coordinates are orthogonal and 
the metric components yield $g_{ij}= \mathrm{diag}(1,r^2)$ respectively 
$g^{ij}=\mathrm{diag}(1,1/r^2)$. For the volume element 
we obtain $\sqrt{|\mathbf{g}|}=r$. 
The contravariant base vectors yield
\begin{eqnarray}
\nonumber 
% \mathbf{e}_r = (\cos\varphi \sin\vartheta,
% \sin\varphi \sin\vartheta,\cos\vartheta) && 
\mathbf{e}^r & = & 
(\sin\vartheta, \cos \vartheta) \\
% \mathbf{e}_{\vartheta} = (r \cos\varphi \cos\vartheta, r \cos\vartheta 
% \sin\varphi, -r \sin \vartheta) &  & 
\mathbf{e}^{\vartheta} & = & \left(
\frac{\cos\vartheta}{r}, -\frac{\sin\vartheta}{r} \right)
\end{eqnarray}
and the non vanishing Christoffel symbols are
\begin{equation}
\label{pol_christof} 
{\Gamma^{\vartheta}}_{r \vartheta} = \frac{1}{r}, 
\quad {\Gamma^{\vartheta}}_{\vartheta r} = \frac{1}{r}, 
\quad {\Gamma^{r}}_{\vartheta \vartheta} = - r .
\end{equation}

\end{description}

\subsection{Continuity Equation}

The strong conservation form of the conservation of mass (\ref{cont_eq_def})
in form of the continuity equation 
\begin{equation}
\partial_t \rho + \mathrm{div} \, (\rho \mathbf{u} ) =
0 
\end{equation}
for time dependend coordinates yields 
\begin{equation} \label{eom_strong_cons}
\partial_t
\Big[\sqrt{|\mathbf{g}|} \rho \Big] + \partial_{i}
\Big[\sqrt{|\mathbf{g}|} \mathbf{{{e}}}^{i} \cdot (\mathbf{u} -
\dot{\mathbf{x}}) \rho \Big] = 0. 
\end{equation} 
Since the conservation of mass is scalar the divergence term may also be
written in components $\partial_t [\sqrt{|\mathbf{g}|} \rho ] + \partial_{i}
[\sqrt{|\mathbf{g}|} ( u^{i} -\dot{x}^{i} ) \rho ] = 0$.  
The explicit form of (\ref{eom_strong_cons}) in spherical coordinates 
yields 
\begin{eqnarray} 
\nonumber \partial_t \Big[ r^2 \sin \vartheta \rho \Big] + 
\partial_{r} \Big[ r^2 \sin \vartheta \left( u^{r} - \dot{x}^{r}
\right) \rho \Big] + && \\
\label{cont_eq_cons_spher} + \partial_{\vartheta}
\Big[ r^2 \sin \vartheta \left( u^{\vartheta} - \dot{x}^{\vartheta} \right)
\rho
\Big] + \partial_{\varphi}
\Big[r^2 \sin \vartheta \left( u^{\varphi} - \dot{x}^{\varphi} \right) \rho
\Big] & = & 0 
\end{eqnarray} 
which we can rewrite via the transformation 
$ \cos \vartheta = \mu $ with differential 
 $d \vartheta = - \sqrt{1-\mu^2} d\mu$ and divide the equation 
by $\sqrt{1-\mu^2}$ in the case of solely radial adaptivity. 
Note that 
even for solely radially adaptive grids the grid velocity $\dot{\mathbf{x}}$
contains angular dependencies, e.g. $\dot{x}^r=  \dot{r} \sin \vartheta
\cos \varphi$ which obviously means $\dot{x} \neq \dot{r}$.
\begin{eqnarray}
% \nonumber 
% \partial_t \Big[ r^2 \sqrt{1- \mu^2} \rho \Big] + \partial_r \Big[
% r^2 \sqrt{1- \mu^2} \left( u^{r} - \dot{x}^{r}
% \right) \rho \Big] - && \\
% \nonumber - \sqrt{1- \mu^2} \partial_{\mu} \Big[ r^2
% \sqrt{1- \mu^2} \left( u^{\vartheta} - \dot{x}^{\vartheta} \right) \rho \Big] 
% + \partial_{\varphi} \Big[ r^2 \sqrt{1- \mu^2} \left( u^{\varphi} -
% \dot{x}^{\varphi} \right) \rho \Big] & = &
% 0 \\
\nonumber 
\partial_t \Big[ r^2 \rho \Big] + \partial_r \Big[
r^2  \left( u^{r} - \dot{x}^{r}
\right) \rho \Big] - && \\
- r^2 \partial_{\mu} \Big[ 
\sqrt{1- \mu^2} \left( u^{\vartheta} - \dot{x}^{\vartheta} \right) \rho \Big] 
+  r^2 \partial_{\varphi} \Big[ \left( u^{\varphi} -
\dot{x}^{\varphi} \right) \rho \Big] & = &
0 
\end{eqnarray}
With the radial grid velocity expanded we obtain the following form 
of the conservation of mass. 
\begin{eqnarray} 
\nonumber \partial_t \Big[ r^2 \rho \Big] + \partial_r \Big[
r^2 \left( u^{r} - \dot{r} \sqrt{1-\mu^2} \cos \varphi \right) \rho \Big] - &&
\\
- r^2 \partial_{\mu} \Big[ 
\sqrt{1- \mu^2} \left( u^{\vartheta} - \dot{r} \sqrt{1-\mu^2} \sin \varphi
\right)
\rho \Big]
+ r^2 \partial_{\varphi} \Big[ \left( u^{\varphi} -
\dot{r} \mu \right) \rho \Big] & = &
0 
\end{eqnarray} 

In the case of 2D polar coordinates the continuity equation yields 
\begin{equation}
\partial_t \Big[ r \rho \Big] + \partial_r \Big[
r \left( u^{r} - \dot{x}^{r}
\right) \rho \Big] - r \sqrt{1- \mu^2} \partial_{\mu} \Big[ 
 \left( u^{\vartheta} - \dot{x}^{\vartheta} \right) \rho \Big] 
= 0
\end{equation}
respectively in case of radially time dependent coordinates
\begin{equation}
 \partial_t \Big[ r \rho \Big] + \partial_r \Big[
r \left( u^{r} - \dot{r} \sqrt{1-\mu^2} \right) \rho \Big] - 
r \sqrt{1- \mu^2}  \partial_{\mu} \Big[ 
\left( u^{\vartheta} - \dot{r} \mu \right) \rho
\Big] = 0. 
\end{equation}

\subsection{Equation of Motion}

The equation of motion that we consider in radiation hydrodynamics contains 
the conservation of moment of plain fluid dynamics (\ref{cons_EOM}) as
well as the radiative flux as coupling term (\ref{rad_coupling_terms}), 
gravitational force as indicated in section \ref{grav_idea} and 
artificial viscosity (\ref{art_visc_corr}). 
\begin{equation}
\partial_t (\rho \mathbf{u}) + \mathrm{div} \, \big( \rho \mathbf{u} 
\mathbf{u} + \mathbf{P} + {\mathbf{Q}} \big) + \rho \mathbf{G} - \frac{4
\pi}{c} \kappa_R \rho 
\mathbf{H}  = \mathbf{0} 
\end{equation}
We investigate the elements of the energy stress tensor a little 
closer before we designate the consistent strong conservation form. 
We define an \emph{effective tensor of gaseous momentum} $\mathbf{R}$
that accounts for 
the motion of the coordinates 
\begin{equation}
\mathbf{R} = r^{ij} \mathbf{{e}}_i \mathbf{{e}}_j
= \rho (\mathbf{u} - \dot{\mathbf{x}} )  \mathbf{u} \\
\end{equation}
with components $r^{ij} = \rho (u^i - \dot{x}^i) u^j $. 
The \emph{isotropic gas pressure tensor} $\mathbf{P}$ as anticipated 
is defined by the scalar pressure and the metric tensor
\begin{equation}
 \mathbf{P} = p \mathbf{g} 
\end{equation}
with obvious decomposition $P^{ij} = p g^{ij}$. The \emph{viscous pressure 
tensor} ${\mathbf{Q}}$ needs some special attention. We emanate from 
the following definition 
\begin{equation}
\mathbf{Q} = - \left( \underbrace{q_1 l_{\text{visc}} \rho c_S}_{\mu_1} +
\underbrace{q_2^2 l_{\text{visc}}^2
\rho}_{\mu_2} \max(-\mathrm{div} \,  \mathbf{u}, 0) \right) \left( \Big[
\nabla \mathbf{u} \Big]_s  - \mathbf{g} \frac{1}{3} \nabla
\cdot \mathbf{u} \right) , 
\end{equation}
and remember that in order to get its contravariant components we need 
to raise them with our metric. 	
\begin{eqnarray}
\nonumber Q_{ij} & = &  \left( \mu_1 + \mu_2 
\max(- \mathrm{div} \, \mathbf{u}, 0)
\right) \left( (\nabla_i u_j + \nabla_j u_i ) - g_{ij} \frac{1}{3} \mathrm{div}
\,  \mathbf{u} \right) \\
\nonumber {Q^i}_j = Q_{lj}g^{li} & = & \dots  \left( g^{li}(\nabla_l u_j + \nabla_j u_l ) - {\delta^i}_j \frac{1}{3}
\mathrm{div}
\,  \mathbf{u} \right) \\
\nonumber 
Q^{ij} = Q_{lm}g^{li}g^{mj} & = & \dots  \left( g^{li}g^{mj}(\nabla_l u_m + \nabla_j u_m ) - g^{ij}
\frac{1}{3} \mathrm{div} \,  \mathbf{u} \right) \\
\end{eqnarray}
We also promised to particularly calculate the trace of $\mathbf{Q}$ in 
spherical coordinates in order to confirm consistency. To do so, we need 
the mixed components which are listed on the following page. 
\newpage

\begin{eqnarray}
\nonumber {Q^r}_r & = & Q_{rr}g^{rr} + Q_{\vartheta r}g^{\vartheta r} +
Q_{\varphi
r}g^{\varphi r} = Q_{rr} \\ 
\nonumber & = & -\frac{1}{3} \mathrm{div} \, \mathbf{u} + \partial_r u_r \\
\nonumber {Q^r}_{\vartheta} & = & Q_{r \vartheta }g^{r r} + Q_{\vartheta
\vartheta}g^{\vartheta r} + Q_{\varphi \vartheta} g^{\varphi r} = Q_{\vartheta r
} \\
\nonumber & = & - \frac{1}{r}u_{\vartheta} +
\frac{1}{2} \partial_r u_{\vartheta} + \frac{1}{2}\partial_{\vartheta} u_r \\
\nonumber {Q^r}_{\varphi} & = & Q_{r \varphi}g^{r r} + Q_{\vartheta
\varphi}g^{\vartheta r} + Q_{\varphi \varphi}g^{\varphi r} =
 Q_{r \varphi} \\ 
\nonumber & = & - \frac{1}{r}u_{\varphi} + \frac{1}{2}\partial_r u_{\varphi} +
\frac{1}{2}\partial_{\varphi} u_r \\
\nonumber {Q^{\vartheta}}_{r} & = & Q_{r r} g^{r \vartheta} + Q_{\vartheta
r}g^{\vartheta
\vartheta} + Q_{\varphi r} g^{\varphi \vartheta} = \frac{1}{r^2}Q_{\vartheta r}
\\
\nonumber & = & \frac{1}{r^2} \left( -\frac{1}{r}u_{\vartheta} + \frac{1}{2}
\partial_r
u_{\vartheta} + \frac{1}{2} \partial_{\vartheta} u_r \right) \\
\nonumber {Q^{\vartheta}}_{\vartheta} & = & Q_{r \vartheta} g^{r \vartheta} +
Q_{\vartheta
\vartheta}g^{\vartheta \vartheta} + Q_{\varphi \vartheta} g^{\varphi \vartheta}
=
\frac{1}{r^2}Q_{\vartheta \vartheta} \\
\nonumber & = & -\frac{1}{3}\mathrm{div} \, \mathbf{u} + \frac{1}{r}u_r +
\frac{1}{r^2} \partial_{\vartheta}u_{\vartheta} \\
\nonumber {Q^{\vartheta}}_{\varphi} & = & Q_{r \varphi} g^{r \vartheta} +
Q_{\vartheta
\varphi}g^{\vartheta \vartheta} + Q_{\varphi \varphi} g^{\varphi \vartheta} =
\frac{1}{r^2}Q_{\vartheta \varphi} \\
\nonumber & = & \frac{1}{r^2} \left( - \cot{\vartheta}u_{\varphi}
+ \frac{1}{2}\partial_{\varphi}u_{\vartheta} +
\frac{1}{2}\partial_{\vartheta}u_{\varphi} \right) \\
\nonumber {Q^{\varphi}}_{r} & = & Q_{r r} g^{r \varphi} + Q_{\vartheta
r}g^{\vartheta \varphi} + Q_{\varphi r} g^{\varphi \varphi} =
\frac{1}{r^2 \sin^2 \vartheta}Q_{\varphi r} \\
\nonumber & = & \frac{1}{r^2 \sin^2 \vartheta} \left( - \frac{1}{r}u_{\varphi} +
\frac{1}{2} \partial_{\varphi} u_r + \frac{1}{2} \partial_r u_{\vartheta}
\right)
\\
\nonumber {Q^{\varphi}}_{\vartheta} & = & Q_{r \vartheta} g^{r \varphi} +
Q_{\vartheta
\vartheta}g^{\vartheta \varphi} + Q_{\varphi \vartheta} g^{\varphi \varphi} =
\frac{1}{r^2 \sin^2 \vartheta}Q_{\varphi \vartheta} \\
\nonumber & = & \frac{1}{r^2 \sin^2 \vartheta} \left( - \cot \vartheta
u_{\varphi} +
\frac{1}{2}
\partial_{\varphi} u_ {\vartheta} + \frac{1}{2} \partial_{\vartheta} u_{\varphi}
\right) \\
\nonumber {Q^{\varphi}}_{\varphi} & = & Q_{r \varphi} g^{r \varphi} +
Q_{\vartheta
\varphi}g^{\vartheta \varphi} + Q_{\varphi \varphi} g^{\varphi \varphi} =
\frac{1}{r^2 \sin^2 \vartheta}Q_{\varphi \varphi} \\
\nonumber & = &  -\frac{1}{3} \mathrm{div} \, \mathbf{u} + \frac{1}{r}u_r +
\frac{1}{r^2}
\cot \vartheta u_{\vartheta} + \frac{1}{r^2 \sin^2 \vartheta} \partial_{\varphi}
u_{\varphi}
\end{eqnarray}
\begin{eqnarray}
\nonumber \mathrm{Tr} \mathbf{Q} = {Q^i}_i & = & -\frac{1}{3} \mathrm{div} \,
\mathbf{u} + \partial_r u_r -
\frac{1}{3}\mathrm{div} \, \mathbf{u} + \frac{1}{r}u_r +
\frac{1}{r^2} \partial_{\vartheta}u_{\vartheta} -\frac{1}{3} \mathrm{div} \, 
\mathbf{u} +
\frac{1}{r}u_r + \\
\nonumber & & + \frac{1}{r^2}
\cot \vartheta u_{\vartheta} + \frac{1}{r^2 \sin^2 \vartheta} \partial_{\varphi}
u_{\varphi} \\
\nonumber & = & -\mathrm{div} \, \mathbf{u} + \frac{2}{r} + \frac{1}{r^2}\cot
\vartheta
u_{\vartheta} + \partial_r u_r + \frac{1}{r^2}\partial_{\vartheta} u_{\vartheta}
+
\frac{1}{r^2 \sin^2 \vartheta}\partial_{\varphi} u_{\varphi} \\
\nonumber & = & -\mathrm{div} \, \mathbf{u} + \underbrace{\frac{2}{r} + 
\cot \vartheta
u^{\vartheta} + \partial_r u^r + \partial_{\vartheta} u^{\vartheta} +
\partial_{\varphi} u^{\varphi}}_{\mathrm{div} \, \mathbf{u}}= 0
\end{eqnarray}
So in fact the viscous pressure tensors trace vanishes in spherical 
coordinates explictly\footnote{The geometrical terms $2/r$ and $\cot \vartheta$ 
are Christoffel symbols of the spherical coordinates (\ref{spher_christof}) 
from $\nabla_i u^i = \partial_i u^i + {\Gamma^i}_{ik} u^k$.}. 

The native decompositions of our tensors $\mathbf{R}$ and $\mathbf{P}$ are 
contravariant hence we will consider also $\mathbf{Q}$'s contravariant 
components for the sake of coherency. 
In the following array we list $Q^{ij}$ for spherical coordiantes. 
\begin{eqnarray}
\nonumber Q^{rr} & = & - \frac{1}{3}\mathrm{div} \mathbf{u} + \partial_r u_r \\
\nonumber Q^{r \vartheta} & = & \frac{1}{r^2} \left( - \frac{1}{r} u_{\vartheta}
+ \frac{1}{2} \partial_r u_{\vartheta} + \frac{1}{2} \partial_{\vartheta} u_r
\right)
\\
\nonumber  Q^{r \varphi} & = & \frac{1}{r^2 \sin^2 \vartheta} \left(
- \frac{1}{r} u_{\varphi} + \frac{1}{2} \partial_{\varphi} u_{r} +
\frac{1}{2} \partial_r u_{\varphi} \right) \\
\nonumber Q^{\vartheta r} & = & \frac{1}{r^2} \left( -\frac{1}{r}u_{\vartheta} +
\frac{1}{2} \partial_r u_{\vartheta} + \frac{1}{2} \partial_{\vartheta} u_r
\right) \\
\nonumber Q^{\vartheta \vartheta} & = & \frac{1}{r^2} \left(
-\frac{1}{3}\mathrm{div}
\mathbf{u} + \frac{1}{r}u_r +
\frac{1}{r^2} \partial_{\vartheta}u_{\vartheta} \right) \\
\nonumber Q^{\vartheta \varphi} & = & \frac{1}{r^4 \sin^2 \vartheta}
\left( - \cot{\vartheta}u_{\varphi}
+ \frac{1}{2}\partial_{\varphi}u_{\vartheta} +
\frac{1}{2}\partial_{\vartheta}u_{\varphi} \right) \\
\nonumber Q^{\varphi r} & = & \frac{1}{r^2 \sin^2 \vartheta} \left( -
\frac{1}{r}u_{\varphi} +
\frac{1}{2} \partial_{\varphi} u_r + \frac{1}{2} \partial_r u_{\vartheta}
\right)
\\
\nonumber Q^{\varphi \vartheta} & = & \frac{1}{r^4 \sin^2 \vartheta}
\left( - \cot \vartheta u_{\varphi} +
\frac{1}{2}
\partial_{\varphi} u_ {\vartheta} + \frac{1}{2} \partial_{\vartheta} u_{\varphi}
\right) \\
\nonumber Q^{\varphi \varphi} & = & \frac{1}{r^2 \sin^2 \vartheta}
\left( -\frac{1}{3} \mathrm{div} \mathbf{u} + \frac{1}{r}u_r +
\frac{1}{r^2}
\cot \vartheta u_{\vartheta} + \frac{1}{r^2 \sin^2 \vartheta} \partial_{\varphi}
u_{\varphi} \right) 
\end{eqnarray}

We will investigate the following strong conservation form of the equation 
of motion 
\begin{eqnarray}
\nonumber 
 \partial_t \Big[ \sqrt{|\mathbf{g}|} \rho \mathbf{u} \Big] + \partial_i \Big[
\sqrt{|\mathbf{g}|} \big( \mathbf{R} + \mathbf{P} + {\mathbf{Q}} \big) 
\cdot \mathbf{{{e}}}^i \Big] \, + && \\ 
+ \, \rho \partial_i \Big[ \sqrt{|\mathbf{g}|} \Phi 
\mathbf{{e}}^i
\Big] - \frac{4 \pi}{c} \kappa_R \sqrt{|\mathbf{g}|} \rho \mathbf{H} & = & 
\mathbf{0} .
\end{eqnarray}
The $k$-th component of the strong conservation equation of motion is given
by the $k$-th cartesic component of our unit vector, e.g. $\mathbf{{{e}}}^r = 
\cos\varphi \sin\vartheta \mathbf{e}^x + \sin\varphi \sin\vartheta \mathbf{e}^y 
+ \cos \vartheta \mathbf{e}^z$  
and its derivatives. The
projection of each physical tensor on the contravariant coordinate lines can be
simplified by its contravariant components with respect to its covariant basis
without losing strong conservation form, i.e. $\mathbf{f}\cdot \mathbf{e}^i = 
f^{in} \mathbf{e}_n$. 
We prefer this form with contravariant 
components since it meets the native design of the stress tensor 
(\ref{visc_stress_def}). 
\begin{eqnarray}
\nonumber \Big\{ \partial_t \Big[ \sqrt{|\mathbf{g}|} 
\rho u^n \mathbf{{e}}_n \Big]
+ \partial_i \Big[
\sqrt{|\mathbf{g}|} \left( r^{in} + p^{in} + Q^{in} \right)
\mathbf{{{e}}}_n \Big] + && \\
 \rho \partial_n \Big[ \sqrt{|\mathbf{g}|} \Phi
\mathbf{{e}}^n
\Big] - \frac{4 \pi}{c} \kappa_R \sqrt{|\mathbf{g}|} 
\rho H^n \mathbf{{e}}_n \Big\}^k  & = &0
\end{eqnarray}
The three components of upper equation in spherical coordinates 
are listed in the following arrays.

{\scriptsize 
\begin{eqnarray}
\nonumber \partial_t \Big[ \rho r^2 \sqrt{1-\mu^2}
 \cos \varphi u^r +  \rho r^3 \mu \rho
\cos \varphi u^{\vartheta} - \rho r^3 \sqrt{1-\mu^2} 
\sin \varphi u^{\varphi} \Big] + &&
\\ \nonumber + \partial_r \Big[ r^2 \sqrt{1-\mu^2} 
\cos \varphi \Big( \rho u^r (u^r -
\dot{x}^r) + p + Q^{rr} \Big) + r^3 \mu \cos \varphi \Big(
\rho
u^{\vartheta}(u^r - \dot{x}^r) + Q^{r \vartheta} \Big) - && \\
\nonumber  - r^3 \sqrt{1-\mu^2} \sin \varphi \Big( \rho
u^{\varphi}(u^r - \dot{x}^r) + Q^{r \varphi} \Big) \Big]  - 
\partial_{\mu} \Big[ r^2 (1-\mu^2) \cos \varphi
\Big( \rho
u^r(u^{\vartheta} - \dot{x}^{\vartheta}) + Q^{\vartheta r} \Big) + 
&& \\
\nonumber  + r^3 \mu \sqrt{1-\mu^2}
\cos \varphi \Big( \rho u^{\vartheta} (u^{\vartheta} - \dot{x}^{\vartheta}
) + \frac{1}{r^2} p + Q^{\vartheta \vartheta} \Big) - r^3 (1-\mu^2) \sin
\varphi \Big(\rho u^{\varphi}(u^{\vartheta} -
\dot{x}^{\vartheta}) + Q^{\vartheta \varphi} \Big) \Big] + && \\
\nonumber + \partial_{\varphi} \Big[ r^2 \sqrt{1-\mu^2} 
\cos \varphi \Big( \rho u^r
(u^{\varphi} - \dot{x}^{\varphi} ) + Q^{\varphi r} \Big) + r^3 \mu
\cos \varphi \Big( \rho u^{\vartheta} (u^{\varphi} -
\dot{x}^{\varphi} + Q^{\varphi \vartheta} ) \Big) 
- &&  \\ 
\nonumber - r^3 \sqrt{1-\mu^2} \sin \varphi \Big( \rho
u^{\varphi}(u^{\varphi} - \dot{x}^{\varphi} ) +
\frac{1}{r^2(1-\mu^2)}p +Q^{\varphi \varphi} \Big) \Big] + && \\
\nonumber + \rho \Big( \partial_r \Big[ r^2 \sqrt{1-\mu^2}
\cos \varphi \Phi \Big] - 
\partial_{\mu} \Big[\mu r \sqrt{1-\mu^2} \cos \varphi \Phi \Big] 
+ \frac{1}{\sqrt{1-\mu^2}}\partial_{\varphi} 
\Big[ - r \sin \varphi \Phi \Big] \Big) - && \\
\nonumber - \frac{4\pi}{c} r^2 
\Big( \sqrt{1-\mu^2} \cos \varphi H^r + \mu r \cos
\varphi H^{\vartheta} - \sqrt{1-\mu^2} r \sin \varphi H^{\varphi} \Big) & = &
0 . \\ \label{3D_sph_EQM_1}
\end{eqnarray}

\begin{eqnarray}
\nonumber \partial_t \Big[ r^2 \sqrt{1-\mu^2} \rho \sin \varphi u^r + r^3 \mu 
\rho \sin \varphi u^{\vartheta} + r^3 \sqrt{1-\mu^2} 
\rho \cos \varphi u^{\varphi} 
\Big] + && \\
\nonumber + \partial_r \Big[ r^2 \sqrt{1-\mu^2} \sin \varphi 
\Big( \rho u^r (u^r - \dot{x}^r) + p + Q^{rr} \Big) 
+ r^3 \mu \sin \varphi \Big( \rho u^{\vartheta} (u^r -
\dot{x}^r ) + Q^{r \vartheta} \Big) + && \\ 
\nonumber + r^3 \sqrt{1-\mu^2} \cos \varphi \Big( \rho
u^{\varphi} (u^r - \dot{x}^r) + Q^{r \varphi}\Big) \Big] - 
\partial_{\mu} \Big[ r^2 (1-\mu^2) \sin \varphi \Big(
\rho  u^r (u^{\vartheta} - \dot{x}^{\vartheta} ) + Q^{\vartheta r} \Big) + && \\
\nonumber +
r^3 \mu \sqrt{1-\mu^2} \sin \varphi  \Big( \rho u^{\vartheta} (u^{\vartheta} -
\dot{x}^{\vartheta}) + \frac{p}{r^2} + Q^{\vartheta \vartheta} \Big) + r^3 
(1-\mu^2)
\cos \varphi \Big( \rho u^{\varphi} (u^{\vartheta} - \dot{x}^{\vartheta} ) +
Q^{\vartheta \varphi} \Big) \Big] + && \\
\nonumber \partial_{\varphi} \Big[ r^2 \sqrt{1-\mu^2} \sin \varphi \Big( 
\rho u^r (u^{\varphi} - \dot{x}^{\varphi}) + Q^{\varphi r} \Big) + 
r^3 \mu \sin \varphi
\Big( \rho u^{\vartheta} (u^{\varphi} - \dot{x}^{\varphi}) + Q^{\varphi
\vartheta} \Big) + && \\
\nonumber + r^3 \sqrt{1-\mu^2} \cos \varphi 
\Big( \rho u^{\varphi} (u^{\varphi} -
\dot{x}^{\varphi}) + \frac{p}{r^2(1-\mu^2)} + Q^{\varphi \varphi} \Big) 
\Big] + && \\
\nonumber + \rho \Big( \partial_r \Big[ r^2 \sqrt{1-\mu^2} 
\sin \varphi \Phi \Big] - 
\partial_{\mu} \Big[\mu r \sqrt{1-\mu^2} \sin \varphi \Phi \Big] 
+ \frac{1}{\sqrt{1-\mu^2}}\partial_{\varphi} 
\Big[ r \cos \varphi \Phi \Big] \Big) 
- && \\
\nonumber  - \frac{4 \pi }{c}r^2 
 \Big( \sqrt{1-\mu^2} \sin \varphi H^r + \mu r \sin
\varphi H^{\vartheta} +\sqrt{1-\mu^2} r \cos \varphi H^{\varphi} \Big) & = & 0 .
\\ \label{3D_sph_EQM_2} 
\end{eqnarray}

\begin{eqnarray}
\nonumber \partial_t \Big[ r^2 \mu \rho u^r - r^3 \sqrt{1-\mu^2} \rho
u^{\vartheta} \Big] + &&  \\
\nonumber + \partial_r \Big[r^2 \mu \Big( \rho u^r 
(u^r - \dot{x}^r) + p + Q^{rr} \Big) - r^3 \sqrt{1-\mu^2} 
\Big( \rho u^{\vartheta}
(u^r - \dot{x}^r ) + Q^{r \vartheta} \Big) \Big] - && \\
\nonumber - \partial_{\mu} \Big[r^2 \mu \sqrt{1-\mu^2} 
\Big( \rho u^r (u^{\vartheta}  -\dot{x}^{\vartheta}) + Q^{\vartheta r} 
\Big) - r^3 (1-\mu^2) \Big( \rho u^{\vartheta} (u^{\vartheta} - 
\dot{x}^{\vartheta})
+ \frac{p}{r^2} + Q^{\vartheta \vartheta}  \Big) \Big] + && \\
\nonumber + \partial_{\varphi} \Big[ \mu r^2 \Big( 
\rho u^r (u^{\varphi} - \dot{x}^{\varphi})+ Q^{\varphi r}\Big)  
- r^3 \sqrt{1-\mu^2} 
\Big( \rho u^{\vartheta} (u^{\varphi} - \dot{x}^{\varphi}) + 
Q^{\varphi \vartheta} \Big) \Big] + && \\
\nonumber + \partial_r \Big[ r^2 \mu \Phi \Big] +  
\partial_{\mu} \Big[ r (1-\mu^2) \Phi \Big] 
- \frac{4\pi}{c} r^2   
\Big( \mu H^r - r \sqrt{1-\mu^2} H^{\vartheta} \Big)  & = & 0 . \\
\label{3D_sph_EQM_3}
\end{eqnarray}
}

When we paste the contravariant components of the artificial 
viscosity explicitly, the $1$-component yields as an example

{\scriptsize  
\begin{eqnarray}
\nonumber \partial_t  \Big[ \rho r^2 (1-\mu^2) \cos \varphi
u^r + \rho r^3 \mu \sqrt{1-\mu^2} \rho
\cos \varphi u^{\vartheta} - \rho r^3 (1-\mu^2) \sin \varphi u^{\varphi} \Big]
+ &&
\\
\nonumber + \partial_r \Big[ r^2 (1-\mu^2) \cos \varphi \Big( \rho u^r (u^r -
\dot{x}^r) + p - \frac{1}{3}\mathrm{div} \, 
\mathbf{u} + \partial_r u_r \Big) + &
& \\
\nonumber r^3 \mu \sqrt{1-\mu^2} \cos \varphi \Big(
\rho
u^{\vartheta}(u^r - \dot{x}^r) +\frac{1}{r^2} \Big( - \frac{1}{r} u_{\vartheta}
+ \frac{1}{2} \partial_r u_{\vartheta} + \frac{1}{2} \partial_{\vartheta} u_r
\Big) \Big) - && \\
\nonumber  - r^3 (1-\mu^2) \sin \varphi \Big( \rho
u^{\varphi}(u^r - \dot{x}^r) + \frac{1}{r^2 (1-\mu^2)} \Big(
- \frac{1}{r} u_{\varphi} + \frac{1}{2} \partial_{\varphi} u_{r} +
\frac{1}{2} \partial_r u_{\varphi} \Big) \Big) \Big] - &&  \\
\nonumber - \sqrt{1-\mu^2} \partial_{\mu} \Big[ r^2 (1-\mu^2) \cos \varphi
\Big( \rho
u^r(u^{\vartheta} - \dot{x}^{\vartheta}) + \frac{1}{r^2} \Big(
-\frac{1}{r}u_{\vartheta} +
\frac{1}{2} \partial_r u_{\vartheta} + \frac{1}{2} \partial_{\vartheta} u_r
\Big) \Big) + 
&& \\
\nonumber  + r^3 \mu \sqrt{1-\mu^2}
\cos \varphi \Big( \rho u^{\vartheta} (u^{\vartheta} - \dot{x}^{\vartheta}
) + \frac{1}{r^2} p + \frac{1}{r^2} \Big(
-\frac{1}{3}\mathrm{div} \, 
\mathbf{u} + \frac{1}{r}u_r +
\frac{1}{r^2} \partial_{\vartheta}u_{\vartheta} \Big) \Big) - && \\
\nonumber - r^3 (1-\mu^2) \sin
\varphi \Big(\rho u^{\varphi}(u^{\vartheta} -
\dot{x}^{\vartheta}) + \frac{1}{r^4 (1-\mu^2)}
\Big( - \frac{\mu}{\sqrt{1-\mu^2}} u_{\varphi}
+ \frac{1}{2}\partial_{\varphi}u_{\vartheta} +
\frac{1}{2}\partial_{\vartheta}u_{\varphi} \Big) \Big) \Big] + && \\
\nonumber + \partial_{\varphi} \Big[ r^2 (1-\mu^2) \cos \varphi \Big( \rho u^r
(u^{\varphi} - \dot{x}^{\varphi} ) + \frac{1}{r^2 (1-\mu^2)} \Big( -
\frac{1}{r}u_{\varphi} +
\frac{1}{2} \partial_{\varphi} u_r + \frac{1}{2} \partial_r u_{\vartheta}
\Big) \Big) + && \\
\nonumber + r^3 \mu
\sqrt{1-\mu^2} \cos \varphi \Big( \rho u^{\vartheta} (u^{\varphi} -
\dot{x}^{\varphi}) + \frac{1}{r^4 (1-\mu^2)}
\Big( - \frac{\mu}{\sqrt{1-\mu^2}} u_{\varphi} +
\frac{1}{2}
\partial_{\varphi} u_ {\vartheta} + \frac{1}{2} \partial_{\vartheta} u_{\varphi}
\Big) \Big) - &&  \\ 
\nonumber - r^3 (1-\mu^2) \sin \varphi \Bigg( \rho
u^{\varphi}(u^{\varphi} - \dot{x}^{\varphi} ) + \frac{1}{r^2(1-\mu^2)}p + && \\
\nonumber  +\frac{1}{r^2 (1-\mu^2)}
\Big( -\frac{1}{3} \mathrm{div} \, \mathbf{u} + \frac{1}{r}u_r +
\frac{1}{r^2}
\frac{\mu}{\sqrt{1-\mu^2}} u_{\vartheta} + \frac{1}{r^2 (1-\mu^2)}
\partial_{\varphi}
u_{\varphi} \Big) \Bigg) \Big] + && \\
\nonumber + \rho \Big( \partial_r \Big[ r^2 (1-\mu^2) \cos \varphi \Phi \Big] - 
\sqrt{1-\mu^2} \partial_{\mu} \Big[\mu r \sqrt{1-\mu^2} \cos \varphi \Phi \Big] 
+ \partial_{\varphi} \Big[ - r \sin \varphi \Phi \Big] \Big) - && \\
\nonumber - \frac{4\pi}{c} r^2 \sqrt{1-\mu^2}  
\Big( \sqrt{1-\mu^2} \cos \varphi H^r + \mu r \cos
\varphi H^{\vartheta} - \sqrt{1-\mu^2} r \sin \varphi H^{\varphi} \Big) & = &
0 . \\
\end{eqnarray} }

The two components of the equation of motion in polar coordinates 
yield 

{\scriptsize 
\begin{eqnarray}
\nonumber \partial_t \Big[ r \sqrt{1-\mu^2} \rho u^r +  r^2 \mu \rho 
u^{\vartheta} \Big] + && \\
\nonumber + \partial_r \Big[ r \sqrt{1-\mu^2} \Big( \rho u^r (u^r - \dot{x}^r)
+ p + Q^{rr} \Big) + r^2 \mu \Big( \rho u^{\vartheta} (u^r - \dot{x}^r) 
+ Q^{r \vartheta} \Big) \Big] - && \\
\nonumber - \sqrt{1- \mu^2} \partial_{\mu} \Big[r \sqrt{1-\mu^2} \Big( 
\rho u^r (u^{\vartheta} -  \dot{x}^{\vartheta}) + Q^{\vartheta r}\Big) +
r^2 \mu \Big( \rho u^{\vartheta} (u^{\vartheta} - \dot{x}^{\vartheta}) 
+ \frac{p}{r^2} + Q^{\vartheta \vartheta} \Big)\Big] + && \\
\nonumber + \rho \Big( \sqrt{1-\mu^2} \partial_r \Big[ r \Phi \Big] - 
\sqrt{1-\mu^2} \partial_{\mu} \Big[ \mu \Phi \Big] \Big) - \frac{4\pi}{c}
r \Big(\sqrt{1-\mu^2} H^r + r \mu H^{\vartheta} \Big)  & = & 0 . \\
\end{eqnarray}
\begin{eqnarray}
\nonumber \partial_t \Big[ r \mu \rho u^r - r^2 \sqrt{1-\mu^2} \rho
u^{\vartheta} \Big] + && \\
\nonumber + \partial_r \Big[ r \mu \Big( \rho u^r (u^r - \dot{x}^r )
+ p + Q^{rr} \Big) - r^2 \sqrt{1-\mu^2} \Big( \rho u^{\vartheta} (u^r - 
\dot{x}^r) + Q^{r \vartheta} \Big) \Big] - && \\
\nonumber - \sqrt{1- \mu^2} \partial_{\mu} \Big[ 
r \mu \Big( \rho u^r (u^{\vartheta} - \dot{x}^{\vartheta}) + Q^{\vartheta r}
\Big) - r^2 \sqrt{1-\mu^2} \Big( \rho u^{\vartheta} (u^{\vartheta} - 
\dot{x}^{\vartheta}) + \frac{p}{r^2} + Q^{\vartheta \vartheta} \Big)
\Big] + && \\
\nonumber + \rho \Big( \mu  \partial_r \Big[ r \Phi \Big] + 
\sqrt{1-\mu^2} \partial_{\mu} \Big[ \sqrt{1 - \mu^2} \Phi \Big] 
\Big) - \frac{4\pi}{c}
r \Big(\mu H^r - r \sqrt{1-\mu^2} H^{\vartheta} \Big)  & = & 0 . \\
\end{eqnarray} }

\subsection{Equation of Internal Energy}

The energy equation we are going to impose contains the 
thermodynamics of the fluid (\ref{energy_eq_def})
 as well as the energy exchange 
with the radiation field (\ref{rad_coupling_terms}) and viscous 
energy dissipation, expressed by the contraction of 
the viscosity with the velocity gradient $\mathbf{Q}:\nabla \mathbf{u}$. 
\begin{equation}
\partial_t (\rho \epsilon) + \mathrm{div} \, (\mathbf{u} \rho
\epsilon) + \mathbf{P} : \nabla\mathbf{u} - 4 \pi \kappa_P 
\rho (J-S) + \mathbf{Q}: \nabla\mathbf{u} = 0 
\end{equation}
Since we assume isotropic gas pressure $\mathbf{P}=p \mathbf{g}$ 
we can reformulate its contribution via the Ricci Lemma (\ref{ricci_lemma}) 
and obtain a very simple scalar expression. 
\begin{eqnarray}
\nonumber \mathbf{P} : \nabla\mathbf{u} & = & g^{ij} p \nabla_i u_j \\
\label{pucontrisopress} & = & p \nabla_i u^i = p \, \mathrm{div} \, \mathbf{u}
\end{eqnarray}
The \emph{viscous energy dissipation} is a little more work to do, since 
both matrices contain a priori no zeros.  
However, for this contraction we do not need to appeal to strong conservation 
differentiation since this term does not have its seeds in a conservation law. 
\begin{eqnarray}
\nonumber \mathbf{Q}: \nabla \mathbf{u} & = & Q^{ij} \nabla_i u_j \\
\nonumber & = & Q^{ij} \left( \partial_i u_j - {\Gamma^k}_{ij} u_k \right) \\
\nonumber & = & 
Q^{rr} \partial_r u_r + 
Q^{r \vartheta} \Big( 
\partial_ru_{\vartheta} - \frac{1}{r} u_{\vartheta} \Big) +  Q^{r \varphi} 
\Big( \partial_r u_{\varphi} - \frac{1}{r} u_{\varphi} \Big) + \dots  \\
\label{gradviscontracted} & = & \mathsf{E}_{\mathrm{diss}}
\end{eqnarray} 
Of course also this quantity from now on denominated 
$\mathsf{E}_{\mathrm{diss}}$ 
is calculated with the aid of {\sf Mathematica}. 

The strong conservative form of the energy equation yields 
\begin{eqnarray}
\nonumber 
\partial_t \Big[\sqrt{|\mathbf{g}|} \rho \epsilon \Big] + 
\partial_i \Big[ \sqrt{|\mathbf{g}|} \rho \epsilon \, 
\mathbf{{{e}}}^{i} \cdot (\mathbf{u} - \dot{\mathbf{x}}) \Big] 
 + \sqrt{|\mathbf{g}|} p \, \mathrm{div} \, \mathbf{u} \, - && \\
\label{EIE_cons} 
- \, 4\pi \sqrt{|\mathbf{g}|} \kappa_P \rho (J-S) + \sqrt{|\mathbf{g}|}
\mathsf{E}_{\mathrm{diss}} & = & 0 .
\end{eqnarray} 
We project the velocities on the covariant 
base, multiply with $\sqrt{|\mathbf{g}|}$ 
and divide by $\sqrt{1-\mu^2}$ to obtain the energy equation 
in spherical 
\begin{eqnarray}
\nonumber 
\partial_t \Big[r^2 \rho \epsilon \Big] + 
\partial_r \Big[ r^2 \rho \epsilon (u^r - \dot{x}^r) \Big] 
- r^2  \partial_{\mu} \Big[ \sqrt{1-\mu^2} 
\rho \epsilon (u^{\vartheta} - \dot{x}^{\vartheta}) \Big] \, + && \\
\nonumber + \, r^2 \partial_{\varphi} 
\Big[ \rho \epsilon (u^{\varphi} - \dot{x}^{\varphi})
\Big] + r^2 p \, \mathrm{div} \, \mathbf{u} - 
4\pi r^2 \kappa_P \rho (J-S) + r^2 \mathsf{E}_{\mathrm{diss}} & = & 0 \\
\label{3D_EIE_sph}
\end{eqnarray} 
and polar coordinates. 
\begin{eqnarray}
\nonumber \partial_t \Big[r \rho \epsilon \Big] + 
\partial_r \Big[ r \rho \epsilon (u^r - \dot{x}^r) \Big] 
- r \sqrt{1- \mu^2}  \partial_{\mu} \Big[ \rho \epsilon (u^{\vartheta} - 
\dot{x}^{\vartheta}) \Big] + && \\
\label{2D_EIE_sph} + \,  r p \, \mathrm{div} \, \mathbf{u} - 
4\pi r \kappa_P \rho (J-S) + r \mathsf{E}_{\mathrm{diss}} & = & 0 
\end{eqnarray}

\subsection{Equation of Radiation Energy}

The simplified frequency integrated radiation energy equation in the 
comoving frame was defined in equation (\ref{rad_en_flux_def}). 
\begin{equation}
\partial_t J + \mathrm{div} \,  
(\mathbf{u} J) + c \, \mathrm{div} \,  \mathbf{H} + 
\mathbf{K} : \nabla\mathbf{u} + c \chi_P (J-S) = 0
\end{equation}
For the scalar energy input of radiative pressure into the material 
we analogously define a new coupling variable 
\begin{eqnarray}
\nonumber \mathbf{K}: \nabla \mathbf{u} & = & K^{ij} \nabla_i u_j \\
\nonumber & = & 
K^{rr} \partial_r u_r + 
K^{r \vartheta} \Big( 
\partial_ru_{\vartheta} - \frac{1}{r} u_{\vartheta} \Big) +  K^{r \varphi} 
\Big( \partial_r u_{\varphi} - \frac{1}{r} u_{\varphi} \Big) + \dots  \\
\label{gradviscontracted} & = & \mathsf{P}_{\mathrm{coup}}
\end{eqnarray}

In strong conservation form it yields 
\begin{eqnarray}
\nonumber 
\partial_t\Big[\sqrt{|\mathbf{g}|} J \Big] + 
\partial_i \Big[ \sqrt{|\mathbf{g}|} \, \mathbf{{{e}}}^{i} \cdot \left( J 
(\mathbf{u}-\dot{\mathbf{x}}) +c \mathbf{H}\right) \Big] + && \\
+ \, \sqrt{|\mathbf{g}|} 
\mathsf{P}_{\mathrm{coup}} + 
\sqrt{|\mathbf{g}|} c \chi_P (J-S) = 0 .
\end{eqnarray}
In spherical coordinates with fixed $\vartheta$ we obtain 
\begin{eqnarray}
\nonumber \partial_t \Big[ r^2 J \Big] + \partial_r \Big[ r^2 \left( 
J ( u^r - \dot{x}^r) + cH^r\right) \Big] - && \\
\nonumber  r^2 \partial_{\mu} \Big[ 
\sqrt{1-\mu^2} \left( J ( u^{\vartheta} - \dot{x}^{\vartheta}) + cH^{\vartheta}
\right)\Big] + && \\
\nonumber  + \, r^2 \partial_{\varphi} \Big[ 
\left( J ( u^{\varphi} - \dot{x}^{\varphi}) + cH^{\varphi} \right)\Big]
+  r^2 \mathsf{P}_{\mathrm{coup}} + 
r^2 c \chi_P (J-S) & = & 0 .\\
\label{3D_ERE_sph}
\end{eqnarray}
and the radiation energy equation in polar coordinates yields 
\begin{eqnarray}
\nonumber \partial_t \Big[ r J \Big] + \partial_r \Big[ r \left( 
J ( u^r - \dot{x}^r) + cH^r\right) \Big] - && \\
\nonumber  \sqrt{1-\mu^2} 
r \partial_{\mu} \Big[ \left( J ( u^{\vartheta} - \dot{x}^{\vartheta}) 
+ cH^{\vartheta} \right)\Big] 
 + \, r \mathsf{P}_{\mathrm{coup}} + 
r c \chi_P (J-S) & = & 0 . \\
\label{2D_ERE_sph}
\end{eqnarray}

\subsection{Radiative Flux Equation}

The radiative flux equation, another 
vectorial conservation law remains to be sketched in strong 
conservation form for general non-steady coordinates. We 
emanate from equation (\ref{rad_en_flux_def}). 
\begin{equation}
\partial_t \mathbf{H} + \mathrm{div} \,  (\mathbf{u} \mathbf{H}) 
+ c \, \mathrm{div} \,  \mathbf{K} + 
\mathbf{H} \cdot \nabla\mathbf{u} + c \chi_R
\mathbf{H} = 0 
\end{equation} 
and for a start define 
an \emph{effective radiative flux tensor} $\mathbf{L}$ analogously 
to the effective tensor of gaseous momentum. 
\begin{eqnarray}
\nonumber (\mathbf{u}-\dot{\mathbf{x}}) 
\mathbf{H} & = & \mathbf{L} = l^{ij}  \mathbf{{e}}_i \mathbf{{e}}_j \\
l^{ij}  & = & (u^i - \dot{x}^i) H^j
\end{eqnarray}
The contribution of radiative momentum to the material 
$\mathbf{H} \cdot \nabla\mathbf{u}$ will 
be abbreviated via another new coupling variable with 
components 
\begin{equation} 
\nonumber \mathsf{F}_{\mathrm{coup}}^j = H^i \nabla_i u^j .
\end{equation}

The geometrically conservative form in non-steady coordinates 
of upper equation will be considered 
in the following shape. 
\begin{eqnarray}
\nonumber 
\partial_t \Big[ \sqrt{|\mathbf{g}|} \mathbf{H} \Big] + \partial_i 
\Big[ \sqrt{|\mathbf{g}|} \, \mathbf{{{e}}}^{i} 
\cdot \left( \mathbf{L} + c \mathbf{K} \right) \Big] + && \\
+\,  \sqrt{|\mathbf{g}|}
 \mathsf{F}_{\mathrm{coup}}
+ \sqrt{|\mathbf{g}|} \kappa_R \rho \mathbf{H} & = & 0 . 
\end{eqnarray}

The three components of the radiative flux equation in spherical 
coordinates yields 
\newpage 

{\scriptsize
\begin{eqnarray}
\nonumber \partial_t\Big[ r^2 \sqrt{1-\mu^2} \cos\varphi H^r + r^3 
\mu \cos\varphi 
H^{\vartheta} - r^3 \sqrt{1-\mu^2} \sin\varphi H^{\varphi} \Big] + && \\
\nonumber + \, \partial_r \Big[ r^2 \sqrt{1-\mu^2} \cos\varphi \Big( H^r (u^r
-\dot{x}^r) + cK^{rr} \Big) + r^3 \mu \cos\varphi 
\Big( H^{\vartheta}
(x^r-\dot{x}^r) + c K^{r \vartheta} \Big) - && \\
\nonumber - \, r^3 \sqrt{1-\mu^2} \sin\varphi \Big( H^{\varphi} (x^r - \dot{x}^r )
\Big)\Big] - \partial_{\mu} \Big[ r^2 (1-\mu^2) 
\cos\varphi \Big( 
H^r(u^{\vartheta} - \dot{x}^{\vartheta}) + cK^{\theta r}\Big) + && \\
\nonumber + \, r^3 \mu \sqrt{1-\mu^2} \cos\varphi 
\Big( H^{\vartheta} (u^{\vartheta} - 
\dot{x}^{\vartheta}) + cK^{\vartheta \vartheta}\Big) - 
r^3 (1-\mu^2) \sin\varphi 
\Big( H^{\varphi} (x^{\vartheta} - \dot{x}^{\vartheta} ) + 
cK^{\vartheta \varphi}\Big) \Big] + && \\
\nonumber \partial_{\varphi} \Big[ r^2 \sqrt{1-\mu^2} \cos\varphi \Big( 
H^r(u^{\varphi} - \dot{x}^{\varphi}) + cK^{\varphi r} \Big) + r^3 
\mu \cos\varphi \Big( H^{\vartheta} (u^{\varphi} - \dot{x}^{\varphi}) 
+ cK^{\varphi \vartheta} \Big) - && \\
\nonumber - \, r^3 \sqrt{1-\mu^2} \sin\varphi \Big( 
H^{\varphi} (u^{\varphi} - \dot{x}^{\varphi}) + cK^{\varphi \varphi} \Big)\Big]
+ && \\ 
\nonumber 
+ \, r^2  \Big( \sqrt{1-\mu^2} \cos\varphi \mathsf{F}_{\mathrm{coup}}^r + 
r \mu \cos \varphi \mathsf{F}_{\mathrm{coup}}^{\vartheta} - r \sqrt{1-\mu^2} \sin\varphi 
\mathsf{F}_{\mathrm{coup}}^{\varphi} \Big) + && \\
\nonumber + \, const r^2 \Big(\sqrt{1-\mu^2} 
\cos\varphi H^r + r \mu \cos\varphi H^{\vartheta} -r \sqrt{1-\mu^2} 
\sin\varphi H^{\varphi}  \Big) & = & 0 .\\
\label{3D_sph_RFE_1}
\end{eqnarray}

\begin{eqnarray}
\nonumber \partial_t \Big[ r^2 \sqrt{1-\mu^2} \sin\varphi H^r + 
r^3 \mu \sin\varphi H^{\vartheta}
+ r^3 \sqrt{1-\mu^2} \cos\varphi H^{\varphi} \Big] + && \\
\nonumber + \, \partial_r \Big[ r^2 \sqrt{1-\mu^2} \sin\varphi \Big( 
H^r (u^r - \dot{x}^r) + cK^{rr}\Big) + r^3 \mu \sin\varphi \Big( 
H^{\vartheta} (u^r - \dot{x}^r) + cK^{r \vartheta}\Big) 
+ && \\
\nonumber + \, r^3 \sqrt{1-\mu^2} \cos\varphi \Big( H^{\varphi} (x^r - \dot{x}^r)
+ c K^{r \varphi}\Big) \Big] - \partial_{\mu} \Big[ 
r^2 (1-\mu^2) \sin\varphi \Big( H^r(u^{\vartheta} - \dot{x}^{\vartheta}) + 
cK^{\vartheta r} \Big) + && \\
\nonumber + \, r^3 \mu \sqrt{1-\mu^2} \sin\varphi \Big( 
H^{\vartheta}(u^{\vartheta} - \dot{x}^{\vartheta}) 
+ c K^{\vartheta \vartheta} \Big) + r^3 (1-\mu^2) \cos\varphi \Big( 
H^{\varphi} (u^{\vartheta} - \dot{x}^{\vartheta}) + cK^{\vartheta \varphi}  
\Big) \Big] + &&\\
\nonumber \partial_{\varphi} \Big[ r^2 \sqrt{1-\mu^2} \sin\varphi \Big( 
H^r (u^{\varphi} - \dot{x}^{\varphi}) + cK^{\varphi r} \Big) + 
r^3 \mu \sin\varphi \Big( H^{\vartheta} (u^{\varphi} -
\dot{x}^{\varphi}) 
+ c K^{\varphi \vartheta} \Big) + && \\ 
\nonumber r^3 \sqrt{1-\mu^2} \cos\varphi \Big(H^{\varphi} (u^{\varphi} - 
\dot{x}^{\varphi}) + cK^{\varphi \varphi} \Big) \Big] + && \\
\nonumber + \, r^2  \Big( 
\sqrt{1-\mu^2} \sin\varphi \mathsf{F}_{\mathrm{coup}}^r + \mu r \sin\varphi \mathsf{F}_{\mathrm{coup}}^{\vartheta}
+ r \sqrt{1-\mu^2} \cos\varphi \mathsf{F}_{\mathrm{coup}}^{\varphi}\Big) + && \\
\nonumber \, + const  r^2 \Big( 
\sqrt{1-\mu^2} \sin\varphi H^r + \sqrt{1-\mu^2} \cos\varphi H^{\varphi}
+ r \mu \sin\varphi H^{\vartheta} \Big) & = & 0 .\\
\label{3D_sph_RFE_2}
\end{eqnarray}

\begin{eqnarray}
\nonumber 
\partial_t \Big[ r^2 \mu H^r - r^3 \sqrt{1-\mu^2} H^{\vartheta} 
\Big] + && \\
\nonumber + \, \partial_r \Big[ r^2 \mu \Big(  
H^r (u^r - \dot{x}^r) + cK^{rr} \Big) - r^3 \sqrt{1-\mu^2} \Big( 
H^{\vartheta} (x^r - \dot{x}^r) + cK^{r \vartheta} \Big) \Big] - && \\
\nonumber - \, \partial{\mu} \Big[ r^2 \mu \sqrt{1-\mu^2} \Big( 
H^r (u^{\vartheta} - \dot{x}^{\vartheta}) + cK^{\vartheta r}\Big) - 
r^3 (1-\mu^2) \Big(
H^{\vartheta} (u^{\vartheta} - \dot{x}^{\vartheta} ) + cK^{\vartheta 
\vartheta} \Big) \Big] + && \\
\nonumber \partial_{\varphi} \Big[ r^2 \mu \Big( 
H^r (u^{\varphi} - \dot{x}^{\varphi}) + cK^{\varphi r} \Big) - 
r^3 \sqrt{1-\mu^2}
\Big( H^{\vartheta} (u^{\varphi} - \dot{x}^{\varphi}) + cK^{\varphi \vartheta}
\Big) \Big] + && \\
\label{3D_sph_RFE_3} 
+ \, r^2  \Big( \mu \mathsf{F}_{\mathrm{coup}}^r - r\sqrt{1-\mu^2} 
\mathsf{F}_{\mathrm{coup}}^{\vartheta} \Big) + const r^2  \Big( \mu H^r - 
r \sqrt{1-\mu^2} H^{\vartheta}\Big) & = & 0 .  
\end{eqnarray} }

The two components in polar coordinates can be found in the following 
arrays. 

{\scriptsize 
\begin{eqnarray}
\nonumber \partial_t \Big[ r \sqrt{1-\mu^2} H^r + r^2 \mu H^{\vartheta} \Big] 
+ && \\ 
\nonumber + \, \partial_r \Big[ r \sqrt{1-\mu^2} \Big( H^r(u^r - \dot{x}^r) 
+c K^{rr} \Big) + r^2 \mu \Big( H^{\vartheta}(u^r-\dot{x}^r) 
+cK{r \vartheta} \Big) \Big] - && \\
\nonumber - \, \sqrt{1-\mu^2} \partial_{\mu} \Big[ r \sqrt{1-\mu^2} 
\Big( H^{r} (u^{\vartheta} - \dot{x}^{\vartheta} ) + cK^{\vartheta r} 
\Big) + r^2 \mu \Big( H^{\vartheta}(u^{\vartheta} - \dot{x}^{\vartheta}) 
+cK^{\vartheta \vartheta} \Big)  \Big] + && \\
\label{2D_sph_RFE_1} + \,  r \Big( \sqrt{1-\mu^2} HradU^r + 
r \mu \mathsf{F}_{\mathrm{coup}}^{\vartheta} \Big) + const r \Big( 
\sqrt{1-\mu^2} H^r + r \mu H^{\vartheta}\Big) & = & 0 . 
\end{eqnarray} 

\begin{eqnarray}
\nonumber \partial_t \Big[ r^2 \mu H^r - r^2  \sqrt{1-\mu^2}
 H^{\vartheta} \Big] 
+ && \\ 
\nonumber + \, \partial_r \Big[ r \mu \Big( H^r(u^r - \dot{x}^r) + 
cK^{r r}\Big) - r^2 \sqrt{1-\mu^2} \Big( H^{\vartheta}(u^r - \dot{r}) + 
cK^{r \vartheta} \Big) \Big] - && \\
\nonumber - \, \sqrt{1-\mu^2} \partial_{\mu} \Big[ r \mu \Big( 
H^{r} (u^{\vartheta} - \dot{x}^{\vartheta}) + cK^{\vartheta r} \Big) - 
r^2 \sqrt{1-\mu^2} \Big( H^{\vartheta}(u^{\vartheta} - \dot{x}^{\vartheta}) 
 + cK^{\vartheta \vartheta} \Big) \Big] + && \\
\label{2D_sph_RFE_2} + \, r \Big( \mu \mathsf{F}_{\mathrm{coup}}^r - r \sqrt{1-\mu^2} 
H^{\vartheta} \Big) + const r \Big( \mu H^r - r \sqrt{1-\mu^2} H^{\vartheta}
\Big)  & = & 0 .
\end{eqnarray} }

% ---------------------------------------------------------------------------- %
% CHAPTER 5 ------------------------------------------------------------------ %
% ---------------------------------------------------------------------------- %

\chapter{Orthogonal and Nonorthogonal Adaptive Grids} \label{Grid_Generation}

In the forthcoming sections we want to motivate and discuss the necessity 
for multi-dimensionally adaptive grids for astrophysical applications 
with radiation hydrodynamics in 2D and 3D and thereby justify our excessive 
mathematics hitherto. At first we will give a brief introduction to 
fundamental concepts in grid generation and then phrase three crucial 
postulations for suitable adaptive grids in implicit RHD numerics.  
We will discover that not all three postulation can be fulfilled 
and therefore occupy ourselves with possible compromises and 
prospective continuation of our ideas and approaches.

\section{Basic Principles of Grid Generation}

Finite difference and finite element methods for numerical solutions of 
partial differential equations require a grid which represents the physical 
domain in discrete terms, often referred to as 
the \emph{logical space}. Generally the grid is a 
set of points, lines, (hyper)surfaces and (hyper)volumes that arise 
from an appropriate transformation between these two domains. 
In many applications the 
points, also called \emph{nodes} and volumes or \emph{cells}
 are to be chosen in a way 
that the underlying problem gets simplified; for instance a transformation to a 
rectangular, an orthogonal or a merely smooth domain. 
Such \emph{structured meshes} can be obtained via coordinate transformations, 
algebraic methods like polynomial interpolation or differential methods
respectively hybrid variants (a textbook giving a 
profound overview is e.g. Knupp and Steinberg  
\cite{KnuppSteinberg}). In chapter \ref{Conservative Numerics} we 
anticipated generating a structured mesh by discretization of coordinates 
we obtained from a well known coordinate transformation even though 
we never explicitly conducted this transition from $(r,\vartheta,\varphi)$ 
to $(r_i,\vartheta_j,\varphi_k)$ and differential to difference operators. 

\emph{Unstructured grids} cover the computational domain with cells of 
arbitrary shape and do rather appear with complex geometrical 
domains. Practical problems of unstructured meshes arise from 
numbering and ordering the 
cells and edges which puts high computational demands even if boundaries do not
move. Clearly also the numerical solution of a partial differential equation on 
such an unstructured is more expensive than on a structured mesh. However, 
one advantage of unstructured grids is a quite handy local mesh refinement by 
dividing grid cells appropriately (Liseikin \cite{LiseikinGridG}). 

Grid generation itself must be understood as an essential 
step of finding suitable methods for solving PDEs 
numerically\footnote{There are popular numerical methods 
working without meshes like the family of spectral methods or 
smoothed-particle hydrodnamics but as the chapter title suggests 
we concern ourselves with grid-based techniques exclusively.}. In the following 
sections we want to point out some possible approaches to grid generation 
in 2 and 3 dimensions. However, for the sake of clearness 2D grids will 
be favored. 

Time dependent problems with steep gradients as they appear in hydrodynamics 
state special challenges for numerical techniques, likewise for numerical 
grid generation. Basically one can distinguish between three main approaches to 
handle non static finite differences or finite elements; the Eulerian, 
the Lagrangian and methods that allow generally adaptive grids. 
As pointed out in section \ref{Adaptive Grids},
Eulerian grids remain static throughout the computation and all physical 
quantities and fluxes are calculated with respect to steady nodes and cells 
whereas Lagrangian grids move with the very velocity of one particular 
physically 
interesting component, in most cases the gas. 
Analysis reveales (Furzeland et al. (1990) \cite{Furzeland1990349})
that {\sl generally adaptive grids} are definitely privileged. 
Connecting the grid with physical quantities respectively some 
measure of their computational error and adjusting its profile iteratively 
provides the best results whenever the number of nodes is fixed 
(i.e. no mesh refinement). 

Fundamental ideas of dynamically adaptive grids in RHD were 
developed by Winkler (1977) \cite{WinklerPHD}, 
Winkler, Norman and Mihalas (1984) \cite{Winkler1984473} and 
\cite{Mihalas1984479} respectively a sophisticated method 
for grid control in 1D 
was found by Dorfi and Drury (1987) \cite{Dorfi1987175}. 
In the following paragraphs we shall investigate some potential  
amplifications to multi-dimensionally adaptive grids.

\section{The Grid We Are Looking For} \label{Postulations}

First and foremost we are looking for a grid that geometrically 
operates radiation hydrodynamics in problem-oriented domains. Since we know 
that all stars rotate and therefore are rather geoids than spheres, polar 
and spherical coordinates do not match optimally. 
In latter geometries, shock waves would not propagate alongside 
coordinate lines respectively within coordinate surfaces but skew to them. 
Immense fluxes in multiple directions would be the consequence 
which is numerically troublesome. 
Hence, an adaptive grid that satisfies our demands ideally would 
have to allow adaptiveness in more than one dimension but remain orthogonal 
all along. 

\begin{Po} 
The grid should be adaptive in more than one dimension but remain 
orthogonal throughout the computation. 
\end{Po}

In the case of strong conservation finite volume methods
in complex geometries the orthogonality of the coordinate lines would 
ensure that the metric tensor (as we have seen a
 function of space and time) remains 
diagonal. With more general metrics, not only the volume element 
$| \mathbf{g} |$ becomes rather elaborate as it is calculated by 
the determinant of the metric. All projections and summations of our physical 
tensors require far more arithmetic operations. For instance, each 
of the nine contravariant component of the viscous pressure 
${\mathbf{Q}}$ 
requires nine arithmetic operations. In discrete terms this implicates 
to execute these 27 $\times$ 2 operations at each node and time step
at worst. 
\[
Q^{ij} = Q_{lm}g^{li}g^{mj} = \dots  
\left( g^{li}g^{mj}(\nabla_l u_m + \nabla_j u_m ) - g^{ij}
\frac{1}{3} \mathrm{div} \,  \mathbf{u} \right)
\]
With orthogonal grids, 
all off diagonal components of the metric are zero 
$g^{ij}=0 \, \forall \, i \neq j$ and in this case 
54 operations reduce to effective six. After all 
this objection is not that relevant practically, 
since in implicit radiation hydrodynamics
most CPU time is consumed by the matrix inversion 
(Dorfi (1999) \cite{Dorfi1999153}). 

Even if orthogonality can not be 
maintained (which indeed will be one of our conclusions), 
the adaptive grids by any means needs some kind of mollifying property 
that avoids excessively twisted, acute-angled or one another overtaking cells. 

\begin{Po}
Grid cells should basically maintain their local geometries.   
\end{Po}

Moreover, the one dimensional special case of simple radial geometry shall 
be inherent in more dimensional grid generation so that one can easily 
compare calculations, e.g. with previous works in implicit RHD 
numerics like Dorfi (1999) \cite{Dorfi1999153} 
respectively Dorfi et al. (2004) \cite{Dorfi2006771}. 

\begin{Po}
The more dimensional grid should contain the one 
dimensional special case of radial geometry. 
\end{Po}

In the upcoming section we will discuss several approaches to grid 
generation in the face of problem oriented geometries and suggest 
feasible approaches to multi-dimensionally adaptive grids in 
2D and 3D radiation hydrodynamics.

\section{Remarks on Grid Generation Methods}

In this section we present some possible approaches to grid generation 
for implicit adaptive mesh techniques and discuss their 
compatibility with our three postulations. 
A common premise is an a priori fixed 
number of grid points that is redistributed during the temporal evolution. 
Again our perspective emphasizes on methods for neither 
fully Eulerian nor fully Lagrangian grid but a hybrid where gas 
velocity and grid velocity are associated but not equal 
as introduced in section \ref{Adaptive Grids}. 

Before we glance at adaptive grid techniques we present some 
fundamental {\sl static} grid generation methods that could
be adapted for time dependent problems.

\subsection{Differential Methods}

Grid generation using differential equation techniques are very popular for 
structured meshes, especially with complex geometries. The idea is to set 
up a system of partial differential equations whose solution defines a 
coordinate transformation. Well known properties of special differential 
equations can be used to control coordinates to some extent and on the 
other hand help to find suitable solution methods. 
These and a lot of 
other methods are well described in several text books, e.g. 
Castillo \cite{Castillo} or Liseikin \cite{LiseikinGridG}. 

Grid generation methods based on solving partial differential equations
are naturally classified elliptic, hyperbolic and parabolic. 
Hyperbolic equations are preferred for meshes around bodies but 
prove inappropriate for internal and closed domains which 
means they naturally disqualify for astrophysical applications. Moreover, 
inherent singularities propagate and grid oscillation and overlapping of 
volumes is encountered frequently. Parabolic techniques combine some 
positive aspects of hyperbolic properties like being able to formulate an 
initial value problem on the one hand and elliptical properties like 
diffusion effects smoothing out singularities on the other hand. 
Numerous references to hyperbolic and parabolic as well as combined 
techniques can also be found in \cite{LiseikinGridG}, p191 ff.  
Mathematical fundamentals as well as research activities in the field of 
differential grid generation are presented in 
Knupp and Steinberg \cite{KnuppSteinberg}. 

Exemplary we want to sketch some basic properties of \emph{elliptic 
systems} using Laplace or Poisson equations. 
Elliptic equations have one important characteristic as they 
ensure smooth solutions in the total 
interior of the domain even if the boundaries are non-smooth. For smooth 
boundaries continuous derivatives of the resulting coordinates are also granted 
on the boundaries. Elliptic equations that obey the extremum principle 
like the Laplace Equation are also known to have very small tendency for 
folding of grid cells (see e.g. Spekreijse (1995) 
\cite{Spekreijse199538}, Zhang et al. (2006) \cite{Zhang2006549}). However, 
we should mention that 
elliptic systems are numerically costly and Laplace-type equations allow 
practically no control of the geometric properties of the grid. In anticipation 
of more sophisticated methods the Laplace system is presented 
briefly.

\subsection{Laplace Systems} 

Many of the commonly used techniques to generate grids via differential 
equations are derivations of a method proposed by Winslow (1966) 
\cite{Winslow1966149} where the nonlinear 
two dimensional Poisson equation is solved. In these approaches the 
constructed meshes can be understood as equipotential surfaces. 

We shall study an archetype of such an elliptical grid generation 
method and consider a computational domain $\Xi^d$, a physical domain $X^d$  
and a coordinate system $\Sigma_{(\alpha)}$. 
\begin{equation}
\Sigma_{(\alpha)} : \Xi^d \rightarrow X^d, 
\quad \mathbf{x}(\mathbf{\xi}) \mapsto ({x_{(\alpha)}}^{i}(\mathbf{\xi})), 
\quad i = 1, \dots, d
\end{equation}
The most obvious elliptic system is the \emph{Laplace system}
that can be formulated 
in the physical 
\begin{equation}
\Delta x^{i} = \frac{\partial^2}{\partial {\xi^j}^2} x^{i} = 0 
\end{equation} 
or computational domain (interchange $x$ and $\xi$) 
which is equivalent with $g^{lm} \partial_l \partial_m x^i = 0$, 
a coupled quasilinear system of partial differential equations in $\Xi$. 
In most cases the transformation $\mathbf{x}(\mathbf{\xi})$ 
will be considered since 
this formulation in the physical space allows direct control of grid spacing 
and orthogonality. 

One main requirement for the transformation is that the 
Jacobian of the transformation $\mathbf{x}(\mathbf{\xi}): \Xi^n \to X^n$ 
is nonzero, i.e. the two domains are diffeomorphic which is 
confirmed by the theorem of Rado for harmonic functions. This leads 
us to the first discrepancies with our postulations made before. 
There is no diffeomorphism between Cartesian and polar respectively
spherical coordinates that includes $r=0$. Moreover, even if we 
cut out the origin, polar and spherical coordinates do not even satisfy 
the Laplace nor a Poisson equation. 
  
Nevertheless we briefly discuss one application and consider 
the transformation $\mathbf{x}(\xi,\eta)$ in a
a two dimensional simply connected bounded domain $X^2$ and let $\Xi^2$ be 
a unit square as commonly imposed. 
In section \ref{Tensor_Calculus} when we introduced 
basic differential geometric relations, we also mentioned that 
the metric tensor $\mathbf{g}$ is symmetric which leads to the following 
boundary value problem\footnote{For our 
orthogonal reference, the polar coordinates we obtain a contradiction 
$(\partial^2_r + \frac{1}{r}\partial^2_{\vartheta}) (r \cos \vartheta, 
r \sin \vartheta) \neq 0 $.}. 
\begin{equation} \label{L2}
\big( g^{\xi \xi}\partial^2_{\xi}  + 2g^{\xi \eta}\partial_{\xi} \partial_{\eta} 
 + g^{\eta \eta} \partial^2_{\eta} \big) x^i = 0 
\end{equation}
Computations as well as differential geometric deliberations reveal that 
grid lines obtained by (\ref{L2}) are attracted to concave boundaries and 
repelled in the vicinity of convex boundaries. In stellar astrophysics 
 this would implicate that the mesh gets tighter towards 
the center of the star automatically. 
The bottom line is that such 
elliptical systems as pure boundary value problems  
disqualify for grid adaptiveness with respect to {\sl inner} domain dynamics.

\subsection{Conformal Mapping}

When thinking about our first postulation of more dimensional adaptivity 
and orthogonality one might consider conformal maps since they are 
angle-preserving. Conformal transformations would remain orthogonal 
throughout the evolution of the grid which is represented 
by a diagonal metric $g_{ij}=\mathrm{diag}(g_{ii})$. Unfortunately coordinate 
caustics and overlapping problems are not the only challenges one would have 
to face. Both domains, $\Xi^n$ and $X^n$ have to be 
{\sl conformally equivalent} which is expressed 
by the fact that the metric tensor is a multiple of the unit 
matrix. 
\begin{equation}
g_{ij} = g(\xi) {\delta^i}_{j}
\end{equation}
Again we come into conflict with our postulations given that 
our reference systems, polar 
and spherical coordinates are not conformally equivalent to Cartesian 
ones\footnote{In polar 
coordinates the metric yields $g_{ij}=\mathrm{diag}
(1,r) \neq g(r,\vartheta) {\delta^i}_j$.}.

\subsection{Algebraic Methods}

\emph{Algebraic grid generation methods} rely on coordinate transformations 
constructed by interpolation techniques. According to concrete 
requirements to the solutions, different interpolation functions 
respectively polynomials are evaluated and transformations are governed 
by the coefficients of these polynomials. 

These schemes prove computationally efficient but lack some convenient 
intrinsic properties we presented in context of differential methods. 
Algebraic grid generation requires significant control techniques to 
obtain practicable meshes otherwise skewness of cells and grid point spacing 
can get messy. There are several ideas to regulate the grid such as 
adding differential calculus elements in order to inhibit that 
coordinate surfaces come too close or even overlap respectively 
degenerate. Other approaches directly influence the interpolation 
coefficients. 

In the following section we present an idea of a 
hybrid method that combines concepts of algebraic methods with 
equidistribution techniques.

\subsection{Equidistribution}

The idea of \emph{equidistribution methods} is to place the grid nodes 
in a way that the 
numerical error is distributed uniformly throughout the computational domain. 
In practice that means to adjust the mesh size locally according to the local 
numerical derivatives - the steeper the gradients the higher the spatial 
resolution required in order to get uniform resolution. In one dimension 
a given physical quantity gets equidistributed along an arc (see e.g. 
Thompson, Warsi, Mastin \cite{ThompsonWarsi}, 
Dorfi and Drury (1987) \cite{Dorfi1987175})
 and the grid is determined uniquely by imposing a 
proper grid equation. 

The approach presented in \cite{Dorfi1987175} is to construct a grid equation 
distributing the nodes $n_i$ proportionally to a desired resolution $R$ 
which is defined in terms of the arc-length along the graph of 
an approximated function $\mathbf{f}$,  
$R=\sqrt{1+(d\mathbf{f}/dx)^2}$
 which originates from the definition 
of the line element $ds^2 = g_{ij}dx^idx^j$. The choice of 
function $\mathbf{f}$ depends on what regime in the calculation shall be the 
focus of resolution, e.g. pressure or temperature. 
A {\sl smoothing} respectively variation limiting 
control of the grid is obtained by imposing constraints 
on grid spacing and the resolution function. 

Generalizations to more dimensions follow different strategies. 
Several attempts have been made to generalize one-dimensional 
equidistribution directly to two and more dimensions by applying 
a series of 1D equidistributions along coordinate arcs. The concept of 
arc equidistribution however can only be satisfied locally but not 
globally in the domain as pointed out by Anderson (1987) 
\cite{Anderson1987211} as well as Hunag and Sloan (1995) \cite{huang:776}. 
Moreover, successive arc equidistributions along {\sl fixed} coordinates 
fails to adapt the geometries of the cells which was one of our 
postulations. Hence such a straight approach is rather comparable 
to adaptive mesh refinement methods. 

As motivated in previous chapters, it is reasonable to perceive 
physical quantities and their numerical pendants in strong 
conservation form in a volume integrated framework 
respectively even in terms of differential forms. So it would seem natural to 
investigate methods that equidistribute a quantity with 
respect to surfaces and volumes, studied e.g. in Delzanno et al. (2008) 
\cite{Delzanno20089841} or 
Sulman, Williams and Russell (2009) \cite{2009AIPC.1168...25S}. 

One of our first attempts to obtain a multi-dimensionally adaptive grid 
that remains orthogonal during temporal evolution was to combine one dimensional
equidistribution techniques and algebraic methods.

\section{Equidistribution and Interpolation Attempt} \label{Algebra}

The intention is to generate a 2D grid that allows certain deformations from 
a polar coordinate system but remain orthogonal 
globally\footnote{For the time being we shall 
blind out that this undertaking is futile altogether 
and refer to posterior findings.}. One family of curves shall represent 
the angular coordinate and a family of normal curves 
would define the radial one. Optimally we would start with a polar 
grid and according to an equidistribution principle, e.g. the angular 
curves would deform and intrinsically define the family of orthogonal 
radial coordinates.

\subsection{Toy Model}
\begin{figure}  \centering 
\includegraphics[width=5cm]{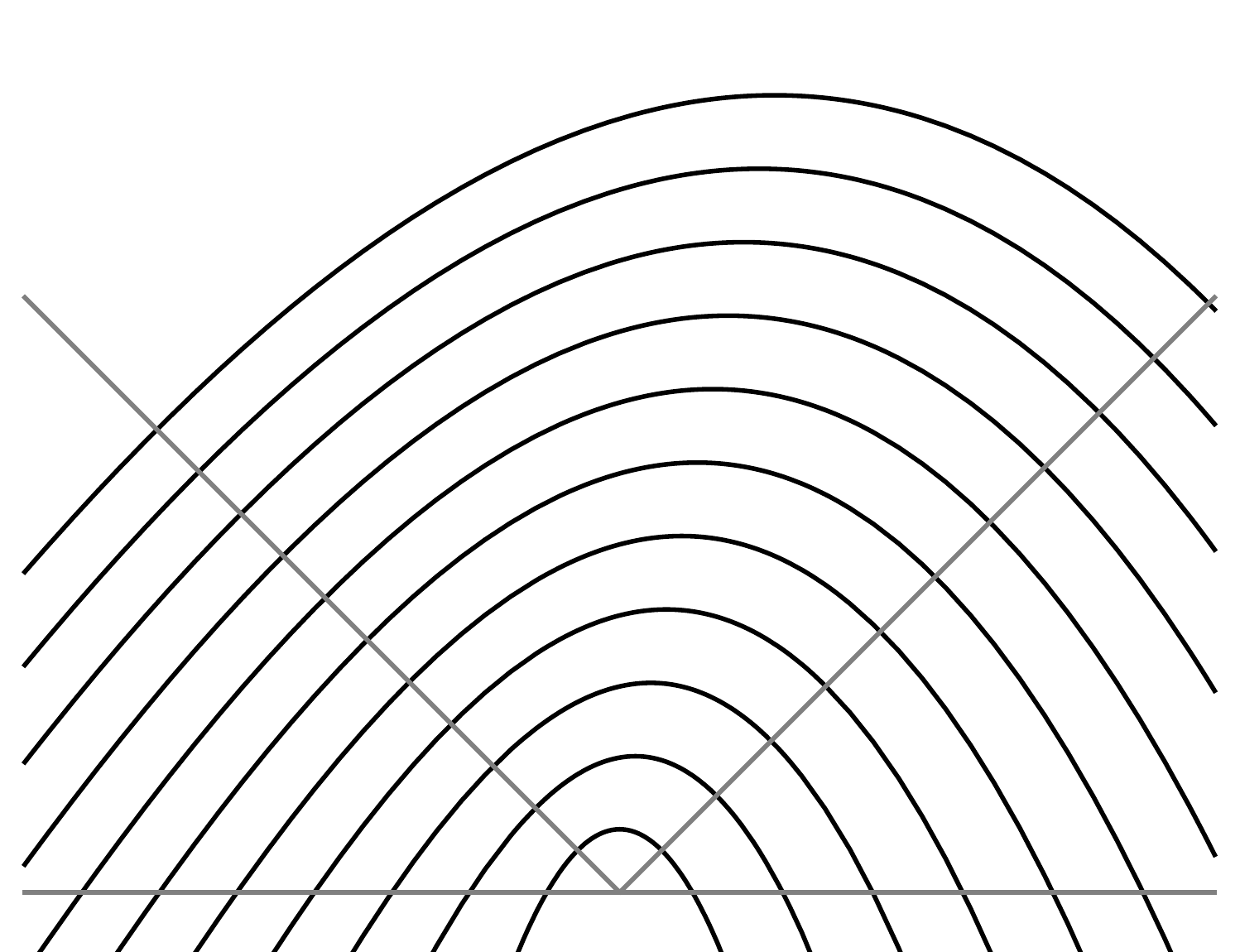} 
\caption{Angular coordinate lines of toy model}
\label{rotated_init_curves}
\end{figure}
Let us consider a transformation $\xi(x, y,t), \eta(x, y,t)$ 
where the coordinates $(\xi,\eta)$ are represented as {\sl angular} 
and a {\sl radial} family of curves. 
In favor of simplification we assume 
azimuthal symmetry and confine our analysis on a quadrant of the real plane. 
The most obvious deformation we would want to be considered is oblateness 
by rotation which means in 2D the deformation from a circle to an 
elliptical shape. As naive toy model we let 
$\eta : \mathbb{R} \times \mathbb{R} \times \mathbb{R}^+ \to \mathbb{R}$ be 
a polynomial with a family of time dependent 
coefficients\footnote{Evidently with this ansatz 
our reference frame, the polar coordinates are not 
included which we also blind out for the present. }. 
\begin{equation}
\eta(x,\tau,t) = \sum_{i<i_{\max}} {a_i}(\tau, t) x^i
\end{equation}
The coefficients $a_i$ are partially determined by boundary conditions as 
we demand normal tangents of $\eta$ at $y=0$ and 
$x=0$ in order to guarantee smooth transitions to the other 
quadrants of the real plane. 
The $\xi$ curve at a point $x_0$ is expectedly 
determined by 
\begin{equation}
\xi(x,\tau,t) = \eta(x_0, \tau,t) - \frac{1}{\eta'(x_0,\tau,t)}(x-x_0) .
\end{equation}
For principal analysis we set $i_{\max}=3$ and studied this system in 
{\sf Mathematica}. This choice determines the angular curves 
with noted boundary conditions uniquely, each further term 
would bring ambiguousness into the grid that could be used 
to include physics respectively grid control mechanisms. 
\begin{figure} 
\includegraphics[width=4.5cm]{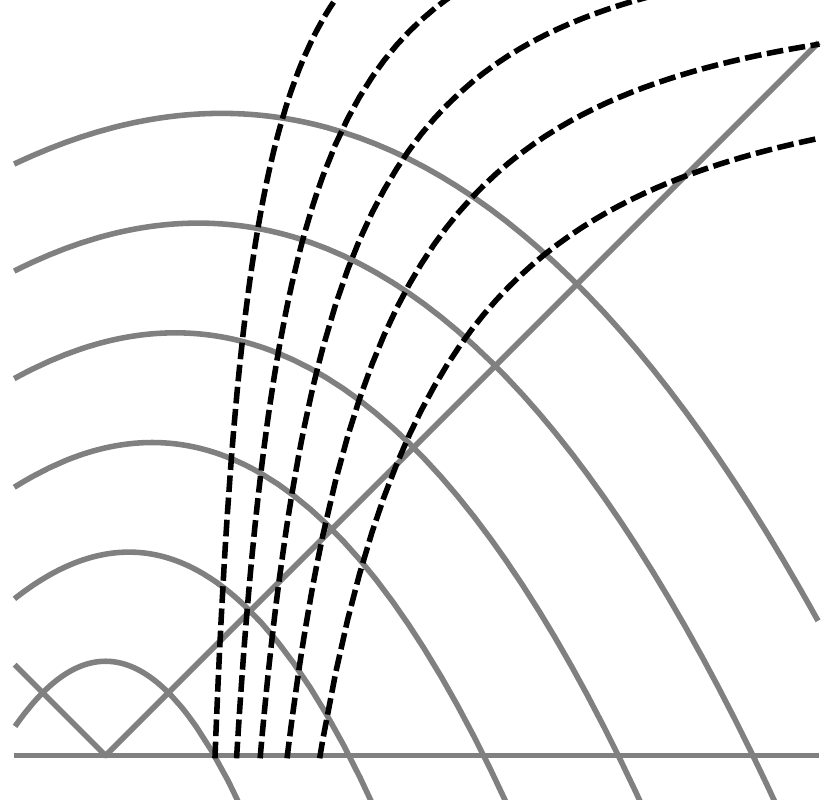} \centering
\caption{Coverage and Caustic}
\label{Caustics}
\end{figure}

\subsection{Analysis}

We define differently equidistributed points at the $x$- and
 the $y$-axis. Since the boundary condition for the $\eta$ curves at 
$y=0$ would imply infinite slope, we rotated the whole setting by
$\alpha = \pi/4$.  
What we get is a set of curves that increases in {\sl ellipticity} with the 
radius which describes the physical situation of a rotating configuration, 
drafted in figure \ref{rotated_init_curves}. 
As already mentioned both the angular coordinate lines and their normal 
curves are determined a priori. 
With determination of one orthogonal 
curve at a point $x_0$, it necessarily would have to be normal 
at all intersections if there existed such a family of curves. 
The most obvious observation we make in 
figure \ref{Caustics} is that 
the radial curves do not intersect with the natural point of origin
nor reproduce the (45 degree rotated) $x$-axis 
and cause coordinate caustics. Of course with growing order of the 
polynomials further boundary conditions can be imposed. 

However, this leads to the second quite evident 
observation that the orthogonal curves to the outermost
 polynomial are not normal 
at all other intersections with inner polynomials. If we wanted to 
impose orthogonality at $n$ intersections, the order of polynomial 
in this ansatz would at least have to match the number of radial nodes 
$i_{\max} \geq n$ which is rather pointless when we consider e.g. $500$ radii. 
\begin{figure}
\includegraphics[width=8cm]{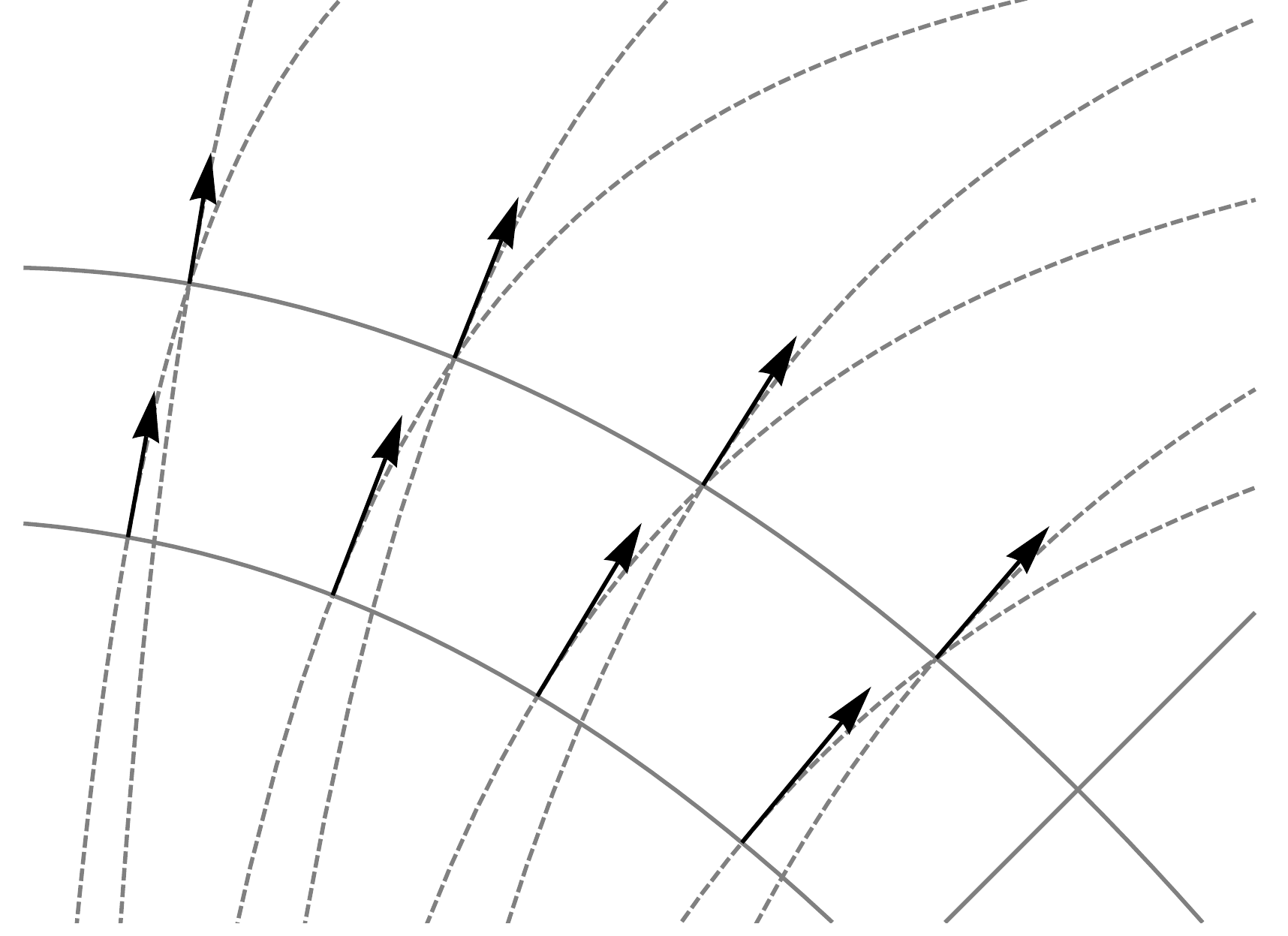} \centering
\caption{Normal vectors}
\end{figure}

This graphical analysis already discloses the limitation of grid generation 
in two dimensions considering our postulations. 
Any deviation from polar coordinates reproduces 
either nonorthogonal grids or fails to reproduce the domains boundaries; hence
postulation one will have to be renounced in its original form. 
The only kind of orthogonal adaptiveness in a polar grid is dislocation 
of nodes alongside the original coordinate lines which corresponds 
to adaptive mesh refinement methods. In the following 
section we want to substantiate our conclusions 
analytically.

\section{On the Peculiarity of Orthogonal Coordinates} \label{Peculiarity}

In section \ref{Tensor_Calculus} we studied differential geometry 
of coordinate transformations such as
the definition of base vectors (\ref{nonlinear_base_def}) and 
the metric tensor (\ref{metric_def}). We noticed that orthogonal 
coordinates imply diagonal metrics, a factor that 
can be used to study this family 
of coordinate transformations. 
Since one feature of coordinate lines 
is that all other coordinates are constant alongside the coordinate, 
transformations have to factorize and the following product ansatz is 
justified\footnote{We confine the analysis on 2D but reasoning and 
results are equivalent in 3D of course.}. 
\begin{eqnarray}
x(\xi, \eta) = a(\xi)\alpha(\eta), \quad y(\xi, \eta) = b(\xi)\beta(\eta)
\end{eqnarray}
With aforesaid definitions we obtain the base vectors 
$\mathbf{e}_{\xi} = (\alpha a',\beta b')$, $\mathbf{e}_{\eta} = 
(a\alpha' b\beta')$ and following covariant metric components. 
\begin{eqnarray}
\nonumber g_{\xi \xi} & = & \alpha^2 a'^2 + \beta^2 b'^2 \\
\nonumber g_{\eta \eta} & = & a^2 \alpha'^2 + b^2 \beta'^2 \\
g_{\xi \eta} = g_{\eta \xi} & = & a \alpha a' \alpha' + 
b \beta b' \beta'
\end{eqnarray}
In order to obtain a PDE we can study and solve analytically, the off 
diagonal entries $g_{\xi \eta}$ are set zero and functions of $\xi$ are 
separated from functions of $\eta$ which yields
\begin{equation}
\frac{aa'}{bb'} = - \frac{\beta \beta'}{\alpha \alpha'} = \lambda 
\end{equation}
where $\lambda \in \mathbb{R}$. With the substitution $A= a^2$ and $B= b^2$ we
obtain $A'/B' = \lambda$ repectively $a^2 = \lambda b^2 + c$
with $c \in \mathbb{R}$ is an arbitrary constant. 
Our boundary conditions come from the 
postulation that the coordinate system shall in principle contain the 
special case of polar coordinates hence we set the point of origin 
for the radial coordinate to 
zero $a(0)=0$, $b(0)=0$ and obtain 
\begin{equation}
a = \sqrt{\lambda} b
\end{equation}
which means that the $a$ and $b$ differ just by an arbitrary factor. 
When we replace $\alpha^2 = \bar{A}$, $\beta^2=\bar{B}$ we obtain 
$\bar{B} = -\lambda \bar{A} + k$ and determine the constant $k$ with 
the boundary conditions that the angular 
coordinate reproduces the $x$-axis at $\eta=0$ and 
the $y$-axis at $\eta=1$ to $k = \lambda$ yielding 
\begin{equation}
\beta = \sqrt{\lambda} \sqrt{1 - \alpha^2} .
\end{equation}
A general orthogonal coordinate transformation which contains the 
reference frame of polar coordinates hence necessarily has to satisfy 
\begin{equation}
x(\xi, \eta) = \sqrt{\lambda} b(\xi) \alpha(\eta), \quad 
y(\xi, \eta) = \sqrt{\lambda} b(\xi) \sqrt{1-\alpha(\eta)^2} .
\end{equation}
The scaling factor $\sqrt{\lambda}$ can evidently be set to $\sqrt{\lambda}=1$
without loss of generality
which leads to our result 
\begin{equation}
a = b, \quad \alpha^2 +\beta^2 = 1 .
\end{equation}
The postulation for orthogonality inevitably leads to radial symmetry and 
the grid can 
never gain an elliptic or even more deformed shape.

\section{Quasi-Polar Coordinates Attempt} \label{Quasi}

Evidently it is necessary to adapt our postulations 
for a multi-dimensionally adaptive grid in the light of our 
previous results. Since orthogonality can 
not be sustained we renounce that strong demand and concentrate on producing 
adaptive grids that include our reference frame. The following ansatz includes 
algebraic elements as well as some differential boundary 
conditions that control the grid properly. 

\subsection{Toy Model}

Again we consider a transformation $\xi(x, y,t), \eta(x, y,t)$ with 
radial and angular coordinates $(\xi,\eta)$ and azimuthal symmetry. 
The product ansatz we suggest is 
\begin{eqnarray} \label{polar-ell_ansatz}
\nonumber && a(\xi,t) = \sum_{i=1}^{i_{\max}} a_i(t) \xi^i, \quad b(\xi,t) = 
\sum_{i=1}^{i_{\max}} b_i(t) \xi^i \\
\nonumber  && \alpha(\eta,t) = \Big( 1+ \sum_{j=1}^{j_{\max}} 
\alpha_j(t) \eta^j \Big)\sin \eta, \quad 
\beta(\eta,t) = \Big( 1+ \sum_{j=1}^{j_{\max}} 
\beta_j(t) \eta^j \Big)\cos \eta \\
\end{eqnarray}
where $(a_i, b_i, \alpha_j, \beta_j)$ are time dependent coefficients 
that govern the adaptivity of the grid. One main feature of this approach 
is that it allows to  {\sl turn off} all deviations from radial 
symmetry by setting $(a_1, b_1)=1$ and $(a_i, b_i)=0 \, 
\forall i \geq 2$ plus $(\alpha_j, \beta_j)=0$ which yields ploar 
coordinates. Basically the unknown coefficients are determined via 
appropriate boundary conditions and whatever coupling with physics 
would be implemented (e.g. equidistribution along coordinate arcs 
or with respect to cell volumes). 

In order to discuss basic characterictics of this 
proposal, we set $i_{\max}=2$ and $j_{\max}=3$ and dissect the system in 
{\sf Mathematica}.

\subsection{Analysis}

For feasible analysis of course we need to constrain the degrees of 
freedom of our toy model but leave some undetermined parameters 
to be manipulated. With boundary conditions that 
on the one hand assure smooth transitions to other quadrants 
(normal derivatives at $x$- and $y$-axis) and on the other hand 
are consistent with the reference frame we 
effectively solve a system of algebraic 
equations in eight unknowns and gain three {\sl free} parameters. 
The {\sf Mathematica} Notebook we used for solving and plotting these 
functions can be found in Appendix \ref{App_Grid_Generation}. 

\begin{figure}
  \centering
  \subfloat[][]{\label{fig:gull}\includegraphics[width=0.25\textwidth]{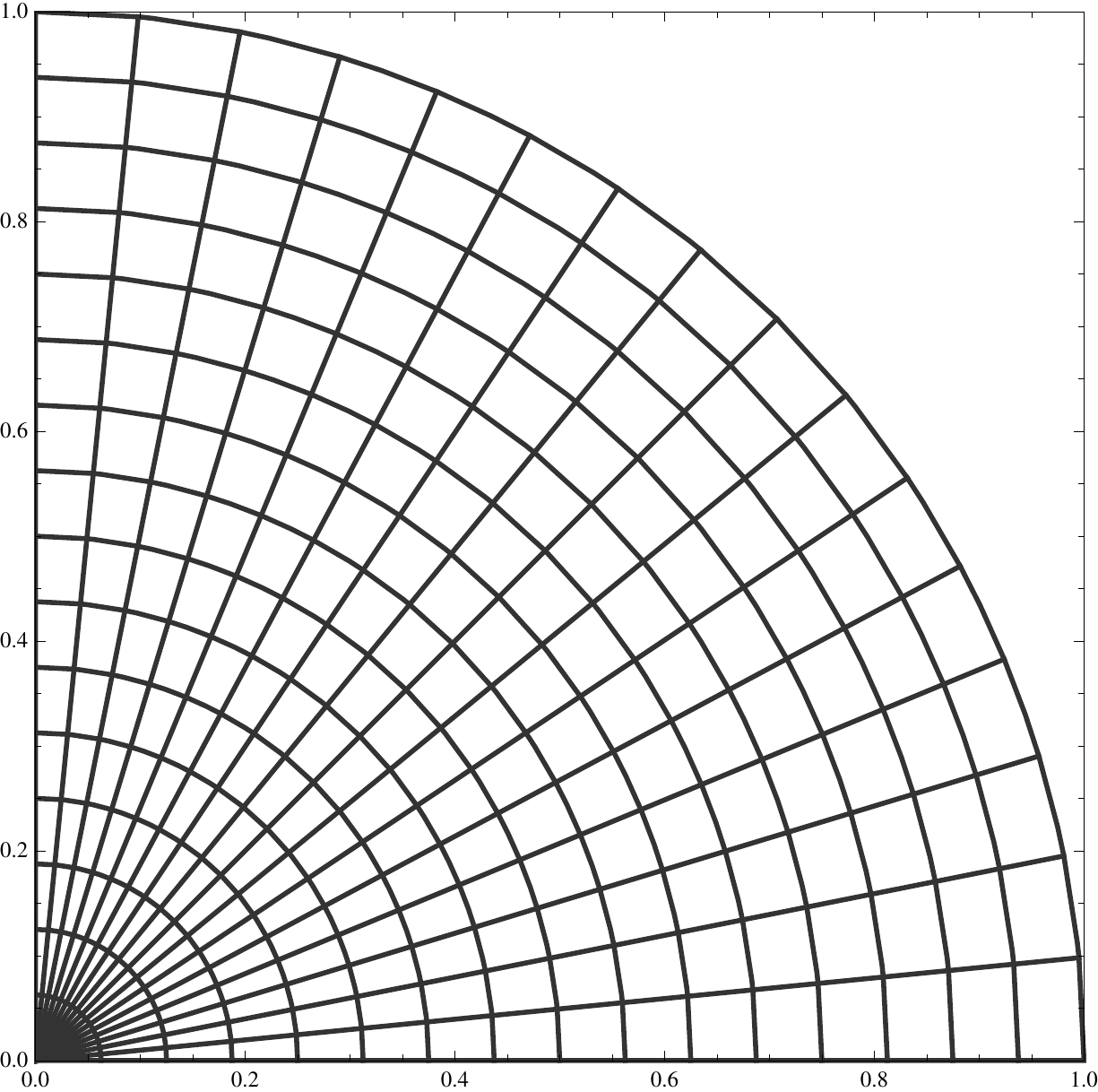}}   \qquad              
  \subfloat[][]{\label{fig:tiger}\includegraphics[width=0.25\textwidth]{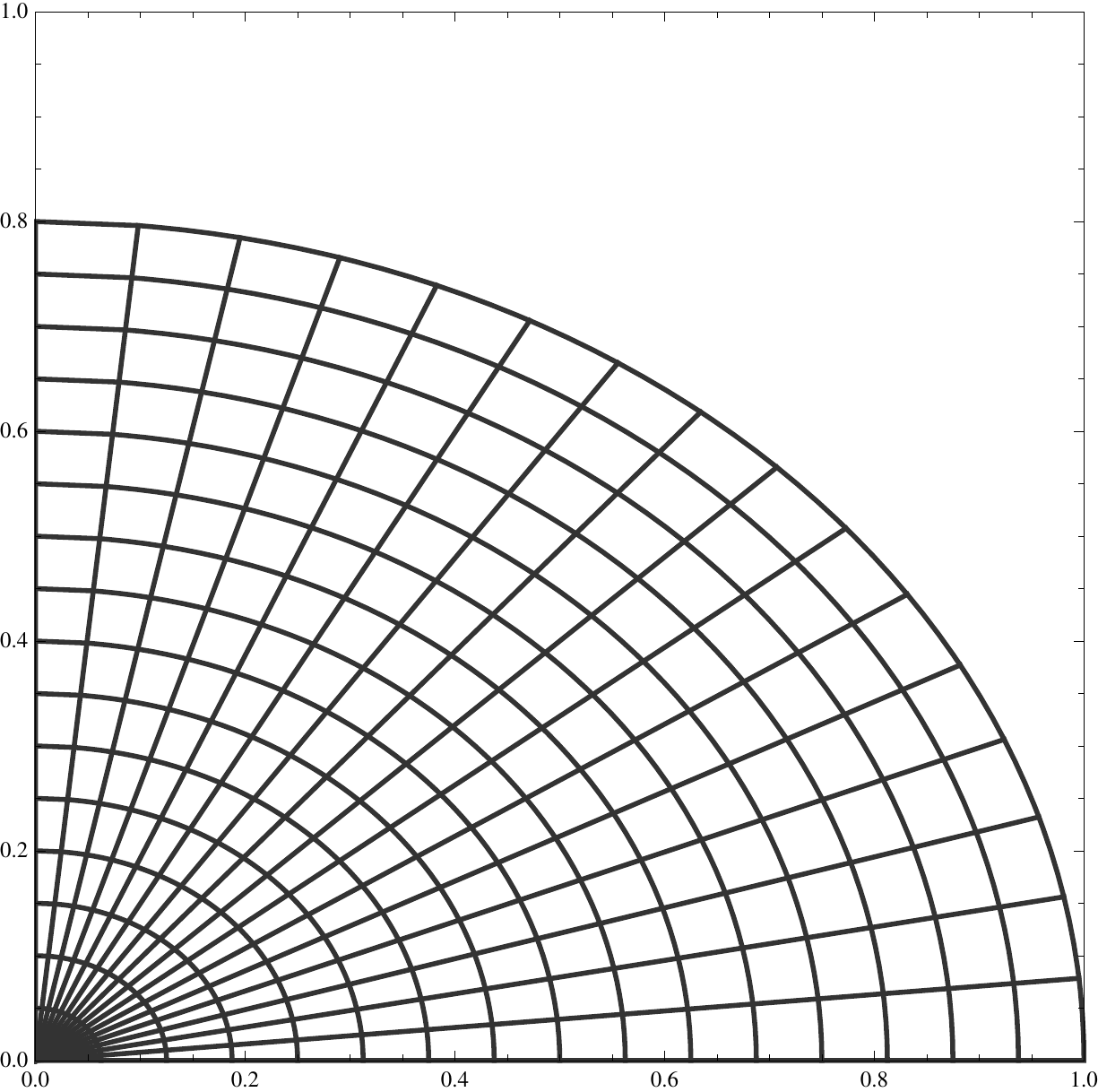}} \\
  \subfloat[][]{\label{fig:gull}\includegraphics[width=0.25\textwidth]{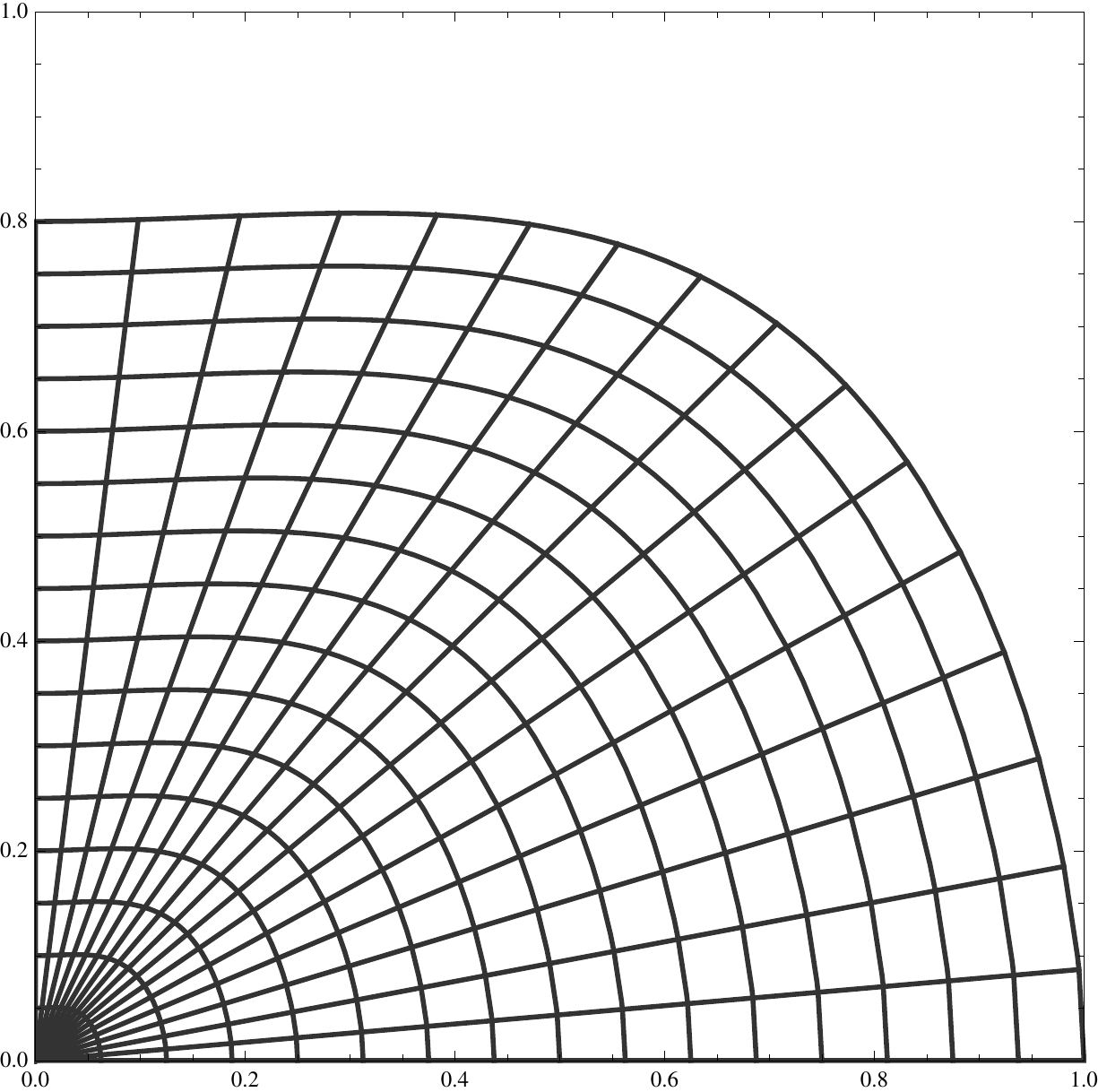}}   \qquad              
  \subfloat[][]{\label{fig:tiger}\includegraphics[width=0.25\textwidth]{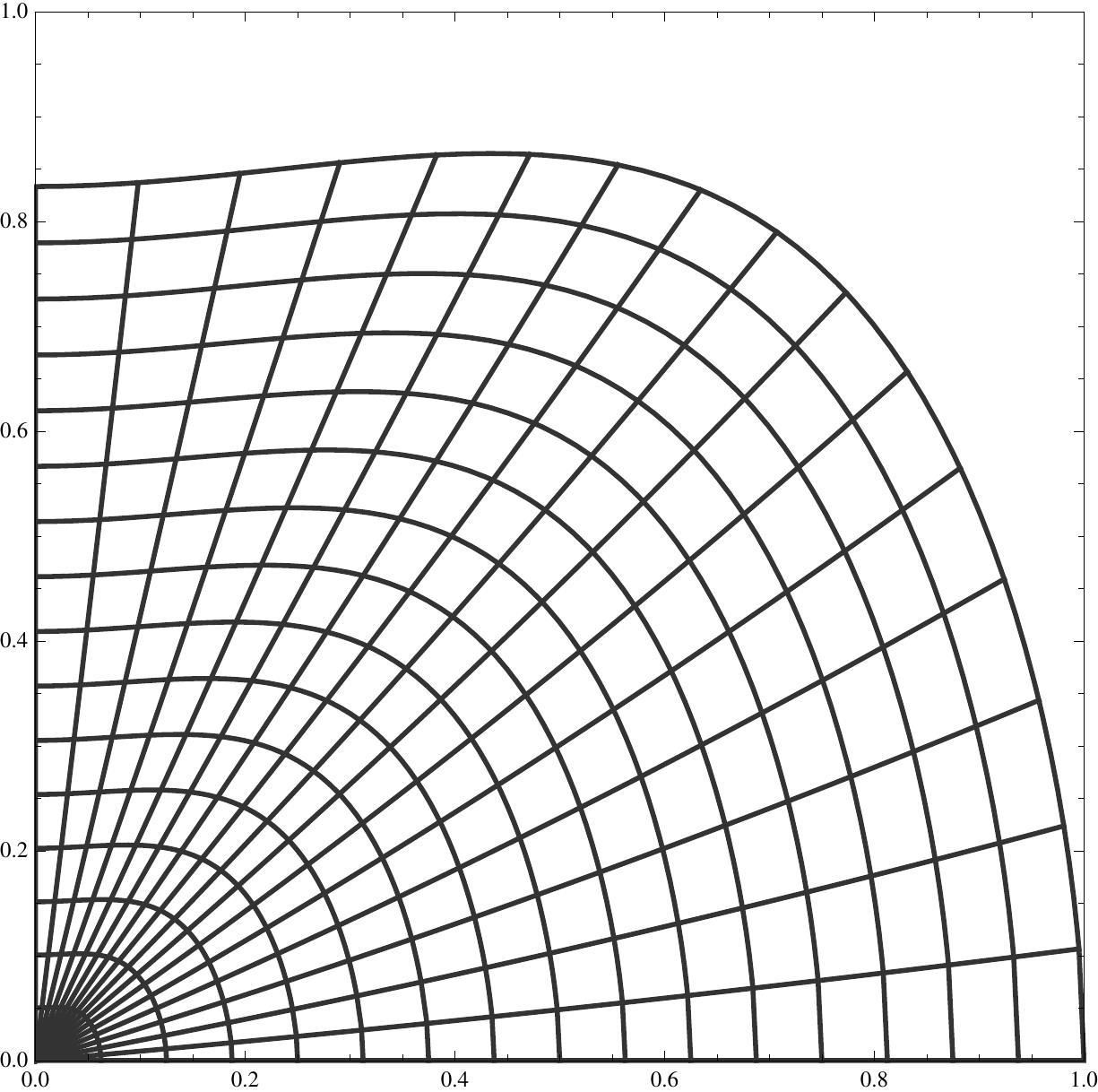}}
   \caption{Toy Model Grids}
  \label{Oblate_Grids}
\end{figure}
One example of a coordinate transformation that in principle satisfies 
all postulations from \ref{Postulations} besides orthogonality 
yields 
\begin{eqnarray}
\nonumber x & = & \xi \cos \eta \\
\nonumber 
%y & = & \frac{\xi  (b_1+b_2 \xi )}{4 \pi ^2 (1+b_2 \xi )}
% \Big(\pi ^2 \left(4+\beta_3 (\pi -2 \eta )^2 \eta \right)+ \\
% && + \, b_2 (\pi -2 \eta )^2 \left(4+\beta_3 \pi ^2 \eta \right) 
%\xi \Big) \sin\eta 
\nonumber y & = & \big(b_1 \xi +b_2 \xi ^2\big) \Big(1+
\frac{ \left(\beta_3 \pi ^3-16 b_2 \xi +b_2 \beta_3 \pi ^3 \xi \right) }{4 \pi  (1+b_2 \xi )}\eta + \\
&& + \, 
\frac{\left(-\beta_3 \pi ^3+4 b_2 \xi -b_2 \beta_3 \pi ^3 \xi \right)}{\pi ^2 (1+b_2 \xi )}\eta ^2
+\beta_3 \eta ^3\Big) \sin \eta
\end{eqnarray} 
where $b_1, b_2$ and $\beta_3 \in \mathbb{R}$ are free parameters. 
Certainly they are not totally free but need to be confined to 
certain bounds that ensure unique covering of the domain. 
Further constraints emerge from differential geometric requirements 
to the grid like a maximal skewness of cells. 

In figure \ref{Oblate_Grids} we visualize grids with varying 
parameters that are listed in table \ref{Table_Parameters} 
along with some characteristics of the grid. The off-diagonal 
metric elements $g_{12}=g_{21} \neq 0$ quantify the skewness of the grid and 
$\sqrt{\mathbf{g}}_{\text{rel}}$ designates the local cell volume 
in relation to polar coordinates. 
Both characteristics were numerically maximized in {\sf Mathematica} in the 
ranges $\xi \in (0,1]$ and $\eta \in [0,\pi/2]$
since the functional expressions get rather bulky. For 
the simplest quasi-polar model (b) however the coordinate transformation 
is $x=\xi \cos \eta, y= 0.8 \, \xi \sin \eta$ and the 
grid characteristics yield explicitly 
$g_{12,\text{(b)}}  =  -0.36 \, \xi  \cos\eta \sin\eta$ and 
$\sqrt{\mathbf{g}}_{\text{rel,(b)}} =  0.64$. 
Due to our boundary conditions the minima are mostly to be found 
at the origin respectively at the axes, for details see 
section \ref{Mathematica_Files}. 
\begin{table} \centering
\caption{Maximaa of geometrical parameters for toy model}
\vspace{3pt}
\begin{tabular}{c|c|c|c|c|c|c|c|c|c} 
\label{Table_Parameters} 
\, & $b_1$ & $b_2$ & $\beta_3$ & $\max |g_{12}(\xi,\eta)/\xi|$ 
& $\max \sqrt{\mathbf{g}}_{\text{rel}}(\xi,\eta)$  \\[2pt]
\hline 
(a) & $1$ & $0$ & $0$ & $0$ & 
$1$ \\
(b) & $0.8$ & $0$ & $0$& $0.18$ \scriptsize{at $(0.85,0.73)$} & $0.64$ \scriptsize{at $(0.9,0.71)$} \\
(c) & $0.8$ & $0$ & $0.5$ & $0.13$ \scriptsize{at $(0.89,0.79)$} & $1.19$ \scriptsize{at $(0.86,0.33)$}  \\
(d)& $0.8$ & $0.2$ & $0.5$ & $0.25$ \scriptsize{at $(1,0.43)$} & $1.62$
 \scriptsize{at $(1,0.29)$}\\
\end{tabular}
\end{table}

\section{Strong Conservation Form in Nonorthogonal Coordinates}

In section \ref{Covariant_RHD_Spher} the equations of RHD were explicitly 
formulated in polar and spherical coordinates and we noticed that even with 
orthogonal coordinates the expressions get rather cumbersome. 
From the view point of tensor calculus, the number of 
operations like raising and lowering of indices with non-diagonal metrics 
increases with the square of the spatial 
dimension considered. Strong conservation form radiation hydrodynamics in 
general curvilinear coordinates hence can be expected to arouse 
extensive equations. 
 
In fact, if we take the full ansatz
(\ref{polar-ell_ansatz}) into our {\sf Mathematica} files that generate 
the equations of radiation hydrodynamics in strong conservation form 
the computing time for just generating the equations gets sizable
and the outputs become huge.  
The question arises at what point of such 
calculations the geometric relations and terms are to be
determined respectively computed ideally. 
A profound answer will have to be given in future investigations.  
However, without a doubt the actual solution of the RHD system which 
is a matrix inversion problem in implicit numerics as 
described e.g. in Dorfi et al. (2004) \cite{Dorfi2006771} will remain the 
most time consuming part even or especially in multiple dimensions.

% ---------------------------------------------------------------------------- %
% CONCLUSIONS ---------------------------------------------------------------- %
% ---------------------------------------------------------------------------- %

\chapter{Conclusions and Prospective}

We present versatile conceptual and methodical
fundamentals for strong conservation 
numerics in general non-steady curvilinear coordinates in 2D and 3D. 
The first chapters \ref{Fund_Math}, \ref{Fund_Phys} and 
\ref{Conservative Numerics}
emphasize on mathematical rigorousness and consistency 
which led to the important finding (\ref{art_visc_corr}) where we 
reformulated the artificial viscosity for curvilinear coordinates.   
The relevance of this finding does not come clear until general 
curvilinear coordinates are used that is why previous calculations 
were not necessarily affected by this inconsistency. 

As {\sl lateral} result we suggest a dynamic equation for the 
self-gravitation in multiple dimensions by considering the 
specific gravitational force and the continuity equation
(\ref{dyn_grav}). We avoid pursuing the Laplace equation in 
2D and 3D where its unique nature as elliptical PDE in our 
otherwise hyperbolic problem evokes 
difficulties in posing adequate boundary conditions. Our non-static 
gravitation equation will have to be studied in future calculations 
whereat its imaginable applications range from stellar physics to 
galactic and cosmological computations where self-gravitation 
plays a role. 

We derive the equations of RHD in strong conservation form 
an exemplarily sketch them in polar and spherical 
coordinates in chapter \ref{Covariant_RHD_Spher}. 
They might be understood as intermediate step to more applicable implementations
in general curvilinear coordinates and play a role as reference frame a
after all. 

In chapter \ref{Grid_Generation} we study adaptive grid generation for 
curvilinear coordinates and find that orthogonality of coordinate 
lines can not be maintained for multi-dimensionally adaptive grids that 
contain the reference frames of polar and spherical coordinates. 
In this sense the polar and polar spherical take an 
exceptional position as orthogonal grids in 2D and 3D. 
However, we propose an interesting ansatz that satisfies 
our main postulations for an astrophysically applicable grid and study 
possible adoptions. 
For practical implementation one has to develop means of controling the 
skewness of the grid respectively the off-diagonal elements of the 
metric tensor adequately. 
A stringent design of such a technique would have 
exceeded the extent of this thesis and will have to be deferred to 
future investigations. Moreover one will have to contemplate feasible 
methods of generating and handling the enormous equations that emerge 
with non-orthogonal coordinates considered. Clearly, one will have to 
pursue automatized techniques like those presented with 
{\sf Mathematica}. Concrete implementations shall also indicate at what point 
in the computations the geometric terms are to be calculated ideally.

% ---------------------------------------------------------------------------- %
% APPENDIX ------------------------------------------------------------------- %
% ---------------------------------------------------------------------------- %

\chapter{Appendix}

\section{Arguments and Details}

\subsection{Symmetry of the Stress Tensor} \label{cons_ang_mom}

In section \ref{physics_of_hd} we argue that the symmetry of the stress tensor 
is due to the conservation of angular momentum. 
If so, the rate of change of the angular momentum of a fluid 
$\rho \mathbf{x} \times\mathbf{v}$ must equal the total applied torque. 
With $d\mathbf{S}$ oriented with normal $\mathbf{n}$, let $\mathbf{t}$ be the 
surface force across this element and the $i$-th component be defined by 
$t^i = T^{ij} n_j$. 
\begin{eqnarray}
\nonumber 
\int\limits_{V(t)} \Big( \mathbf{x} \times \partial_t \big( \rho  \mathbf{u} 
\big) \Big) \, dV + 
\int\limits_{\partial V(t)} \Big( \mathbf{x} \times \mathbf{t} \Big) \cdot
d\mathbf{S} & = & 0 \\
\nonumber \int\limits_{V(t)} \partial_t \big( \rho \varepsilon_{ijk} x^j u^k 
\big)  \, dV + \int\limits_{\partial V(t)} \varepsilon_{ijk} x^j T^{lk} n_l 
\, dS & = & 0 \\
\nonumber \int\limits_{V(t)} \Big( \partial_t \big( \rho \varepsilon_{ijk} 
x^j u^k \big) +
\nabla_l \varepsilon_{ijk} x^j T^{lk} \Big) \, dV & = & 0 
\end{eqnarray}
Since the volume of our conservation law is arbitrary, we can focus on the 
integrand.  
\begin{eqnarray}
\nonumber 
\underbrace{\partial_t \big( \rho \varepsilon_{ijk} 
x^j u^k \big) + \varepsilon_{ijk} x^j 
{T^{lk}}_{,l}}_{= \varepsilon_{ijk}x^j(\partial_t \rho u^k + {T^{lk}}_{,l}) = 0} + 
\varepsilon_{ijk} {x^j}_{,l} T^{lk}  & = & 0 \\
\nonumber \varepsilon_{ijk} {\delta^j}_l T^{lk}  & = & 0 \\
\label{symm_stress_tensor} \varepsilon_{ijk} T^{jk} & = & 0 
\end{eqnarray}
The left hand side of equation (\ref{symm_stress_tensor}) is zero, 
if $T^{jk}=T^{kj} \quad \forall \, j,k$. 

% ---------------------------------------------------------------------------- %
% ---------------------------------------------------------------------------- %

\subsection{Rankine-Hugoniot Conditions} \label{app_rankine_hugoniot}

We consider a one-dimensional Riemann-problem for 
\ref{cauchy_problem_integral} with discontinuous initial condition 
\begin{equation} 
d_0(x) = \left\{ \begin{array}{cl}
d_l, & x<0 \\
d_r, & x>0
\end{array} \right. 
\end{equation}
and look for a solution $d(x,t)$ for the left and the right hand side of 
the shock
\begin{equation} 
d(x,t) = \left\{\begin{array}{cl}
d_l, & x<u_s t \\
d_r, & x>u_s t
\end{array} \right. \end{equation}
with positive velocity $u_s$. To do so, we split the integrals in 
(\ref{cauchy_weak_formulation}) cleverly at the 
discontinuity and get for the first term 
\begin{eqnarray}
\nonumber \int\limits_0^{\infty}\int\limits_{-\infty}^{\infty} d \, \partial_t
\gamma \, dx dt & = & 
\int\limits_{-\infty}^0\int\limits_{0}^{\infty}  d \, \partial_t 
\gamma \, dx dt +
\int\limits_{0}^{\infty}\int\limits_{0}^{x/u_s} d \, \partial_t 
\gamma \, dx dt + \\
\nonumber && + \int\limits_{0}^{\infty}\int\limits_{u_s/x}^{\infty} 
d \, \partial_t \gamma \, dx dt \\
\nonumber & = & -d_l \int\limits_{-\infty}^0 \gamma(x,0) \, dx - d_r 
\int\limits_{0}^{\infty}\gamma(x,0) \, dx + \\
\nonumber && + \int\limits_0^{\infty}\gamma(x,x/u_s) \Big( d_r-d_l \Big) 
\, dx \\
 & = & -\int\limits_{-\infty}^{\infty}\gamma(x,0) d_0(x) \, dx 
-\int\limits_0^{\infty}\gamma(x,x/u_s) \Big( d_l-d_r \Big) \, dx .
\end{eqnarray}
For the flux term on the right hand side we proceed analogously and 
rewrite the integral over $dx$ with $t=x/u_s$. 
\begin{eqnarray} 
\nonumber \int\limits_0^{\infty}\int\limits_{-\infty}^{\infty} f(d) \,
\partial_x \gamma \, dx dt & = & 
\int\limits_{0}^{\infty}\left( \int\limits_{-\infty}^{u_st}
\gamma \, \partial_x f(d) \, dx + 
\int\limits_{u_st}^{\infty} \partial_x f(d) \, dx \right) dt \\
\nonumber & = & \int\limits_{0}^{\infty} \gamma(u_st,t)
\, \Big( f(d_l)-f(d_r) \Big) \, dt \\
& = & \frac{1}{u_s} \int\limits_{0}^{\infty} \gamma(x,x/u_s)
\, \Big( f(d_l)-f(d_r) \Big) \, dx
\end{eqnarray}

\begin{eqnarray}
\nonumber \int\limits_0^{\infty}\int\limits_{-\infty}^{\infty} 
\left( d \, \partial_t \gamma + f(d) \, \partial_x
\gamma \right) \, dx dt  =  -\int\limits_{-\infty}^{\infty} 
\gamma(x,0)\, d_0(x) \, dx +\\
\label{rankine_1} + \int\limits_0^{\infty} \gamma(x,x/u_s) 
\, \left( \frac{f(d_l)-f(d_r)}{u_s}-\Big( d_l-d_r \Big) \right) \, dx 
\end{eqnarray}
(\ref{rankine_1}) is weak solution of the Riemann-problem, if 
\begin{equation} \frac{f(d_l)-f(d_r)}{u_s}-\Big(d_l-d_r \Big)=0 \end{equation}
or  
\begin{equation} \label{ranking_hugoniot_conditions_m} 
u_s = \frac{f(d_l)-f(d_r)}{d_l-d_r} .
\end{equation}
We call $u_s$ shock velocity and (\ref{ranking_hugoniot_conditions_m})
Rankine-Hugoniot condition. 

% ---------------------------------------------------------------------------- %
% ---------------------------------------------------------------------------- %

\subsection{Diffusion Approximation} \label{diffusion_approx_appendix}

If the radiation field does not depend on the direction i.e. it is isotropic 
$I_{\gamma,\text{diff}} = J_{\gamma,\text{diff}} 
\neq J_{\gamma,\text{diff}}(\mathbf{n})$, 
the integration over the solid angle can be executed straightforward. 
Let be ${\mathbf{e}}_r $ the normal vector $\mathbf{n}$ in 
local spherical coordinates. 
\begin{eqnarray}
\nonumber 
\mathbf{F}_{\gamma, \text{diff}} & = & 
\int\limits_0^{2 \pi}\int\limits_0^{\pi} I_{\gamma} {\mathbf{e}}_r 
\sin\vartheta \, d\vartheta d\varphi = I_{\gamma} 
\int\limits_0^{2 \pi}\int\limits_0^{\pi} \sin\vartheta
\left( \begin{array}{c} \cos\varphi \sin\vartheta \\ \sin\varphi \sin\vartheta
\\ \cos\vartheta \end{array} \right) \, d\vartheta d\varphi = \mathbf{0} \\
\nonumber 
\mathbf{P}_{\gamma ,\text{diff}} & = & 
\frac{1}{c}\int\limits_0^{2 \pi}\int\limits_0^{\pi} I_{\gamma}
{{\mathbf{e}}_r}^{\,\!} {{\mathbf{e}}_r} 
\sin\vartheta \, d\vartheta d\varphi =  \frac{I_{\gamma}}{c} 
\int\limits_0^{2 \pi}\int\limits_0^{\pi}  \sin\vartheta \, d\vartheta 
d\varphi \cdot \\
\nonumber &  & \cdot 
\left( \begin{array}{ccc}
\cos\varphi^2 \sin\vartheta^2 &\cos\varphi \sin\varphi \sin\vartheta^2 
&\cos\varphi\cos\vartheta \sin\vartheta \\
\cos\varphi \sin\varphi \sin\vartheta^2 & \sin\varphi^2 \sin\vartheta^2 
&\cos\vartheta \sin\varphi \sin\vartheta \\
\cos\varphi\cos\vartheta \sin\vartheta &\cos\vartheta \sin\varphi 
\sin\vartheta &\cos\vartheta^2
\end{array} \right) = \\
& = & \frac{I_{\gamma}}{c} \left( \begin{array}{ccc}
\frac{4\pi}{3} & 0 & 0 \\
0 & \frac{4\pi}{3} & 0 \\
0 & 0 & \frac{4\pi}{3} 
\end{array} \right) = \frac{1}{3} \mathbf{1} J_{\gamma}
\end{eqnarray}

% ---------------------------------------------------------------------------- %
% ---------------------------------------------------------------------------- %

%\subsection{Minkowski Geometry and Lorentz Transfromation} 
%\label{minko_lorentz_app}
%
%In order to clarify terms and quantities that occur in section 
%\ref{Fluid_Frame} when we perform the transformation of radiative 
%quantities to the fluid frame, a very brief compendium to Minkowski 
%geometry and Lorentz transformation is presented (we refer 
%to Ray d'Invernos textbook \cite{RayD} for details). 
%
%
%...in terms of an appropriate Lorentz metric $\eta_{\alpha \beta}$ \footnote{In 
%problem oriented spherical 
%coordinates $(r,\varphi,\vartheta)$ the corresponding 
%line element yields 
%$ ds^2 = \eta_{\alpha \beta} \, dx^{\alpha} dx^{\beta} = 
%-c^2 \, dt^2 + dr^2 + r^2(d\vartheta^2 + \sin^2 \vartheta \, 
%d\varphi^2)$. }. 
%Here the metric is the mathematical description of the spacetime and thus 
%has not solely geometrical but physical relevance. Without forceful 
%deduction we 

% ---------------------------------------------------------------------------- %
% ---------------------------------------------------------------------------- %

\subsection{Strong Conservation Form on Adaptive Grids} \label{app_adaprive}

In section \ref{Adaptive Grids} we outlined the strong conservation 
form for time dependent coordinates. Following  
Thompson, Warsi, Mastin \cite{ThompsonWarsi} we present the full 
deduction of this conclusion. 

The time derivative of a physical quantity $\mathbf{d}$ in the 
non-steady coordinate system $\Sigma_{(\beta)}$ relative to a 
coordinate system $\Sigma_{(\alpha)}$ is given by $
{\partial_t \mathbf{d}}_{(\beta)} = {\partial _t\mathbf{d}}_{(\alpha)} + 
\mathbf{\nabla}_{(\alpha)}\mathbf{d} {\partial_t \mathbf{x}}_{(\beta)}
$ which is basically the Reynolds transport theorem. 
In system $(\alpha)$ the time derivative of $\mathbf{d}$ yields 
${\partial_t \mathbf{d}}_{(\alpha)} = {\partial _t\mathbf{d}}_{(\beta)} - 
\dot{\mathbf{x}} \mathbf{\nabla}_{(\alpha)}\mathbf{d}$. 
With $\dot{\mathbf{x}}$ we introduced the grid velocity,  
the second term on the right hand side we also call grid advection. 
An advection term in a hyperbolic conservation law as we 
consider them usually gains the form 
\begin{equation}
K = \mathbf{d}_t + \mathrm{div} \, (\mathbf{u} \mathbf{d})
\end{equation}
respectively in the transformed system 
\begin{equation}
K = \mathbf{d}_t - \dot{\mathbf{x}} \cdot \mathbf{\nabla}\mathbf{d} +
 \mathrm{div} \, (\mathbf{u}\mathbf{d})  . 
\end{equation}
Applying the strong conservation form of our spatial differential operators 
we developed in section \ref{Strong_Cons_Form}, this convection term yields 
\begin{equation} \label{conv_cons}
K = \mathbf{d}_t - \dot{\mathbf{x}} \cdot \frac{1}{\sqrt{|\mathbf{g}|}}\partial_{i}
\Big[
\sqrt{|\mathbf{g}|} \mathbf{{e}}^{i} \mathbf{d} \Big] + 
\frac{1}{\sqrt{|\mathbf{g}|}}\partial_{i} \Big[
\sqrt{|\mathbf{g}|} \mathbf{u} \cdot \mathbf{{e}}^{i} \mathbf{d} \Big] .
\end{equation}

Since we deal with non-steady coordinates both the base vectors 
and the metric are functions of time. We resort to an identity 
from tensor analysis for the metric tensor to consider 
the temporal derivative of $\sqrt{|\mathbf{g}|}$
\begin{eqnarray}
\nonumber \partial_t \sqrt{|\mathbf{g}|} & = & \partial_t
\Big[\mathbf{{e}}_{1}
\cdot (\mathbf{{e}}_{2} \times \mathbf{{e}}_{3}) \Big] \\
\nonumber & = & \partial_t {\mathbf{{e}}_{1}} \cdot 
(\mathbf{{e}}_{2} \times \mathbf{{e}}_{3}) + \partial_t 
{\mathbf{{e}}_{2}} \cdot 
(\mathbf{{e}}_{3} \times \mathbf{{e}}_{1}) + \dots \\
& = &  \sqrt{|\mathbf{g}|} \partial_t \, \mathbf{{e}}_{i} \cdot
\mathbf{{e}}^{i}
\end{eqnarray}
using $\mathbf{{e}}^{i} = {|\mathbf{g}|}^{-1/2} 
(\mathbf{{e}}_{j} \times \mathbf{{e}}_{k})$. 
Moreover, we remember the definition of the covariant base vector  
\begin{equation} \label{def_cov_basis}
\mathbf{{e}}_{(\alpha) i} = \frac{\partial \mathbf{x}}
{\partial x_{(\alpha)}^{i}}
\end{equation}
which leads to an interesting conclusion for non-steady coordinates 
respectively adaptive grids. 
\begin{equation}
\partial_t \sqrt{|\mathbf{g}|} =  \sqrt{|\mathbf{g}|}  
 \partial_{i} \dot{\mathbf{x}} \cdot \mathbf{{e}}^{i} 
\end{equation} 
The variation of the volume element in time hence is connected with 
the derivatives of the grid velocity in all directions. 
Again we mention that even for solely radially adaptive grids the 
term $\dot{\mathbf{x}}$ contains angular contributions, to wit 
$\dot{\mathbf{x}} \neq \dot{r}$. 

We consider following expression in order to asses the temporal 
change of a {\sl geometrically weighted} in comparison to a 
plain density function $\mathbf{d}$ 
\begin{eqnarray}
\nonumber  \frac{1}{\sqrt{|\mathbf{g}|}} \partial_t \Big[\sqrt{|\mathbf{g}|} \mathbf{d}
\Big] & =
&  \frac{1}{\sqrt{|\mathbf{g}|}} \Big[  \partial_t \sqrt{|\mathbf{g}|} \mathbf{d} +
\sqrt{|\mathbf{g}|} \partial_t
\mathbf{d} \Big] \\
\nonumber & = & \frac{1}{\sqrt{|\mathbf{g}|}} \Big[ \sqrt{|\mathbf{g}|}  
 \partial_{i} \dot{\mathbf{x}} \cdot \mathbf{{e}}^{i} \mathbf{d} +
\sqrt{|\mathbf{g}|} 
\partial_t \mathbf{d} \Big] 
\end{eqnarray}
and obtain 
\begin{equation}
\partial_t \mathbf{d} = \frac{1}{\sqrt{|\mathbf{g}|}} \partial_t
\Big[\sqrt{|\mathbf{g}|} \mathbf{d} \Big] -   \partial_{i} \dot{\mathbf{x}} \cdot
\mathbf{{e}}^{i} \mathbf{d} .
\end{equation}
We compare that result with our convection term  (\ref{conv_cons}) 
to gain the strong conservation form of the 
advection term in non-steady coordinates.  
\begin{eqnarray}
\nonumber K & = & \frac{1}{\sqrt{|\mathbf{g}|}} \partial_t
\Big[\sqrt{|\mathbf{g}|} \mathbf{d} \Big] - \partial_{i}
\dot{\mathbf{x}} \cdot
\mathbf{{e}}^{i} \mathbf{d} -  \dot{\mathbf{x}} \cdot
\frac{1}{\sqrt{|\mathbf{g}|}}\partial_{i} \Big[
\sqrt{|\mathbf{g}|} \mathbf{{e}}^{i} \mathbf{d} \Big]  + 
\frac{1}{\sqrt{|\mathbf{g}|}}\partial_{i} \Big[
\sqrt{|\mathbf{g}|} \mathbf{u} \cdot \mathbf{{e}}^{i} \mathbf{d} \Big] \\
\nonumber & = & \frac{1}{\sqrt{|\mathbf{g}|}} \partial_t
\Big[\sqrt{|\mathbf{g}|} \mathbf{d} \Big] - \frac{1}{\sqrt{|\mathbf{g}|}}
\partial_{i}
\Big[\sqrt{|\mathbf{g}|} \dot{\mathbf{x}} \cdot \mathbf{{{e}}}^{i} \mathbf{d}
\Big]  + 
\frac{1}{\sqrt{|\mathbf{g}|}}\partial_{i} \Big[
\sqrt{|\mathbf{g}|} \mathbf{u} \cdot \mathbf{{e}}^{i} \mathbf{d} \Big] \\
\label{flux_cons} \sqrt{|\mathbf{g}|} K & = & \partial_t
\Big[\sqrt{|\mathbf{g}|} \mathbf{d} \Big] + \partial_{i}
\Big[\sqrt{|\mathbf{g}|} \mathbf{{{e}}}^{i} \cdot (\mathbf{u} -
\dot{\mathbf{x}}) \mathbf{d} \Big]
\end{eqnarray}
If we define the contravariant velocity components relative to the moving 
grid by 
\[
U^{i} \equiv \mathbf{{{e}}}^{i} \cdot (\mathbf{u} - \dot{\mathbf{x}})
\] 
the upper equations yields
\begin{equation}
\sqrt{|\mathbf{g}|} K = \partial_t
\Big[ \sqrt{|\mathbf{g}|} \mathbf{d} \Big] + \partial_{i}
\Big[ \sqrt{|\mathbf{g}|} U^{i} \mathbf{d} \Big] .
\end{equation}

% ---------------------------------------------------------------------------- %
% ---------------------------------------------------------------------------- %
\newpage 
\section{Mathematica Files} \label{Mathematica_Files}
\label{App_Grid_Generation}

On the following pages some examples of {\sf Mathematica Notebooks} are 
presented that were used 
for tensor analytical computations in several coordinates 
and generation of strong conservation form equations. 
The analysis of the grid generation toy models presented in sections 
\ref{Algebra} and \ref{Quasi} conclude the Appendix. 

Technical remark: in order to ensure that {\sl integral} 
respectively conserved quantities are not divided automatically by the 
computer algebra system, we make use of Patterns. In this sense the partial
derivatives act on braced expressions like 
$\partial_i \left[ \dots \right]$ that  would be processed for 
the discretization. 

\begin{description}
\item[Spherical Coordinates] ({\tt spherical\_coordinates.nb}) 
Differential geometric definitions and 
relations for base vectors, metric components and Christoffel symbols, 
exemplarily executed with spherical coordinates. 
If the initial definition of the coordinate transformation gets modified 
to a nonorthogonal coordinate system, minor changes have to be made. 
\item[Artificial Viscosity in Spherical Coordinates] 
({\tt artificial\_viscosity.nb}) 
Using results from {\tt spherical\_coordinates.nb} we compute 
co- and contravariant components of the viscous pressure tensor
in spherical coordinates. 
Symmetry of $\mathbf{Q}$ and vanishing trace are evaluated explicitly. 
\item[Equation of Motion in Spherical Coordiantes] 
({\tt EOM\_3D\_sph.nb}) Using the previous definitions for geometric 
quantities and the viscous pressure we generate the equation of motion 
in spherical coordinates in strong conservation form for 
non-steady grids. 
\item[Polynomial Coordinates Ansatz] ({\tt poly\_coordinates.nb})
Graphical analysis of a polynomial ansatz for orthogonal family of curves.  
\item[Polar Non-Orthogonal Grid Ansatz]  ({\tt quasi\_polar\_grid.nb}) 
Generation of a non-orthogonal grid that 
in principle satisfies 
all postulations from \ref{Postulations} besides orthogonality. With the 
coordinate transformation determined one could go run through upper 
files and generate strong conservation form equations on non-steady 
non-orthogonal grids. 
\end{description}



```
In[13]:= E_theta = 1 / rootg * Cross[e_phi, e_r] // FullSimplify
```

$$\text{Out[13]= } \left\{ \frac{\cos[phi]\cos[theta]}{r}, \frac{\cos[theta]\sin[phi]}{r}, -\frac{\sin[theta]}{r} \right\}$$

```
In[14]:= E_phi = 1 / rootg * Cross[e_r, e_theta] // FullSimplify
```

$$\text{Out[14]= } \left\{ -\frac{\csc[theta]\sin[phi]}{r}, \frac{\cos[phi]\csc[theta]}{r}, 0 \right\}$$

### ■ Covariant, Mixed and Contravariant Components of the Metric Tensor

```
In[15]:= Table[metricg[i, j] = e_i.e_j // FullSimplify, {i, {r, theta, phi}}, {j, {r, theta, phi}}]
```

$$\text{Out[15]= } \left\{ \{1, 0, 0\}, \{0, r^2, 0\}, \{0, 0, r^2 \sin[theta]^2\} \right\}$$

```
In[16]:= Table[Metricg[i, j] = E_i.E_j // FullSimplify, {i, {r, theta, phi}}, {j, {r, theta, phi}}]
```

$$\text{Out[16]= } \left\{ \{1, 0, 0\}, \left\{0, \frac{1}{r^2}, 0\right\}, \left\{0, 0, \frac{\csc[theta]^2}{r^2}\right\} \right\}$$

```
In[17]:= Table[delta[i, j] = E_i.e_j // FullSimplify, {i, {r, theta, phi}}, {j, {r, theta, phi}}]
```

Out[17]= {{1, 0, 0}, {0, 1, 0}, {0, 0, 1}}

### ■ Christoffel Symbols

```
In[18]:= Table[Table[Table[Gammer[sigma, mu, nu] =
        1 / 2 * Sum[(E_sigma.E_rho) * (D[(e_nu.e_rho), mu] + D[(e_mu.e_rho), nu] - D[(e_mu.e_nu), rho]),
          {rho, {r, theta, phi}}] // FullSimplify,
        {sigma, {r, theta, phi}}], {mu, {r, theta, phi}}], {nu, {r, theta, phi}}]
```

$$\text{Out[18]= } \left\{ \left\{ \{0, 0, 0\}, \left\{0, \frac{1}{r}, 0\right\}, \left\{0, 0, \frac{1}{r}\right\} \right\}, \left\{ \left\{0, \frac{1}{r}, 0\right\}, \{-r, 0, 0\}, \{0, 0, \cot[theta]\} \right\}, \right.$$
$$\left. \left\{ \left\{0, 0, \frac{1}{r}\right\}, \{0, 0, \cot[theta]\}, \{-r\sin[theta]^2, -\cos[theta]\sin[theta], 0\} \right\} \right\}$$

# Artificial Viscosity

## Viscous Pressure Tensor

### ■ Definitions of Viscous Pressure

In order to proof consistency of our modified artificial viscosity we use different definitions and relations.

```
In[19]:= Table[
          qschlange[i, j] =
           1 / 2 * (d[i][u[j]] + d[j][u[i]] - Sum[Gammer[k, i, j] * u[k], {k, {r, theta, phi}}] -
                 Sum[Gammer[k, j, i] * u[k], {k, {r, theta, phi}}]) - 1 / 3 * metricg[i, j] * div[u],
          {i, {r, theta, phi}}, {j, {r, theta, phi}}]; (* covariant componentes *)

In[20]:= Table[qschlangeoben[m, n] = Sum[(Metricg[m, k] * Metricg[n, l]) *
              (1 / 2 * (d[k][u[l]] + d[l][u[k]] - Sum[Gammer[o, k, l] * u[o], {o, {r, theta, phi}}] -
                    Sum[Gammer[o, l, k] * u[o], {o, {r, theta, phi}}]) -
                 1 / 3 * metricg[m, n] * div[u]), {k, {r, theta, phi}}, {l, {r, theta, phi}}],
          {m, {r, theta, phi}}, {n, {r, theta, phi}}]; (* contravariant components *)

In[21]:= Table[
          qoben[m, n] = Sum[(Metricg[m, i] * Metricg[n, j]) qschlange[i, j],
             {i, {r, theta, phi}}, {j, {r, theta, phi}}],
          {m, {r, theta, phi}}, {n, {r, theta, phi}}]; (* contravariant components *)

In[22]:= Table[
          qmix[i, j] = Sum[Metricg[l, i] * qschlange[l, j], {l, {r, theta, phi}}],
          {i, {r, theta, phi}}, {j, {r, theta, phi}}]; (* mixed components *)

In[23]:= Table[
          qmix2[i, n] = Sum[metricg[j, n] * qschlangeoben[i, j], {j, {r, theta, phi}}],
          {i, {r, theta, phi}}, {n, {r, theta, phi}}]; (* mixed components *)
```

■ **Components Explicitly**

In[24]:= `Table[qmix[m, n], {m, {r, theta, phi}}, {n, {r, theta, phi}}] // FullSimplify // Expand`

Out[24]= $\left\{\left\{-\frac{\text{div}[u]}{3}+d[r][u[r]],\ -\frac{u[\text{theta}]}{r}+\frac{1}{2}d[r][u[\text{theta}]]+\frac{1}{2}d[\text{theta}][u[r]],\right.\right.$
$\left.-\frac{u[\text{phi}]}{r}+\frac{1}{2}d[\text{phi}][u[r]]+\frac{1}{2}d[r][u[\text{phi}]]\right\},$
$\left\{-\frac{u[\text{theta}]}{r^3}+\frac{d[r][u[\text{theta}]]}{2r^2}+\frac{d[\text{theta}][u[r]]}{2r^2},\ -\frac{\text{div}[u]}{3}+\frac{u[r]}{r}+\frac{d[\text{theta}][u[\text{theta}]]}{r^2},\right.$
$\left.-\frac{\text{Cot}[\text{theta}]\,u[\text{phi}]}{r^2}+\frac{d[\text{phi}][u[\text{theta}]]}{2r^2}+\frac{d[\text{theta}][u[\text{phi}]]}{2r^2}\right\},$
$\left\{-\frac{\text{Csc}[\text{theta}]^2\,u[\text{phi}]}{r^3}+\frac{\text{Csc}[\text{theta}]^2\,d[\text{phi}][u[r]]}{2r^2}+\frac{\text{Csc}[\text{theta}]^2\,d[r][u[\text{phi}]]}{2r^2},\right.$
$-\frac{\text{Cot}[\text{theta}]\,\text{Csc}[\text{theta}]^2\,u[\text{phi}]}{r^2}+$
$\frac{\text{Csc}[\text{theta}]^2\,d[\text{phi}][u[\text{theta}]]}{2r^2}+\frac{\text{Csc}[\text{theta}]^2\,d[\text{theta}][u[\text{phi}]]}{2r^2},$
$\left.\left.-\frac{\text{div}[u]}{3}+\frac{u[r]}{r}+\frac{\text{Cot}[\text{theta}]\,u[\text{theta}]}{r^2}+\frac{\text{Csc}[\text{theta}]^2\,d[\text{phi}][u[\text{phi}]]}{r^2}\right\}\right\}$

In[25]:= `Do[Print[Q[m, n], " = ", qschlange[m, n] // FullSimplify // Expand],`
`   {m, {r, theta, phi}}, {n, {r, theta, phi}}]`

$Q[r, r] = -\frac{\text{div}[u]}{3}+d[r][u[r]]$

$Q[r, \text{theta}] = -\frac{u[\text{theta}]}{r}+\frac{1}{2}d[r][u[\text{theta}]]+\frac{1}{2}d[\text{theta}][u[r]]$

$Q[r, \text{phi}] = -\frac{u[\text{phi}]}{r}+\frac{1}{2}d[\text{phi}][u[r]]+\frac{1}{2}d[r][u[\text{phi}]]$

$Q[\text{theta}, r] = -\frac{u[\text{theta}]}{r}+\frac{1}{2}d[r][u[\text{theta}]]+\frac{1}{2}d[\text{theta}][u[r]]$

$Q[\text{theta}, \text{theta}] = -\frac{1}{3}r^2\,\text{div}[u]+r\,u[r]+d[\text{theta}][u[\text{theta}]]$

$Q[\text{theta}, \text{phi}] = -\text{Cot}[\text{theta}]\,u[\text{phi}]+\frac{1}{2}d[\text{phi}][u[\text{theta}]]+\frac{1}{2}d[\text{theta}][u[\text{phi}]]$

$Q[\text{phi}, r] = -\frac{u[\text{phi}]}{r}+\frac{1}{2}d[\text{phi}][u[r]]+\frac{1}{2}d[r][u[\text{phi}]]$

$Q[\text{phi}, \text{theta}] = -\text{Cot}[\text{theta}]\,u[\text{phi}]+\frac{1}{2}d[\text{phi}][u[\text{theta}]]+\frac{1}{2}d[\text{theta}][u[\text{phi}]]$

$Q[\text{phi}, \text{phi}] =$
$-\frac{1}{3}r^2\,\text{div}[u]\,\text{Sin}[\text{theta}]^2+r\,\text{Sin}[\text{theta}]^2\,u[r]+\text{Cos}[\text{theta}]\,\text{Sin}[\text{theta}]\,u[\text{theta}]+d[\text{phi}][u[\text{phi}]]$

In[26]:= `Do[Print[Qcontra[m, n], " = ", qschlangeoben[m, n] // FullSimplify // Expand],`
`   {m, {r, theta, phi}}, {n, {r, theta, phi}}]`

$$Qcontra[r, r] = -\frac{div[u]}{3} + d[r][u[r]]$$

$$Qcontra[r, theta] = -\frac{u[theta]}{r^3} + \frac{d[r][u[theta]]}{2r^2} + \frac{d[theta][u[r]]}{2r^2}$$

$$Qcontra[r, phi] = -\frac{Csc[theta]^2 u[phi]}{r^3} + \frac{Csc[theta]^2 d[phi][u[r]]}{2r^2} + \frac{Csc[theta]^2 d[r][u[phi]]}{2r^2}$$

$$Qcontra[theta, r] = -\frac{u[theta]}{r^3} + \frac{d[r][u[theta]]}{2r^2} + \frac{d[theta][u[r]]}{2r^2}$$

$$Qcontra[theta, theta] = -\frac{div[u]}{3r^2} + \frac{u[r]}{r^3} + \frac{d[theta][u[theta]]}{r^4}$$

$$Qcontra[theta, phi] =$$
$$-\frac{Cot[theta] Csc[theta]^2 u[phi]}{r^4} + \frac{Csc[theta]^2 d[phi][u[theta]]}{2r^4} + \frac{Csc[theta]^2 d[theta][u[phi]]}{2r^4}$$

$$Qcontra[phi, r] = -\frac{Csc[theta]^2 u[phi]}{r^3} + \frac{Csc[theta]^2 d[phi][u[r]]}{2r^2} + \frac{Csc[theta]^2 d[r][u[phi]]}{2r^2}$$

$$Qcontra[phi, theta] =$$
$$-\frac{Cot[theta] Csc[theta]^2 u[phi]}{r^4} + \frac{Csc[theta]^2 d[phi][u[theta]]}{2r^4} + \frac{Csc[theta]^2 d[theta][u[phi]]}{2r^4}$$

$$Qcontra[phi, phi] = -\frac{Csc[theta]^2 div[u]}{3r^2} + \frac{Csc[theta]^2 u[r]}{r^3} +$$
$$\frac{Cot[theta] Csc[theta]^2 u[theta]}{r^4} + \frac{Csc[theta]^4 d[phi][u[phi]]}{r^4}$$

### ■ Symmetry

In fact Q is symmetric in its co- and contravariant components.

In[27]:= `Table[qschlangeoben[m, n] === qschlangeoben[n, m], {m, {r, theta, phi}}, {n, {r, theta, phi}}]`

Out[27]= {{True, True, True}, {True, True, True}, {True, True, True}}

In[28]:= `Table[qschlange[m, n] === qschlange[n, m], {m, {r, theta, phi}}, {n, {r, theta, phi}}]`

Out[28]= {{True, True, True}, {True, True, True}, {True, True, True}}

In[29]:= `Table[qmix[m, n] === qmix[n, m], {m, {r, theta, phi}}, {n, {r, theta, phi}}]`

Out[29]= {{True, False, False}, {False, True, False}, {False, False, True}}

### ■ Trace of Q

The trace vanishes as expected.

In[30]:= `Sum[qmix[m, m], {m, {r, theta, phi}}]`

Out[30]= $-\dfrac{\text{div}[u]}{3} + \dfrac{1}{r^2}\text{Csc}[\text{theta}]^2 \left(-\dfrac{1}{3} r^2 \text{div}[u] \text{Sin}[\text{theta}]^2 + \dfrac{1}{2}\left(2 r \text{Sin}[\text{theta}]^2 u[r] + 2 \text{Cos}[\text{theta}] \text{Sin}[\text{theta}] u[\text{theta}] + 2 d[\text{phi}][u[\text{phi}]]\right)\right) + d[r][u[r]] + \dfrac{-\dfrac{1}{3} r^2 \text{div}[u] + \dfrac{1}{2}(2 r u[r] + 2 d[\text{theta}][u[\text{theta}]])}{r^2}$

In[31]:= `Sum[qmix[m, m], {m, {r, theta, phi}}] /.`
`{div[u] -> `$\dfrac{2 u[r]}{r} + \dfrac{\text{Cot}[\text{theta}] u[\text{theta}]}{r^2} + \dfrac{\text{Csc}[\text{theta}]^2 d[\text{phi}][u[\text{phi}]]}{r^2} +$
`d[r][u[r]] + `$\dfrac{d[\text{theta}][u[\text{theta}]]}{r^2}$`} // FullSimplify // Expand`

Out[31]= 0

In[32]:= `Sum[qmix2[m, m], {m, {r, theta, phi}}] /.`
`{div[u] -> `$\dfrac{2 u[r]}{r} + \dfrac{\text{Cot}[\text{theta}] u[\text{theta}]}{r^2} + \dfrac{\text{Csc}[\text{theta}]^2 d[\text{phi}][u[\text{phi}]]}{r^2} +$
`d[r][u[r]] + `$\dfrac{d[\text{theta}][u[\text{theta}]]}{r^2}$`} // FullSimplify // Expand`

Out[32]= 0

# Equation of Motion in Spherical Coordinates

## Generation of Strong Conservation Form Equation

- **Definition of Contravariant Components of Tensors**

```
Table[rschlange[i, j] = rho * (u[i] - xdot[i]) * u[j] // FullSimplify,
 {i, {r, theta, phi}}, {j, {r, theta, phi}}]
Table[p[i, j] = p * Metricg[i, j] // FullSimplify, {i, {r, theta, phi}}, {j, {r, theta, phi}}]
Table[q[i, j] = Sum[(Metricg[i, m] * Metricg[j, n])
     (1 / 2 * (d[n][u[m]] + d[n][u[m]] - Sum[Gammer[k, m, m] * u[k], {k, {r, theta, phi}}] -
        Sum[Gammer[k, m, n] * u[k], {k, {r, theta, phi}}]) - 1 / 3 * metricg[n, m] * div[u]),
    {n, {r, theta, phi}}, {m, {r, theta, phi}}] // FullSimplify,
 {i, {r, theta, phi}}, {j, {r, theta, phi}}]
(* note that there emerge covariant velocity components in Q due to its definition *)
```

Out[33]= {{rho u[r] (u[r] - xdot[r]), rho u[theta] (u[r] - xdot[r]), rho u[phi] (u[r] - xdot[r])},
 {rho u[r] (u[theta] - xdot[theta]), rho u[theta] (u[theta] - xdot[theta]),
  rho u[phi] (u[theta] - xdot[theta])}, {rho u[r] (u[phi] - xdot[phi]),
  rho u[theta] (u[phi] - xdot[phi]), rho u[phi] (u[phi] - xdot[phi])}}

Out[34]= $\left\{\{p, 0, 0\}, \left\{0, \frac{p}{r^2}, 0\right\}, \left\{0, 0, \frac{p \, \text{Csc}[\text{theta}]^2}{r^2}\right\}\right\}$

Out[35]= $\left\{\left\{-\frac{\text{div}[u]}{3} + d[r][u[r]],\right.\right.$
 $-\frac{u[\text{theta}] - 2 r \, d[\text{theta}][u[r]]}{2 r^3}, -\frac{\text{Csc}[\text{theta}]^2 \, (u[\text{phi}] - 2 r \, d[\text{phi}][u[r]])}{2 r^3}\bigg\},$
 $\left\{\frac{r^2 u[r] - u[\text{theta}] + 2 r \, d[r][u[\text{theta}]]}{2 r^3}, \frac{-r^2 \, \text{div}[u] + 3 \, (r \, u[r] + d[\text{theta}][u[\text{theta}]])}{3 r^4},\right.$
 $\left.\frac{\text{Csc}[\text{theta}]^2 \, (-\text{Cot}[\text{theta}] \, u[\text{phi}] + r \, u[r] + 2 \, d[\text{phi}][u[\text{theta}]])}{2 r^4}\right\},$
 $\left\{\frac{r^2 u[r] + r \, \text{Cot}[\text{theta}] \, u[\text{theta}] - \text{Csc}[\text{theta}]^2 \, (u[\text{phi}] - 2 r \, d[r][u[\text{phi}]])}{2 r^3},\right.$
 $\frac{-\text{Cot}[\text{theta}] \, \text{Csc}[\text{theta}]^2 \, u[\text{phi}] + r \, u[r] + \text{Cot}[\text{theta}] \, u[\text{theta}] + 2 \, \text{Csc}[\text{theta}]^2 \, d[\text{theta}][u[\text{phi}]]}{2 r^4},$
 $\left.\left.\frac{\text{Csc}[\text{theta}]^2 \, \left(-r^2 \, \text{div}[u] + 3 \, \left(r \, u[r] + \text{Cot}[\text{theta}] \, u[\text{theta}] + \text{Csc}[\text{theta}]^2 \, d[\text{phi}][u[\text{phi}]]\right)\right)}{3 r^4}\right\}\right\}$

- **Equation of Motion in Spherical Coordinates and Strong Conservation Form**

In[36]:= ```
Table[d[t][Sum[Sqrt[g] * rho * u[n] * e_n[[k]], {n, {r, theta, phi}}]] + Sum[
    Sum[d[i][Sqrt[g] * (rschlange[i, j] + p[i, j] + Qkomp[i, j]) * e_j[[k]]], {i, {r, theta, phi}}],
   {j, {r, theta, phi}}] + rho * Sum[d[l][Sqrt[g] * Phi * E_l[[k]]], {l, {r, theta, phi}}] -
  4 Pi / c * Sqrt[g] * (Sum[H[o] * e_o[[k]], {o, {r, theta, phi}}]) /.
  {Sin[theta] → Sqrt[1 - mu^2], d[theta] → -Sqrt[1 - mu^2] d[mu],
   Csc[theta] → 1 / Sqrt[1 - mu^2], Cos[theta] → mu}, {k, {1, 2, 3}}]
```

Out[36]= $\Big\{-\dfrac{4\pi\sqrt{(1-mu^2)\,r^4}\,\left(\sqrt{1-mu^2}\,\text{Cos}[phi]\,H[r]+mu\,r\,\text{Cos}[phi]\,H[theta]-\sqrt{1-mu^2}\,r\,H[phi]\,\text{Sin}[phi]\right)}{c}+$

$\left(-\sqrt{1-mu^2}\,d[mu]\right)\Big[-\sqrt{1-mu^2}\,r\,\sqrt{(1-mu^2)\,r^4}\,\text{Sin}[phi]$

$(\text{Qkomp}[theta, phi]+\text{rho}\,u[phi]\,(u[theta]-xdot[theta]))\Big]+\left(-\sqrt{1-mu^2}\,d[mu]\right)\Big[$

$\sqrt{1-mu^2}\,\sqrt{(1-mu^2)\,r^4}\,\text{Cos}[phi]\,(\text{Qkomp}[theta, r]+\text{rho}\,u[r]\,(u[theta]-xdot[theta]))\Big]+$

$\left(-\sqrt{1-mu^2}\,d[mu]\right)\Big[mu\,r\,\sqrt{(1-mu^2)\,r^4}\,\text{Cos}[phi]$

$\left(\dfrac{p}{r^2}+\text{Qkomp}[theta, theta]+\text{rho}\,u[theta]\,(u[theta]-xdot[theta])\right)\Big]+d[phi]\Big[-\sqrt{1-mu^2}\,r$

$\sqrt{(1-mu^2)\,r^4}\,\text{Sin}[phi]\,\left(\dfrac{p}{(1-mu^2)\,r^2}+\text{Qkomp}[phi, phi]+\text{rho}\,u[phi]\,(u[phi]-xdot[phi])\right)\Big]+$

$d[phi]\Big[\sqrt{1-mu^2}\,\sqrt{(1-mu^2)\,r^4}\,\text{Cos}[phi]\,(\text{Qkomp}[phi, r]+\text{rho}\,u[r]\,(u[phi]-xdot[phi]))\Big]+$

$d[phi]\Big[mu\,r\,\sqrt{(1-mu^2)\,r^4}\,\text{Cos}[phi]\,(\text{Qkomp}[phi, theta]+\text{rho}\,u[theta]\,(u[phi]-xdot[phi]))\Big]+$

$\text{rho}\,\Bigg(\left(-\sqrt{1-mu^2}\,d[mu]\right)\Big[\dfrac{mu\,\Phi\,\sqrt{(1-mu^2)\,r^4}\,\text{Cos}[phi]}{r}\Big]+$

$d[phi]\Big[-\dfrac{\Phi\,\sqrt{(1-mu^2)\,r^4}\,\text{Sin}[phi]}{\sqrt{1-mu^2}\,r}\Big]+d[r]\Big[\sqrt{1-mu^2}\,\Phi\,\sqrt{(1-mu^2)\,r^4}\,\text{Cos}[phi]\Big]\Bigg)+$

$d[r]\Big[-\sqrt{1-mu^2}\,r\,\sqrt{(1-mu^2)\,r^4}\,\text{Sin}[phi]\,(\text{Qkomp}[r, phi]+\text{rho}\,u[phi]\,(u[r]-xdot[r]))\Big]+$

$d[r]\Big[\sqrt{1-mu^2}\,\sqrt{(1-mu^2)\,r^4}\,\text{Cos}[phi]\,(p+\text{Qkomp}[r, r]+\text{rho}\,u[r]\,(u[r]-xdot[r]))\Big]+$

$d[r]\Big[mu\,r\,\sqrt{(1-mu^2)\,r^4}\,\text{Cos}[phi]\,(\text{Qkomp}[r, theta]+\text{rho}\,u[theta]\,(u[r]-xdot[r]))\Big]+$

$d[t]\Big[-\sqrt{1-mu^2}\,r\,\sqrt{(1-mu^2)\,r^4}\,\text{rho}\,\text{Sin}[phi]\,u[phi]+$

$\sqrt{1-mu^2}\,\sqrt{(1-mu^2)\,r^4}\,\text{rho}\,\text{Cos}[phi]\,u[r]+mu\,r\,\sqrt{(1-mu^2)\,r^4}\,\text{rho}\,\text{Cos}[phi]\,u[theta]\Big],$

$-\dfrac{4\pi\sqrt{(1-mu^2)\,r^4}\,\left(\sqrt{1-mu^2}\,r\,\text{Cos}[phi]\,H[phi]+\sqrt{1-mu^2}\,H[r]\,\text{Sin}[phi]+mu\,r\,H[theta]\,\text{Sin}[phi]\right)}{c}+$

$\left(-\sqrt{1-mu^2}\,d[mu]\right)\Big[\sqrt{1-mu^2}\,r\,\sqrt{(1-mu^2)\,r^4}\,\text{Cos}[phi]$

$(\text{Qkomp}[\text{theta}, \text{phi}] + \text{rho}\, u[\text{phi}]\, (u[\text{theta}] - \text{xdot}[\text{theta}]))\Big] + \left(-\sqrt{1-\text{mu}^2}\, d[\text{mu}]\right)\Big[$

$\sqrt{1-\text{mu}^2}\, \sqrt{(1-\text{mu}^2)\, r^4}\, \text{Sin}[\text{phi}]\, (\text{Qkomp}[\text{theta}, r] + \text{rho}\, u[r]\, (u[\text{theta}] - \text{xdot}[\text{theta}]))\Big] +$

$\left(-\sqrt{1-\text{mu}^2}\, d[\text{mu}]\right)\Big[\text{mu}\, r\, \sqrt{(1-\text{mu}^2)\, r^4}\, \text{Sin}[\text{phi}]$

$\left(\frac{p}{r^2} + \text{Qkomp}[\text{theta}, \text{theta}] + \text{rho}\, u[\text{theta}]\, (u[\text{theta}] - \text{xdot}[\text{theta}])\right)\Big] + d[\text{phi}]\Big[\sqrt{1-\text{mu}^2}\, r$

$\sqrt{(1-\text{mu}^2)\, r^4}\, \text{Cos}[\text{phi}]\, \left(\frac{p}{(1-\text{mu}^2)\, r^2} + \text{Qkomp}[\text{phi}, \text{phi}] + \text{rho}\, u[\text{phi}]\, (u[\text{phi}] - \text{xdot}[\text{phi}])\right)\Big] +$

$d[\text{phi}]\Big[\sqrt{1-\text{mu}^2}\, \sqrt{(1-\text{mu}^2)\, r^4}\, \text{Sin}[\text{phi}]\, (\text{Qkomp}[\text{phi}, r] + \text{rho}\, u[r]\, (u[\text{phi}] - \text{xdot}[\text{phi}]))\Big] +$

$d[\text{phi}]\Big[\text{mu}\, r\, \sqrt{(1-\text{mu}^2)\, r^4}\, \text{Sin}[\text{phi}]\, (\text{Qkomp}[\text{phi}, \text{theta}] + \text{rho}\, u[\text{theta}]\, (u[\text{phi}] - \text{xdot}[\text{phi}]))\Big] +$

$\text{rho}\, \Bigg(\left(-\sqrt{1-\text{mu}^2}\, d[\text{mu}]\right)\Big[\frac{\text{mu}\, \Phi\, \sqrt{(1-\text{mu}^2)\, r^4}\, \text{Sin}[\text{phi}]}{r}\Big] +$

$d[\text{phi}]\Big[\frac{\Phi\, \sqrt{(1-\text{mu}^2)\, r^4}\, \text{Cos}[\text{phi}]}{\sqrt{1-\text{mu}^2}\, r}\Big] + d[r]\Big[\sqrt{1-\text{mu}^2}\, \Phi\, \sqrt{(1-\text{mu}^2)\, r^4}\, \text{Sin}[\text{phi}]\Big]\Bigg) +$

$d[r]\Big[\sqrt{1-\text{mu}^2}\, r\, \sqrt{(1-\text{mu}^2)\, r^4}\, \text{Cos}[\text{phi}]\, (\text{Qkomp}[r, \text{phi}] + \text{rho}\, u[\text{phi}]\, (u[r] - \text{xdot}[r]))\Big] +$

$d[r]\Big[\sqrt{1-\text{mu}^2}\, \sqrt{(1-\text{mu}^2)\, r^4}\, \text{Sin}[\text{phi}]\, (p + \text{Qkomp}[r, r] + \text{rho}\, u[r]\, (u[r] - \text{xdot}[r]))\Big] +$

$d[r]\Big[\text{mu}\, r\, \sqrt{(1-\text{mu}^2)\, r^4}\, \text{Sin}[\text{phi}]\, (\text{Qkomp}[r, \text{theta}] + \text{rho}\, u[\text{theta}]\, (u[r] - \text{xdot}[r]))\Big] +$

$d[t]\Big[\sqrt{1-\text{mu}^2}\, r\, \sqrt{(1-\text{mu}^2)\, r^4}\, \text{rho}\, \text{Cos}[\text{phi}]\, u[\text{phi}] +$

$\sqrt{1-\text{mu}^2}\, \sqrt{(1-\text{mu}^2)\, r^4}\, \text{rho}\, \text{Sin}[\text{phi}]\, u[r] + \text{mu}\, r\, \sqrt{(1-\text{mu}^2)\, r^4}\, \text{rho}\, \text{Sin}[\text{phi}]\, u[\text{theta}]\Big],$

$-\dfrac{4\, \pi\, \sqrt{(1-\text{mu}^2)\, r^4}\, \left(\text{mu}\, H[r] - \sqrt{1-\text{mu}^2}\, r\, H[\text{theta}]\right)}{c} + \left(-\sqrt{1-\text{mu}^2}\, d[\text{mu}]\right)[0] +$

$\left(-\sqrt{1-\text{mu}^2}\, d[\text{mu}]\right)\Big[\text{mu}\, \sqrt{(1-\text{mu}^2)\, r^4}\, (\text{Qkomp}[\text{theta}, r] + \text{rho}\, u[r]\, (u[\text{theta}] - \text{xdot}[\text{theta}]))\Big] +$

$\left(-\sqrt{1-\text{mu}^2}\, d[\text{mu}]\right)\Big[$

$-\sqrt{1-\text{mu}^2}\, r\, \sqrt{(1-\text{mu}^2)\, r^4}\, \left(\frac{p}{r^2} + \text{Qkomp}[\text{theta}, \text{theta}] + \text{rho}\, u[\text{theta}]\, (u[\text{theta}] - \text{xdot}[\text{theta}])\right)\Big] +$

$d[\text{phi}][0] + d[\text{phi}]\Big[\text{mu}\, \sqrt{(1-\text{mu}^2)\, r^4}\, (\text{Qkomp}[\text{phi}, r] + \text{rho}\, u[r]\, (u[\text{phi}] - \text{xdot}[\text{phi}]))\Big] + d[\text{phi}]\Big[$

$$-\sqrt{1-mu^2}\ r\ \sqrt{(1-mu^2)\ r^4}\ (\text{Qkomp}[\text{phi, theta}] + \text{rho u}[\text{theta}]\ (u[\text{phi}] - \text{xdot}[\text{phi}]))\Big] + d[r][0] +$$

$$\text{rho}\left(\left(-\sqrt{1-mu^2}\ d[mu]\right)\left[-\frac{\sqrt{1-mu^2}\ \text{Phi}\ \sqrt{(1-mu^2)\ r^4}}{r}\right] + d[\text{phi}][0] + d[r]\left[mu\ \text{Phi}\ \sqrt{(1-mu^2)\ r^4}\ \right]\right) +$$

$$d[r]\left[mu\ \sqrt{(1-mu^2)\ r^4}\ (p + \text{Qkomp}[r, r] + \text{rho u}[r]\ (u[r] - \text{xdot}[r]))\right] +$$

$$d[r]\left[-\sqrt{1-mu^2}\ r\ \sqrt{(1-mu^2)\ r^4}\ (\text{Qkomp}[r, \text{theta}] + \text{rho u}[\text{theta}]\ (u[r] - \text{xdot}[r]))\right] +$$

$$d[t]\left[mu\ \sqrt{(1-mu^2)\ r^4}\ \text{rho u}[r] - \sqrt{1-mu^2}\ r\ \sqrt{(1-mu^2)\ r^4}\ \text{rho u}[\text{theta}]\right]\Big\}$$

### ■ EOM with Q explicitly

```
In[37]:= Table[d[t][Sum[Sqrt[g] * rho * u[n] * e_n[[k]], {n, {r, theta, phi}}]] +
    Sum[Sum[d[i][Sqrt[g] * ( rschlange[i, j] + p[i, j] + q[i, j]) * e_j[[k]]], {i, {r, theta, phi}}],
     {j, {r, theta, phi}}] - +rho * Sum[d[l][Sqrt[g] * Phi * E_l[[k]]], {l, {r, theta, phi}}]
     4 Pi / c * Sqrt[g] * (Sum[H[o] * e_o[[k]], {o, {r, theta, phi}}]) /.
    {Sin[theta] → Sqrt[1 - mu^2], d[theta] → - Sqrt[1 - mu^2] d[mu],
     Csc[theta] → 1 / Sqrt[1 - mu^2], Cos[theta] → mu}, {k, {1, 2, 3}}]
```

$$\text{Out[37]}= \Bigg\{\left(-\sqrt{1-mu^2}\ d[mu]\right)\Bigg[mu\ r\ \sqrt{(1-mu^2)\ r^4}\ \text{Cos}[\text{phi}]\ \Bigg(\frac{p}{r^2} + \text{rho u}[\text{theta}]\ (u[\text{theta}] - \text{xdot}[\text{theta}]) +$$

$$\frac{-r^2\ \text{div}[u] + 3\ \left(r\ u[r] + \left(-\sqrt{1-mu^2}\ d[mu]\right)[u[\text{theta}]]\right)}{3\ r^4}\Bigg)\Bigg] +$$

$$\left(-\sqrt{1-mu^2}\ d[mu]\right)\Bigg[-\sqrt{1-mu^2}\ r\ \sqrt{(1-mu^2)\ r^4}\ \text{Sin}[\text{phi}]$$

$$\left(\text{rho u}[\text{phi}]\ (u[\text{theta}] - \text{xdot}[\text{theta}]) + \frac{-\text{Cot}[\text{theta}]\ u[\text{phi}] + r\ u[r] + 2\ d[\text{phi}][u[\text{theta}]]}{2\ (1-mu^2)\ r^4}\right)\Bigg] +$$

$$\left(-\sqrt{1-mu^2}\ d[mu]\right)\Bigg[\sqrt{1-mu^2}\ \sqrt{(1-mu^2)\ r^4}\ \text{Cos}[\text{phi}]$$

$$\left(\text{rho u}[r]\ (u[\text{theta}] - \text{xdot}[\text{theta}]) + \frac{r^2\ u[r] - u[\text{theta}] + 2\ r\ d[r][u[\text{theta}]]}{2\ r^3}\right)\Bigg] +$$

$$d[phi]\left[mu\, r\, \sqrt{(1-mu^2)\, r^4}\, Cos[phi]\left(rho\, u[theta]\, (u[phi] - xdot[phi]) + \right.\right.$$

$$\left.\left.\frac{-\frac{Cot[theta]\, u[phi]}{1-mu^2} + r\, u[r] + Cot[theta]\, u[theta] + \frac{2\left(-\sqrt{1-mu^2}\, d[mu]\right)[u[phi]]}{1-mu^2}}{2\, r^4}\right)\right] +$$

$$d[phi]\left[-\sqrt{1-mu^2}\, r\, \sqrt{(1-mu^2)\, r^4}\, Sin[phi]\left(\frac{p}{(1-mu^2)\, r^2} + rho\, u[phi]\, (u[phi] - xdot[phi]) + \right.\right.$$

$$\left.\left.\frac{-r^2\, div[u] + 3\left(r\, u[r] + Cot[theta]\, u[theta] + \frac{d[phi][u[phi]]}{1-mu^2}\right)}{3\,(1-mu^2)\, r^4}\right)\right] +$$

$$d[phi]\left[\sqrt{1-mu^2}\, \sqrt{(1-mu^2)\, r^4}\, Cos[phi]\left(rho\, u[r]\, (u[phi] - xdot[phi]) + \right.\right.$$

$$\left.\left.\frac{r^2\, u[r] + r\, Cot[theta]\, u[theta] - \frac{u[phi] - 2\, r\, d[r][u[phi]]}{1-mu^2}}{2\, r^3}\right)\right] - \frac{1}{c}4\,\pi\,\sqrt{(1-mu^2)\, r^4}$$

$$rho\left(\sqrt{1-mu^2}\, Cos[phi]\, H[r] + mu\, r\, Cos[phi]\, H[theta] - \sqrt{1-mu^2}\, r\, H[phi]\, Sin[phi]\right)$$

$$\left(\left(-\sqrt{1-mu^2}\, d[mu]\right)\left[\frac{mu\, Phi\, \sqrt{(1-mu^2)\, r^4}\, Cos[phi]}{r}\right] + d[phi]\left[-\frac{Phi\, \sqrt{(1-mu^2)\, r^4}\, Sin[phi]}{\sqrt{1-mu^2}\, r}\right] + \right.$$

$$d[r]\left[\sqrt{1-mu^2}\, Phi\, \sqrt{(1-mu^2)\, r^4}\, Cos[phi]\right] + d[r]\left[mu\, r\, \sqrt{(1-mu^2)\, r^4}\, Cos[phi]\right.$$

$$\left.\left(rho\, u[theta]\, (u[r] - xdot[r]) - \frac{u[theta] - 2\, r\left(-\sqrt{1-mu^2}\, d[mu]\right)[u[r]]}{2\, r^3}\right)\right] + d[r]\left[\right.$$

$$\left.-\sqrt{1-mu^2}\, r\, \sqrt{(1-mu^2)\, r^4}\, Sin[phi]\left(rho\, u[phi]\, (u[r] - xdot[r]) - \frac{u[phi] - 2\, r\, d[phi][u[r]]}{2\,(1-mu^2)\, r^3}\right)\right] +$$

$$d[r]\left[\sqrt{1-\mathrm{mu}^2}\ \sqrt{(1-\mathrm{mu}^2)\ r^4}\ \mathrm{Cos}[\mathrm{phi}]\ \left(p - \frac{\mathrm{div}[u]}{3} + \mathrm{rho}\ u[r]\ (u[r] - \mathrm{xdot}[r]) + d[r][u[r]]\right)\right] +$$

$$d[t]\left[-\sqrt{1-\mathrm{mu}^2}\ r\ \sqrt{(1-\mathrm{mu}^2)\ r^4}\ \mathrm{rho}\ \mathrm{Sin}[\mathrm{phi}]\ u[\mathrm{phi}] + \right.$$

$$\left. \sqrt{1-\mathrm{mu}^2}\ \sqrt{(1-\mathrm{mu}^2)\ r^4}\ \mathrm{rho}\ \mathrm{Cos}[\mathrm{phi}]\ u[r] + \mathrm{mu}\ r\ \sqrt{(1-\mathrm{mu}^2)\ r^4}\ \mathrm{rho}\ \mathrm{Cos}[\mathrm{phi}]\ u[\mathrm{theta}]\right],$$

$$\left(-\sqrt{1-\mathrm{mu}^2}\ d[\mathrm{mu}]\right)\left[\mathrm{mu}\ r\ \sqrt{(1-\mathrm{mu}^2)\ r^4}\ \mathrm{Sin}[\mathrm{phi}]\ \left(\frac{p}{r^2} + \mathrm{rho}\ u[\mathrm{theta}]\ (u[\mathrm{theta}] - \mathrm{xdot}[\mathrm{theta}]) + \right.\right.$$

$$\left.\left. \frac{-r^2\ \mathrm{div}[u] + 3\left(r\ u[r] + \left(-\sqrt{1-\mathrm{mu}^2}\ d[\mathrm{mu}]\right)[u[\mathrm{theta}]]\right)}{3\ r^4}\right)\right] +$$

$$\left(-\sqrt{1-\mathrm{mu}^2}\ d[\mathrm{mu}]\right)\left[\sqrt{1-\mathrm{mu}^2}\ r\ \sqrt{(1-\mathrm{mu}^2)\ r^4}\ \mathrm{Cos}[\mathrm{phi}]\right.$$

$$\left. \left(\mathrm{rho}\ u[\mathrm{phi}]\ (u[\mathrm{theta}] - \mathrm{xdot}[\mathrm{theta}]) + \frac{-\mathrm{Cot}[\mathrm{theta}]\ u[\mathrm{phi}] + r\ u[r] + 2\ d[\mathrm{phi}][u[\mathrm{theta}]]}{2\ (1-\mathrm{mu}^2)\ r^4}\right)\right] +$$

$$\left(-\sqrt{1-\mathrm{mu}^2}\ d[\mathrm{mu}]\right)\left[\sqrt{1-\mathrm{mu}^2}\ \sqrt{(1-\mathrm{mu}^2)\ r^4}\ \mathrm{Sin}[\mathrm{phi}]\right.$$

$$\left. \left(\mathrm{rho}\ u[r]\ (u[\mathrm{theta}] - \mathrm{xdot}[\mathrm{theta}]) + \frac{r^2\ u[r] - u[\mathrm{theta}] + 2\ r\ d[r][u[\mathrm{theta}]]}{2\ r^3}\right)\right] +$$

$$d[\mathrm{phi}]\left[\mathrm{mu}\ r\ \sqrt{(1-\mathrm{mu}^2)\ r^4}\ \mathrm{Sin}[\mathrm{phi}]\ \left(\mathrm{rho}\ u[\mathrm{theta}]\ (u[\mathrm{phi}] - \mathrm{xdot}[\mathrm{phi}]) + \right.\right.$$

$$\left.\left. \frac{-\frac{\mathrm{Cot}[\mathrm{theta}]\ u[\mathrm{phi}]}{1-\mathrm{mu}^2} + r\ u[r] + \mathrm{Cot}[\mathrm{theta}]\ u[\mathrm{theta}] + \frac{2\left(-\sqrt{1-\mathrm{mu}^2}\ d[\mathrm{mu}]\right)[u[\mathrm{phi}]]}{1-\mathrm{mu}^2}}{2\ r^4}\right)\right] +$$

$$d[\mathrm{phi}]\left[\sqrt{1-\mathrm{mu}^2}\ r\ \sqrt{(1-\mathrm{mu}^2)\ r^4}\ \mathrm{Cos}[\mathrm{phi}]\ \left(\frac{p}{(1-\mathrm{mu}^2)\ r^2} + \mathrm{rho}\ u[\mathrm{phi}]\ (u[\mathrm{phi}] - \mathrm{xdot}[\mathrm{phi}]) + \right.\right.$$

$$\left.\left. \frac{-r^2\ \mathrm{div}[u] + 3\left(r\ u[r] + \mathrm{Cot}[\mathrm{theta}]\ u[\mathrm{theta}] + \frac{d[\mathrm{phi}][u[\mathrm{phi}]]}{1-\mathrm{mu}^2}\right)}{3\ (1-\mathrm{mu}^2)\ r^4}\right)\right] +$$

$$\begin{aligned}
&d[phi]\left[\sqrt{1-mu^2}\;\sqrt{(1-mu^2)\;r^4}\;\text{Sin}[phi]\left(\text{rho}\;u[r]\;(u[phi]-\text{xdot}[phi])+\right.\right.\\
&\left.\left.\frac{r^2\;u[r]+r\;\text{Cot}[\text{theta}]\;u[\text{theta}]-\frac{u[phi]-2\;r\;d[r][u[phi]]}{1-mu^2}}{2\;r^3}\right)\right]-\frac{1}{c}4\pi\sqrt{(1-mu^2)\;r^4}\\
&\text{rho}\left(\sqrt{1-mu^2}\;r\;\text{Cos}[phi]\;H[phi]+\sqrt{1-mu^2}\;H[r]\;\text{Sin}[phi]+\text{mu}\;r\;H[\text{theta}]\;\text{Sin}[phi]\right)\\
&\left(\left(-\sqrt{1-mu^2}\;d[mu]\right)\left[\frac{\text{mu}\;\text{Phi}\;\sqrt{(1-mu^2)\;r^4}\;\text{Sin}[phi]}{r}\right]+d[phi]\left[\frac{\text{Phi}\;\sqrt{(1-mu^2)\;r^4}\;\text{Cos}[phi]}{\sqrt{1-mu^2}\;r}\right]+\right.\\
&\left.d[r]\left[\sqrt{1-mu^2}\;\text{Phi}\;\sqrt{(1-mu^2)\;r^4}\;\text{Sin}[phi]\right]\right)+d[r]\left[\text{mu}\;r\;\sqrt{(1-mu^2)\;r^4}\;\text{Sin}[phi]\right.\\
&\left.\left(\text{rho}\;u[\text{theta}]\;(u[r]-\text{xdot}[r])-\frac{u[\text{theta}]-2\;r\;\left(-\sqrt{1-mu^2}\;d[mu]\right)[u[r]]}{2\;r^3}\right)\right]+d[r]\left[\right.\\
&\sqrt{1-mu^2}\;r\;\sqrt{(1-mu^2)\;r^4}\;\text{Cos}[phi]\left(\text{rho}\;u[phi]\;(u[r]-\text{xdot}[r])-\frac{u[phi]-2\;r\;d[phi][u[r]]}{2\;(1-mu^2)\;r^3}\right)\right]+\\
&d[r]\left[\sqrt{1-mu^2}\;\sqrt{(1-mu^2)\;r^4}\;\text{Sin}[phi]\left(p-\frac{\text{div}[u]}{3}+\text{rho}\;u[r]\;(u[r]-\text{xdot}[r])+d[r][u[r]]\right)\right]+\\
&d[t]\left[\sqrt{1-mu^2}\;r\;\sqrt{(1-mu^2)\;r^4}\;\text{rho}\;\text{Cos}[phi]\;u[phi]+\sqrt{1-mu^2}\;\sqrt{(1-mu^2)\;r^4}\;\text{rho}\;\text{Sin}[phi]\;u[r]+\right.\\
&\left.\text{mu}\;r\;\sqrt{(1-mu^2)\;r^4}\;\text{rho}\;\text{Sin}[phi]\;u[\text{theta}]\right],\left(-\sqrt{1-mu^2}\;d[mu]\right)[0]+\\
&\left(-\sqrt{1-mu^2}\;d[mu]\right)\left[-\sqrt{1-mu^2}\;r\;\sqrt{(1-mu^2)\;r^4}\left(\frac{p}{r^2}+\text{rho}\;u[\text{theta}]\;(u[\text{theta}]-\text{xdot}[\text{theta}])+\right.\right.\\
&\left.\left.\frac{-r^2\;\text{div}[u]+3\left(r\;u[r]+\left(-\sqrt{1-mu^2}\;d[mu]\right)[u[\text{theta}]]\right)}{3\;r^4}\right)\right]+\left(-\sqrt{1-mu^2}\;d[mu]\right)\left[\right.\\
&\text{mu}\;\sqrt{(1-mu^2)\;r^4}\left(\text{rho}\;u[r]\;(u[\text{theta}]-\text{xdot}[\text{theta}])+\frac{r^2\;u[r]-u[\text{theta}]+2\;r\;d[r][u[\text{theta}]]}{2\;r^3}\right)\right]+
\end{aligned}$$

$$d[phi][0] + d[phi]\left[-\sqrt{1-mu^2}\ r\ \sqrt{(1-mu^2)\ r^4}\ \left(rho\ u[theta]\ (u[phi] - xdot[phi]) + \right.\right.$$

$$\left.\left.\frac{-\frac{Cot[theta]\ u[phi]}{1-mu^2} + r\ u[r] + Cot[theta]\ u[theta] + \frac{2\left(-\sqrt{1-mu^2}\ d[mu]\right)[u[phi]]}{1-mu^2}}{2\ r^4}\right)\right] +$$

$$d[phi]\left[mu\ \sqrt{(1-mu^2)\ r^4}\ \left(rho\ u[r]\ (u[phi] - xdot[phi]) + \right.\right.$$

$$\left.\left.\frac{r^2\ u[r] + r\ Cot[theta]\ u[theta] - \frac{u[phi] - 2\ r\ d[r][u[phi]]}{1-mu^2}}{2\ r^3}\right)\right] +$$

$$d[r][0] - \frac{1}{c} 4\ \pi\ \sqrt{(1-mu^2)\ r^4}\ rho\ \left(mu\ H[r] - \sqrt{1-mu^2}\ r\ H[theta]\right)$$

$$\left(\left(-\sqrt{1-mu^2}\ d[mu]\right)\left[-\frac{\sqrt{1-mu^2}\ Phi\ \sqrt{(1-mu^2)\ r^4}}{r}\right] + d[phi][0] + d[r]\left[mu\ Phi\ \sqrt{(1-mu^2)\ r^4}\right]\right) +$$

$$d[r]\left[-\sqrt{1-mu^2}\ r\ \sqrt{(1-mu^2)\ r^4}\right.$$

$$\left.\left(rho\ u[theta]\ (u[r] - xdot[r]) - \frac{u[theta] - 2\ r\left(-\sqrt{1-mu^2}\ d[mu]\right)[u[r]]}{2\ r^3}\right)\right] +$$

$$d[r]\left[mu\ \sqrt{(1-mu^2)\ r^4}\ \left(p - \frac{div[u]}{3} + rho\ u[r]\ (u[r] - xdot[r]) + d[r][u[r]]\right)\right] +$$

$$d[t]\left[mu\ \sqrt{(1-mu^2)\ r^4}\ rho\ u[r] - \sqrt{1-mu^2}\ r\ \sqrt{(1-mu^2)\ r^4}\ rho\ u[theta]\right]\right\}$$

# Polynomial Coordinates Ansatz

## Grid Generation via Orthogonal Curves

- **Ansatz for Polynomial Curves**

In[1]:= η[x_, r_, t_] = a0[r, t] + a1[r, t] * x + a2[r, t] * x^2 + a3[r, t] * x^3
      (*+a4[r,t]*x^4+a5[r,t]*x^5*);

First derivative to determine slope

In[2]:= g[x_, r_, t_] = D[η[x, r, t], x];

Second derivative

In[3]:= h[x_, r_, t_] = D[η[x, r, t], x, x];

- **Rotation Matrix to Rotate Coordinate System**

Since infinite slope cannot be requiremeent for linear equation system, the whole coorindate system gets rotated by 45 degree. So the slopes are + 1 and - 1.

In[4]:= A[α_] = {{Cos[α], -Sin[α]}, {Sin[α], Cos[α]}};

In[5]:= coeff = Table[
     P1 = A[Pi / 4].{0.1 + n * 0.13, 0};
     P2 = A[Pi / 4].{0, 0.1 + n * 0.1};
     Solve[{η[P1[[1]], r, t] == P1[[2]], η[P2[[1]], r, t] == P2[[2]],
        g[P1[[1]], r, t] == -1, g[P2[[1]], r, t] == 1(*,f[0,r,t]==0.1+n*0.12,g[0,r,t]==0*)},
       {a0[r, t], a1[r, t], a2[r, t], a3[r, t](*,a4[r,t],a5[r,t]*)}], {n, 0, 10}]
     (* generation of polynomials intersecting with two equidistanced points at x-
      and y-axis *)

Out[5]= {{{a0[r, t] → 0.106066, a1[r, t] → 0., a2[r, t] → -7.07107, a3[r, t] → 0.}},
  {{a0[r, t] → 0.226564, a1[r, t] → 0.173909, a2[r, t] → -3.24084, a3[r, t] → -1.5093}},
  {{a0[r, t] → 0.346169, a1[r, t] → 0.226146, a2[r, t] → -2.08962, a3[r, t] → -0.834794}},
  {{a0[r, t] → 0.465575, a1[r, t] → 0.251258, a2[r, t] → -1.54026, a3[r, t] → -0.510661}},
  {{a0[r, t] → 0.584904, a1[r, t] → 0.266012, a2[r, t] → -1.21921, a3[r, t] → -0.341655}},
  {{a0[r, t] → 0.704197, a1[r, t] → 0.27572, a2[r, t] → -1.00877, a3[r, t] → -0.243865}},
  {{a0[r, t] → 0.823468, a1[r, t] → 0.282592, a2[r, t] → -0.860221, a3[r, t] → -0.182541}},
  {{a0[r, t] → 0.942727, a1[r, t] → 0.287713, a2[r, t] → -0.749781, a3[r, t] → -0.141659}},
  {{a0[r, t] → 1.06198, a1[r, t] → 0.291675, a2[r, t] → -0.664457, a3[r, t] → -0.113079}},
  {{a0[r, t] → 1.18122, a1[r, t] → 0.294833, a2[r, t] → -0.59656, a3[r, t] → -0.0923307}},
  {{a0[r, t] → 1.30046, a1[r, t] → 0.297408, a2[r, t] → -0.541248, a3[r, t] → -0.0768}}}

```
In[6]:= Rotate[
    Show[Plot[{Evaluate[η[x, r, t] /. {coeff}](*,Evaluate[f[x,r,t]/.{ergmore2}]*), Abs[x], 0},
       {x, -1.2, 1.2}, PlotRange → {-0.1, 1.5}, AspectRatio → Automatic, Background → White,
       Axes → None, PlotStyle → {{Black, Thick}, {Gray, Thick}, {Gray, Thick}}](*,
       Graphics[{Thick,Red,Arrow[{{0,0},{1,0}}],Thick,Blue,Arrow[{{0,0},{0,1}}],
         Thick,Black,Dashed,Arrow[{{0,0},1*{1,0}+1*{0,1}}]}]*)], -45 Degree]
```

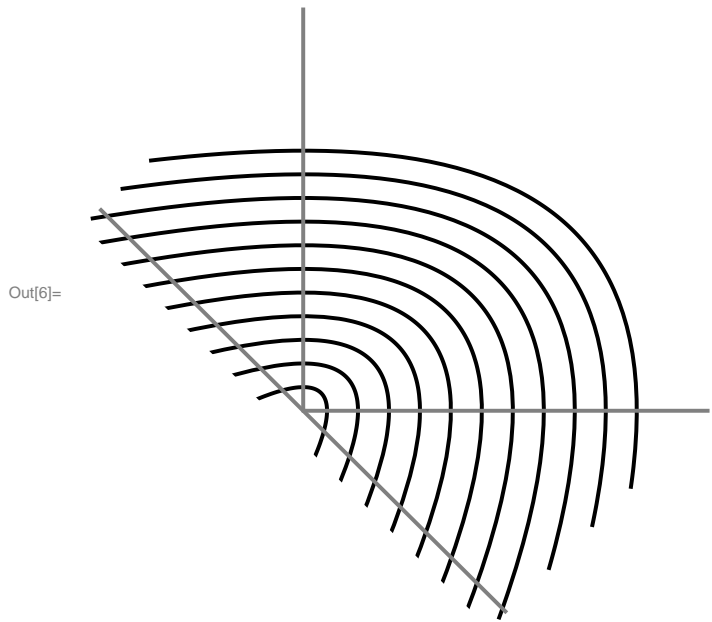

### ■ Orthogonal Curves

```
In[7]:= po1[x_] = η[x, r, t] /. {coeff[[1]]}; (* define polynomial functions *)
     po2[x_] = η[x, r, t] /. {coeff[[2]]};
     po3[x_] = η[x, r, t] /. {coeff[[3]]};
     po4[x_] = η[x, r, t] /. {coeff[[4]]};
     po5[x_] = η[x, r, t] /. {coeff[[5]]};
     po6[x_] = η[x, r, t] /. {coeff[[6]]};
     po7[x_] = η[x, r, t] /. {coeff[[7]]};
     po8[x_] = η[x, r, t] /. {coeff[[8]]};

In[15]:= ort4[x0_, x_] = -1 / (D[po4[x], x]) * (x - x0) + po7[x0];
     (* orthogonal curves to polynomial # *)
     ort7[x0_, x_] = -1 / (D[po7[x], x]) * (x - x0) + po7[x0];
     ort8[x0_, x_] = -1 / (D[po8[x], x]) * (x - x0) + po8[x0];

In[18]:= Table[
      intersectort8po7[m] =
       FindInstance[{Evaluate[ort8[0.25 + 0.1 * m, x]] == po7[x], x > 0.18, x < 0.8}, x, Reals][[1]][[
         1]] (* determine intersections *)
      ,
      {m,
       0,
       4}]

Out[18]= {x → 0.2382, x → 0.320295, x → 0.403032, x → 0.487458, x → 0.574101}
```

```
In[19]:= Table[
    intersectort7po8old[n] =
     FindInstance[{Evaluate[ort7[0.25 + 0.1 * n, x]] == po8[x], x > 0.18, x < 0.8}, x, Reals][[1]][[
      1]]
    ,
    {n,
     0,
     4}]
```

Out[19]= $\{x \to 0.279988, x \to 0.409555, x \to 0.530964, x \to 0.645, x \to 0.754393\}$

```
In[20]:= Show[Plot[{(*neu8[x],neu7[x],*)po7[x], po6[x], po5[x], po4[x],
     po3[x], po2[x], po1[x], Abs[x], 0}, {x, 0, 0.8}, PlotRange → {-0.05, 1.},
    AspectRatio → Automatic, Background → White, PlotStyle → {{Gray, Thick}}, Axes → False],
   Table[Plot[Evaluate[ort7[0.2 + 0.09 * m, x]], {x, 0.17, 0.8},
     AspectRatio → Automatic, Background → White, PlotRange → {0., 1.},
     PlotStyle → {Black, Dashed, Thick}, Axes → False], {m, 0, 8}]
   (*Table[Plot[Evaluate[y8[x/.Schnitty7mitneu8[n],x]],{x,0.2,0.8},AspectRatio→Automatic,
     Background→White,PlotRange→{0.4,1.2},PlotStyle→{Blue},Axes→False],{n,0,3}],*)
   ]
```

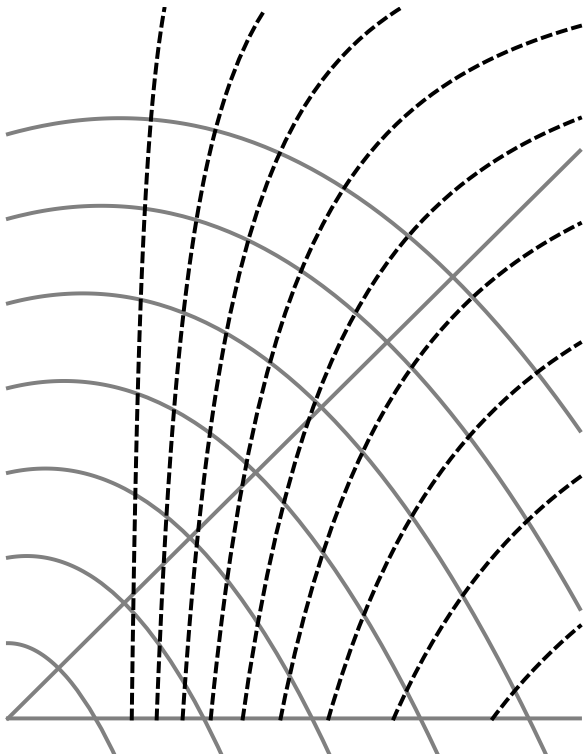

### ■ Normal Vectors

In[21]:= `slope7[x_] = -1 / (D[po7[x], x])`

Out[21]= $\left\{\left\{-\dfrac{1}{0.282592 - 1.72044\,x - 0.547624\,x^2}\right\}\right\}$

```
In[22]:= vecend[n_] = {0.25 + 0.1 * n, po7[0.25 + 0.1 * n][[1]][[1]]} +
    {0.25 + 0.1 * n + 1, po7[0.25 + 0.1 * n][[1]][[1]] + slope7[0.25 + 0.1 * n][[1]][[1]]}
```

$$\text{Out[22]}= \left\{1.5 + 0.2\,n,\; 1.64694 - \frac{1}{0.282592 - 1.72044\,(0.25 + 0.1\,n) - 0.547624\,(0.25 + 0.1\,n)^2} + \right.$$
$$\left. 0.565184\,(0.25 + 0.1\,n) - 1.72044\,(0.25 + 0.1\,n)^2 - 0.365083\,(0.25 + 0.1\,n)^3\right\}$$

```
In[23]:= slope8[x_] = -1 / (D[po8[x], x])
```

$$\text{Out[23]}= \left\{\left\{-\frac{1}{0.287713 - 1.49956\,x - 0.424976\,x^2}\right\}\right\}$$

```
In[24]:= Table[
    intersectort7po8[n] = x /. FindInstance[
        {Evaluate[ort7[0.25 + 0.1 * n, x]] == po8[x], x > 0.18, x < 0.8}, x, Reals][[1]][[1]]
    ,
    {n,
     0,
     4}]
```

Out[24]= {0.279988, 0.409555, 0.530964, 0.645, 0.754393}

```
In[25]:= Table[vecend8[m] =
    {intersectort7po8[m], po8[intersectort7po8[m]][[1]][[1]]} + {intersectort7po8[m] + 1,
      po8[intersectort7po8[m]][[1]][[1]] + slope8[intersectort7po8[m]][[1]][[1]]]}, {m, 0, 3}]
```

Out[25]= {{1.55998, 7.9665}, {1.81911, 4.36444}, {2.06193, 3.31738}, {2.29, 2.72453}}

```
In[26]:= Show[Plot[{po8[x], po7[x], po6[x], po5[x], Abs[x]}, {x, 0.1, 0.9}, PlotRange → {0.5, 1.3},
    AspectRatio → Automatic, Background → White, PlotStyle → {{Gray, Thick}}, Axes → False],
   Table[Plot[Evaluate[ort7[0.25 + 0.1 * m, x]], {x, 0.2, 0.9},
     AspectRatio → Automatic, Background → White, PlotRange → {0.5, 1.2},
     PlotStyle → {Gray, Dashed, Thick}, Axes → False], {m, 0, 3}],
   Table[Plot[Evaluate[ort8[x /. intersectort7po8old[n], x]], {x, 0.2, 0.8},
     AspectRatio → Automatic, Background → White, PlotRange → {0.5, 1.2},
     PlotStyle → {Gray, Dashed, Thick}, Axes → False], {n, 0, 3}],
   Table[Graphics[{Thick, Black, Arrow[{{0.25 + 0.1 * n, po7[0.25 + 0.1 * n][[1]][[1]]},
       {0.25 + 0.1 * n + 0.7 / Norm[vecend[n]]^2 * 1, po7[0.25 + 0.1 * n][[1]][[1]] +
         0.7 / Norm[vecend[n]]^2 * slope7[0.25 + 0.1 * n][[1]][[1]]}}]}], {n, 0, 3}],
   Table[Graphics[{Thick, Black, Arrow[{{intersectort7po8[n],
        po8[intersectort7po8[n]][[1]][[1]]},
       {intersectort7po8[n] + 0.7 / Norm[vecend8[n]]^2 * 1, po8[intersectort7po8[n]][[1]][[1]] +
         0.7 / Norm[vecend8[n]]^2 * slope8[intersectort7po8[n]][[1]][[1]]}}]}], {n, 0, 3}]
  ]
```

Out[26]=

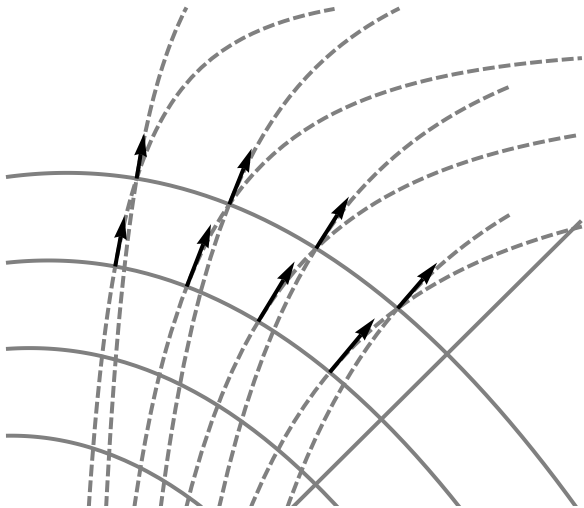

# Polar Non-Orthogonal Grid Ansatz

Generation of Quasi-Polar Grid in 2D

## Ansatz for Quasi-Polar Coordinates

### Coordinate Transformation with Product Ansatz

We define a coordinate transformation with product ansatz, which is reasonable for any kind of coordinate system.

```
In[1]:= x = X[ξ, η];
       y = Y[ξ, η];

In[3]:= X[ξ_, η_] = a[ξ] * α[η];
       Y[ξ_, η_] = b[ξ] * β[η];

In[5]:= α[η_] = (1 + alpha1 * η + alpha2 * η^2 + alpha3 * η^3) * Cos[η];
       β[η_] = (1 + beta1 * η + beta2 * η^2 + beta3 * η^3) * Sin[η];
       a[ξ_] = (a1 * ξ + a2 * ξ^2) ;
       b[ξ_] = (b1 * ξ + b2 * ξ^2) ;
```

We calculate the Jacobian of this transformation to get the base vectors in the new coordinate system.

```
In[9]:= lij = {{D[x, ξ], D[x, η]}, {D[y, ξ], D[y, η]}};

In[10]:= lijtransposed := Transpose[lij]
```

### Covariant Components of Metric Tensor

For trivial signatures, the metric tensor is simply given by the Jacobian times its transpositon. The off diagonal elements contain the information about the non-orthogonality (angles), the diagonal elements are measures of lenght in the coordinate directions.

```
In[11]:= gij = lijtransposed.lij;

In[12]:= g11[ξ_, η_] = gij[[1, 1]];
       g12[ξ_, η_] = gij[[1, 2]];
       g21[ξ_, η_] = gij[[2, 1]];
       g22[ξ_, η_] = gij[[2, 2]];
```

### Covariant Base Vectors

The base vectors yield

```
In[16]:= e_ξ[ξ_, η_] = lijtransposed[[1]]
```

Out[16]= $\{(1 + \text{alpha1}\,\eta + \text{alpha2}\,\eta^2 + \text{alpha3}\,\eta^3)\,(a1 + 2\,a2\,\xi)\,\text{Cos}[\eta],$
$(1 + \text{beta1}\,\eta + \text{beta2}\,\eta^2 + \text{beta3}\,\eta^3)\,(b1 + 2\,b2\,\xi)\,\text{Sin}[\eta]\}$

In[17]:= $\mathbf{e}_\eta[\xi\_, \eta\_] = \mathtt{lijtransposed}[[2]]$

Out[17]= $\{(\text{alpha1} + 2\,\text{alpha2}\,\eta + 3\,\text{alpha3}\,\eta^2)\,(\text{a1}\,\xi + \text{a2}\,\xi^2)\,\cos[\eta] -$
$\quad(1 + \text{alpha1}\,\eta + \text{alpha2}\,\eta^2 + \text{alpha3}\,\eta^3)\,(\text{a1}\,\xi + \text{a2}\,\xi^2)\,\sin[\eta],$
$\quad(1 + \text{beta1}\,\eta + \text{beta2}\,\eta^2 + \text{beta3}\,\eta^3)\,(\text{b1}\,\xi + \text{b2}\,\xi^2)\,\cos[\eta] +$
$\quad(\text{beta1} + 2\,\text{beta2}\,\eta + 3\,\text{beta3}\,\eta^2)\,(\text{b1}\,\xi + \text{b2}\,\xi^2)\,\sin[\eta]\}$

## PDEs + Boundary Conditions

### ■ Controling the Grid

In order to control the shape of the grid by a manageable number of parameters and for reasons of symmetry we impose geometrically motivated boundary conditions. Since some of them are either redundant or exclusive, the following list is somewhat heuristic.

In[18]:= ```
ESSol = Solve[{
    (*g12[ξ,η]==0,*) (* orthogonality *)
    e_ξ[ξ, 0][[1]] == 1, (* scaling *)
    e_ξ[ξ, 0][[2]] == 0, (* in "x"-axis *)
    e_ξ[ξ, Pi / 2][[1]] == 0, (* in "y"-axis *)
    e_ξ[ξ, Pi / 2][[2]] == 1, (* scaling *)
    e_η[ξ, 0][[1]] == 0, (* normal to "x"-axis *)
    e_η[ξ, 0][[2]] == ξ, (* scaling *)
    e_η[ξ, Pi / 2][[1]] == -ξ, (* orientation *)
    e_η[ξ, Pi / 2][[2]] == 0, (* normal to "y"-axis *)
    X[0, 0] == 0, (* set origin *)
    Y[0, 0] == 0, (* set origin *)
    X[ξ, 0] == X[ξ, 2 * Pi], (* periodicity *)
    Y[ξ, 0] == Y[ξ, 2 * Pi], (* periodicity *)
    X[ξ, 0] == ξ * X[1, 0] (* scaling *)
    },
    {alpha1, alpha2, beta1, beta2, , alpha3, beta3, a1, a2}]
```

Solve::svars : Equations may not give solutions for all "solve" variables. >>

Out[18]= $\left\{\left\{\text{beta1} \to \dfrac{\text{beta3}\,\pi^3 - 16\,\text{b2}\,\xi + \text{b2}\,\text{beta3}\,\pi^3\,\xi}{4\,\pi\,(1 + \text{b2}\,\xi)},\right.\right.$
$\left.\text{beta2} \to \dfrac{-\text{beta3}\,\pi^3 + 4\,\text{b2}\,\xi - \text{b2}\,\text{beta3}\,\pi^3\,\xi}{\pi^2\,(1 + \text{b2}\,\xi)},\,\text{alpha2} \to 0,\,\text{alpha1} \to 0,\,\text{alpha3} \to 0,\,\text{a1} \to 1,\,\text{a2} \to 0\right\}\right\}$

In[19]:= $\mathtt{Xnew}[\xi\_, \eta\_] = \mathtt{Evaluate}[X[\xi, \eta]\ /.\ \mathtt{ESSol}[[1]]]\ //\ \mathtt{Simplify}$
$\mathtt{Ynew}[\xi\_, \eta\_] = \mathtt{Evaluate}[Y[\xi, \eta]\ /.\ \mathtt{ESSol}[[1]]]\ //\ \mathtt{Simplify}$

Out[19]= $\xi\,\cos[\eta]$

Out[20]= $\dfrac{1}{4\,\pi^2\,(1 + \text{b2}\,\xi)}\,\xi\,(\text{b1} + \text{b2}\,\xi)$
$\left(-16\,\text{b2}\,\pi\,\eta\,\xi + 16\,\text{b2}\,\eta^2\,\xi + \text{beta3}\,\pi^4\,\eta\,(1 + \text{b2}\,\xi) - 4\,\text{beta3}\,\pi^3\,\eta^2\,(1 + \text{b2}\,\xi) + 4\,\pi^2\,(1 + \text{beta3}\,\eta^3)\,(1 + \text{b2}\,\xi)\right)$
$\sin[\eta]$

### ■ Plots

This example for a quasi-polar mesh is being controled by three parameters.

```
In[21]:= Manipulate[
          ParametricPlot[{ξ Cos[η], 1/(4 π² (1 + b2 ξ)) ξ (b1 + b2 ξ) (-16 b2 π η ξ + 16 b2 η² ξ + beta3 π⁴ η (1 + b2 ξ) -
              4 beta3 π³ η² (1 + b2 ξ) + 4 π² (1 + beta3 η³) (1 + b2 ξ)) Sin[η]},
           {ξ, 0, 1}, {η, 0, Pi/2}, AspectRatio → 1, PlotRange → {0, 1}, Axes → None,
           BoundaryStyle → {RGBColor[.2, .2, .2], Thick},
           MeshStyle → {{RGBColor[.2, .2, .2], Thick}}, MeshFunctions → Automatic,
           MeshShading -> {{White, White}, {White, White}}, Background → White(*,LabelStyle→*)],
          {{b1, 1}, -1, 1}, {{b2, 0}, -1, 1}, {{beta3, 0}, -1, 1}]
```

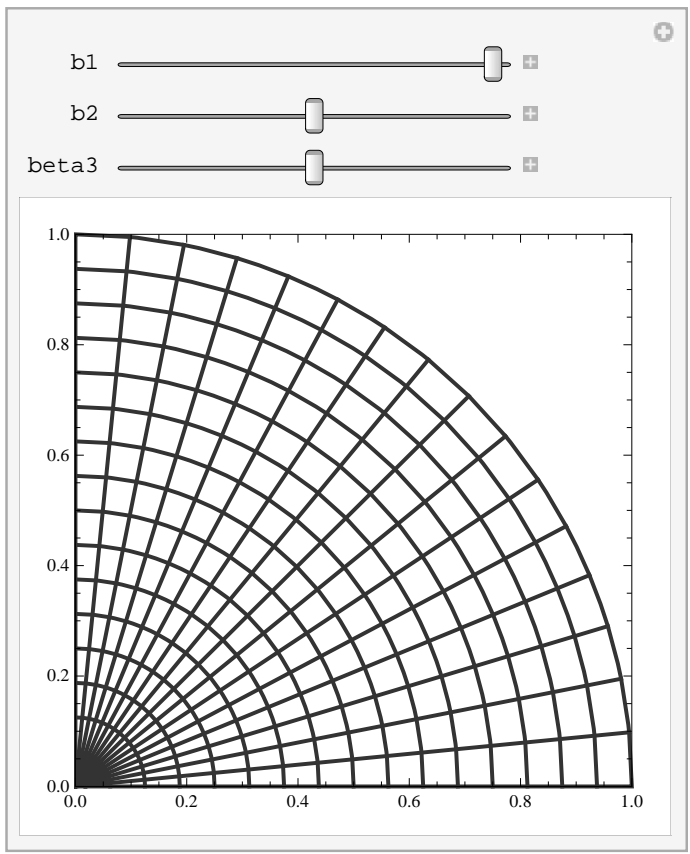

## Geometrical Characteristics of Toy Models

Determination of maximal relative volumes respectively off-diagonal metric elements.

```
In[22]:= lijnew = {{D[Xnew[ξ, η], ξ], D[Xnew[ξ, η], η]}, {D[Ynew[ξ, η], ξ], D[Ynew[ξ, η], η]}};

In[23]:= gijnew = Transpose[lijnew].lijnew;

In[24]:= grida = {b1 → 1, b2 → 0, beta3 → 0};
        gridb = {b1 → 0.8, b2 → 0, beta3 → 0};
        gridc = {b1 → 0.8, b2 → 0, beta3 → 0.5};
        gridd = {b1 → 0.8, b2 → 0.2, beta3 → 0.5};
```

```
In[28]:= deta[ξ_, η_] = Evaluate[Det[gijnew] /. grida];
        detb[ξ_, η_] = Evaluate[Det[gijnew] /. gridb];
        detc[ξ_, η_] = Evaluate[Det[gijnew] /. gridc];
        detd[ξ_, η_] = Evaluate[Det[gijnew] /. gridd];

In[32]:= offa[ξ_, η_] = Evaluate[gijnew[[1, 2]] /. grida];
        offb[ξ_, η_] = Evaluate[gijnew[[1, 2]] /. gridb];
        offc[ξ_, η_] = Evaluate[gijnew[[1, 2]] /. gridc];
        offd[ξ_, η_] = Evaluate[gijnew[[1, 2]] /. gridd];
```

### ■ Maxima

```
In[36]:= FindMaximum[{deta[ξ, η] / ξ^2, ξ ≥ 0 && η ≥ 0 && ξ ≤ 1 && η ≤ Pi / 4}, {ξ, η}]

Out[36]= {1., {ξ → 0.899989, η → 0.706847}}

In[37]:= FindMaximum[{Abs[offa[ξ, η] / ξ], ξ ≥ 0 && η ≥ 0 && ξ ≤ 1 && η ≤ Pi / 4}, {ξ, η}]

Out[37]= {0., {ξ → 0., η → 0.}}

In[38]:= FindMaximum[{detb[ξ, η] / ξ^2, ξ ≥ 0 && η ≥ 0 && ξ ≤ .8 && η ≤ Pi / 4}, {ξ, η}]

Out[38]= {0.64, {ξ → 0.719994, η → 0.706852}}

In[39]:= FindMaximum[{Abs[offb[ξ, η] / ξ], ξ ≥ 0 && η ≥ 0 && ξ ≤ 1 && η ≤ Pi / 4}, {ξ, η}]

Out[39]= {0.179999, {ξ → 0.851429, η → 0.783879}}

In[40]:= FindMaximum[{detc[ξ, η] / ξ^2, ξ ≥ 0 && η ≥ 0 && ξ ≤ 1 && η ≤ Pi / 4}, {ξ, η}]

Out[40]= {1.19161, {ξ → 0.860374, η → 0.327134}}

In[41]:= FindMaximum[{Abs[offc[ξ, η] / ξ], ξ ≥ 0 && η ≥ 0 && ξ ≤ 1 && η ≤ Pi / 4}, {ξ, η}]

Out[41]= {0.128795, {ξ → 0.899702, η → 0.785398}}

In[42]:= FindMaximum[{detd[ξ, η] / ξ^2, ξ ≥ 0 && η ≥ 0 && ξ ≤ 1 && η ≤ Pi / 4}, {ξ, η}]

Out[42]= {1.62084, {ξ → 1., η → 0.293574}}

In[43]:= FindMaximum[{Abs[offd[ξ, η] / ξ], ξ ≥ 0 && η ≥ 0 && ξ ≤ 1 && η ≤ Pi / 4}, {ξ, η}]

Out[43]= {0.248654, {ξ → 0.999999, η → 0.424638}}
```

### ■ Minima

```
In[44]:= FindMinimum[{deta[ξ, η] / ξ^2, ξ ≥ 0 && η ≥ 0 && ξ ≤ 1 && η ≤ Pi / 4}, {ξ, η}]

Out[44]= {1., {ξ → 0.899989, η → 0.706847}}

In[45]:= FindMinimum[{Abs[offa[ξ, η] / ξ], ξ ≥ 0 && η ≥ 0 && ξ ≤ 1 && η ≤ Pi / 4}, {ξ, η}]

Out[45]= {0., {ξ → 0., η → 0.}}

In[46]:= FindMinimum[{detb[ξ, η] / ξ^2, ξ ≥ 0 && η ≥ 0 && ξ ≤ .8 && η ≤ Pi / 4}, {ξ, η}]

Out[46]= {0.64, {ξ → 0.719994, η → 0.706852}}
```

In[47]:= `FindMinimum[{Abs[offb[ξ, η] / ξ], ξ ≥ 0 && η ≥ 0 && ξ ≤ 1 && η ≤ Pi / 4}, {ξ, η}]`

Out[47]= $\{8.29014 \times 10^{-8}, \{\xi \to 0.772027, \eta \to 2.30282 \times 10^{-7}\}\}$

In[48]:= `FindMinimum[{detc[ξ, η] / ξ^2, ξ ≥ 0 && η ≥ 0 && ξ ≤ 1 && η ≤ Pi / 4}, {ξ, η}]`

Out[48]= $\{0.757631, \{\xi \to 0.897993, \eta \to 0.785396\}\}$

In[49]:= `FindMinimum[{Abs[offc[ξ, η] / ξ], ξ ≥ 0 && η ≥ 0 && ξ ≤ 1 && η ≤ Pi / 4}, {ξ, η}]`

Out[49]= $\{2.46448 \times 10^{-8}, \{\xi \to 0.734207, \eta \to 6.84577 \times 10^{-8}\}\}$

In[50]:= `FindMinimum[{detd[ξ, η] / ξ^2, ξ ≥ 0 && η ≥ 0 && ξ ≤ 1 && η ≤ Pi / 4}, {ξ, η}]`

Out[50]= $\{0.75763, \{\xi \to 5.89009 \times 10^{-7}, \eta \to 0.785398\}\}$

In[53]:= `FindMinimum[{Abs[offd[ξ, η] / ξ], ξ ≥ 0 && η ≥ 0 && ξ ≤ 1 && η ≤ Pi / 4}, {ξ, η}]`

FindMinimum::eit :
 The algorithm does not converge to the tolerance of 4.806217383937354`*^-6
    in 500 iterations. The best estimated solution, with
    feasibility residual, KKT residual, or complementary residual
    of {0.000658608, 0.100761, 0.000246563}, is returned. ≫

Out[53]= $\{3.42416 \times 10^{-6}, \{\xi \to 0.709494, \eta \to 0.716048\}\}$



% ---------------------------------------------------------------------------- %
% THEBIBLIOGRAPHY ------------------------------------------------------------ %
% ---------------------------------------------------------------------------- %

\bibliographystyle{abbrv.bst}
\bibliography{dipl}

\newpage
\thispagestyle{empty}
\section*{Abstract}

We study conservative numerical methods for implicit radiation hydrodynamics
(RHD) in general non-steady coordinates in 2D and 3D. 

We discuss fundamental 
mathematics and physics of RHD with special focus on
differential geometric consistency and study numerical methods for nonlinear 
conservation laws. 
We apply Vinokurs theorem to obtain the strong conservation form
for conservation laws in general curvilinear coordinates. 
Differential geometric derivations in this context lead to a reformulation of 
artificial viscosity in such general coordinates. 

A non-static 
gravitation equation in order to avoid solving the Poisson equation 
for the gravitational potential is suggested. 

It is shown that there is no coordinate system in 2D and 3D that 
contains the polar respectively spherical coordinates and allows 
multi-dimensionally adaptiveness. We suggest an ansatz to generate 
non-orthogonal non-steady coordinates that contain these 
orthogonal reference frames.

\section*{Zusammenfassung}

Wir untersuchen konservative, implizite numerische Methoden im Rahmen der 
Strahlungshydrodynamik (SHD) in allgemeinen Koordinaten. 

Grundlegende mathematische und physikalische Konzepte werden mit speziellem 
Fokus auf geometrische Konsistenz diskutiert und numerische Methoden 
f\"ur nichtlineare Erhaltungss\"atze untersucht. Mit Hilfe des Vinokurschen 
Theorems l\"asst sich die konservative Form der Erhaltungss\"atze auf 
allgemeinen Koordinaten formulieren. Differentialgeometrische \"Uberlegungen 
in diesem Zusammenhang f\"uhren zu einer Umformulierung der k\"unstlichen 
Viskosit\"at f\"ur solch allgemeine Koordinatensysteme. 

Wir schlagen eine nicht-statische Gleichung f\"ur die Eigengravitation vor, um 
die L\"osung der Poisson-Gleichung f\"ur das Gravitationspotential zu umgehen. 

Es wird gezeigt, dass kein Koordinatensystem in 2D und 3D existiert, das 
die Polar- bzw. Kugelkoordinaten beinhaltet und mehrdimensionale Adaptivit\"at 
des Gitters erlaubt. Wir schlagen einen Ansatz f\"ur nicht-orthogonale 
zeitabh\"angige Koordinaten vor, der diese Referenzgitter enth\"alt.

\newpage 
\thispagestyle{empty}

 \begin{center}
\Large
\textbf{\sf Curriculum Vitae}
 \\
 \vspace*{2cm}
\end{center}

\begin{tabular}{l l}
Name: & \textbf{Harald H\"OLLER Bakk.rer.nat} \\
Datum und Ort der Geburt: & \textbf{04. M\"arz 1983} in Villach, \"Osterreich \\
Eltern: & Anna H\"OLLER und Engelbert Viktor H\"OLLER \\
& \\
Adresse: & Alszeile 119, 1170 Wien \\
E-Mail: &  \texttt{harald.hoeller@univie.ac.at} \\
Homepage: & \texttt{http://homepage.univie.ac.at/harald.hoeller/}
\end{tabular}
\\[20pt]

\section*{Stationen}
\begin{tabular}{r p{11cm}}
{\bf 2009, 2010} & Lektor an der Fakult\"at f\"ur Physik - Lehrveranstaltungen aus
Theoretischer Physik und Mathematischer Methoden.  \\
{\bf 08/2007-01/2010}  & Projektmitarbeiter bei \emph{eLearnPhysik} - E-Learning an 
der Fakult\"at f\"ur Physik. Schwerpunkte waren die redaktionelle und technische 
Betreuung des Wiki der Fakult\"at f\"ur Physik, 
Content und Usability Fragen, sowie E-Learning im Bereich Theoretische Physik 
und der Einsatz von Computeralgebra in der Lehre. \\ 
{\bf 2007, 2008} & eTutor an der Fakult\"at f\"ur Physik - Lehrveranstaltungen aus
Experimentalphysik sowie Theoretischer Physik und Mathematischer Methoden.  \\
{\bf 2007, 2008, 2009} & Seminare und Workshops im Rahmen der KinderUni Wien, des 
Wissenschaftsclub f\"ur Jugendliche sowie der StauneLaune Forschungswochen. \\
{\bf Sommer 2005, 2006} & Ferialpraktikant bei Infineon Technologies Austria AG 
am Standort Villach im Bereich Forschung und Entwicklung. 
\end{tabular}

\section*{Schule und Studium}
\begin{tabular}{r p{11cm}}
{\bf seit 2005} & Magisterstudium Astronomie. \\
{\bf seit 2003} & Diplomstudium Physik (Zweitstudium) an der Universit\"at Wien. \\
{\bf 2002, 2005} & Zuerkennung von Leistungsstipendium. \\
{\bf 2005} & Abschluss des Bakkalaureatsstudiums Astronomie mit ausgezeichnetem Erfolg. \\
{\bf 2005} & Unterstellung in neuen Studienplan; Bakkalaureatsstudium Astronomie. \\
{\bf 2001-2005} & Diplomstudium Astronomie an der Universit\"at Wien. \\
{\bf 2001} & Matura am BORG Klagenfurt mit ausgezeichnetem Erfolg. \\
\end{tabular}
 
\thispagestyle{empty}
\section*{Publikation}
\begin{itemize}
\item 
{\bf Wiki Based Teaching and Learning Scenarios at the University of Vienna}, 
Harald H\"oller und Peter Reisinger, Vortrag auf der \emph{World Conference on 
Educational Multimedia, Hypermedia \& Telecommunications} 
(ED-MEDIA 30.06.-04.07.08, Technische Universit\"at Wien), in den Proceedings 
erschienen. 
\end{itemize}
\thispagestyle{empty}

\section*{Eingeladene Vortr\"age}
\begin{itemize}
\item {\bf Hearing zum Finale des \emph{Medida Prix 2009}}, 
Franz Embacher, Harald H\"oller, Christian Primetshofer, 
Peter Reisinger, Brigitte Wolny - Vorstellung des Projekts 
\emph{eLearnPhysik}, 15.09.2009, Berlin. 
\item {\bf Phaidra, Wiki und die Content-Strategie der Fakult\"at f\"ur Physik}, 
Harald H\"oller, Peter Reisinger im Rahmen des \emph{Phaidra Day an der Fakult\"at 
f\"ur Physik} am 01.10.2009 .
\item {\bf eLearnPhysik}, Franz Embacher, Harald H\"oller, Christian Primetshofer, 
Peter Reisinger - Vorstellung des Projekts im Rahmen der \emph{Friday Lectures} am 
Projektzentrum Lehrentwicklung am 07.12.2007.
\end{itemize}

\end{document}